\documentclass[11pt,preprint,tightenlines,
showpacs,preprintnumbers,amsmath,amssymb,
a4paper,nofootinbib]{revtex4-2}

\usepackage{multirow}

\usepackage{nicefrac}
\usepackage{natbib}
\usepackage{amssymb,amsbsy,amsmath,amsfonts}
\usepackage{graphicx}% Include figure files
\usepackage{epsf,epsfig,float,latexsym,amsthm,fancyhdr,rotating}
\usepackage{graphics,psfrag,longtable}
\usepackage{slashed}
\usepackage[colorlinks,citecolor=blue,linktoc=all,linkcolor=cyan,urlcolor=blue]{hyperref}
\usepackage{upgreek}
 
 \usepackage{setspace}

\def\hpm{\hphantom{-}}
\def\beq{\begin{equation}}
\def\eeq{\end{equation}}
\def\bea{\begin{eqnarray}}
\def\eea{\end{eqnarray}}
\def\beqa{\begin{equation}\begin{array}{l}}
\def\eeqa{\end{array}\end{equation}}
\def\eqlab#1{\label{eq:#1}}

\def\seclab#1{\label{sec:#1}}
\def\eref#1{(\ref{eq:#1})}
\def\Eqref#1{Eq.~(\ref{eq:#1})}

\def\barr{\left(\begin{array}{c}}
\def\earr{\end{array}\right)}
\def\bmat{\left(\begin{array}{cc}}
\def\emat{\end{array}\right)}
\def\al{\alpha}

\def\ga{\gamma} 
\def\de{\delta} \def\De{\Delta}
  \def\eps{\epsilon}

\def\si{\sigma}

\def\nn{\nonumber}
\def\dd{\mathrm{d}}

\DeclareMathOperator\arctanh{arctanh}
\DeclareMathOperator\im{Im}
\def\3d{3-D}

\def\ol#1{\overline{#1}}

\begin{document}
\title {Forward doubly-virtual 
Compton scattering off the nucleon in chiral perturbation theory: II. Spin polarizabilities and moments of polarized structure functions}
\author{Jose Manuel Alarc\'on}
\affiliation{Departamento de F\'isica Te\'orica \& IPARCOS, Universidad Complutense de Madrid, 28040
Madrid, Spain}
\author{Franziska Hagelstein}
\affiliation{Albert Einstein Center for Fundamental Physics, Institute for Theoretical Physics, University of Bern, Sidlerstrasse 5, CH-3012 Bern, Switzerland}
\author{Vadim Lensky}
\affiliation{Institut f\"ur Kernphysik \&  Cluster of Excellence PRISMA,
 Johannes Gutenberg-Universit\"at  Mainz,  D-55128 Mainz, Germany}
\author{Vladimir Pascalutsa}
\affiliation{Institut f\"ur Kernphysik \&  Cluster of Excellence PRISMA,
 Johannes Gutenberg-Universit\"at  Mainz,  D-55128 Mainz, Germany}
 \email{vladipas@kph.uni-mainz.de}

\begin{abstract}
We examine the polarized doubly-virtual 
Compton scattering (VVCS) off the nucleon using  chiral perturbation
theory ($\chi$PT). The polarized VVCS contains a wealth of information on the spin structure of the nucleon which is 
relevant to the calculation
of the two-photon-exchange effects in atomic spectroscopy and electron scattering. We report on a complete next-to-leading-order (NLO)
calculation of the polarized VVCS amplitudes $S_1(\nu, Q^2)$
and $S_2(\nu, Q^2)$, and the corresponding polarized spin structure functions
$g_1(x, Q^2)$ and $g_2(x,Q^2)$. Our results for the moments of polarized structure functions, partially related to different spin polarizabilities, are compared to other theoretical predictions and ``data-driven'' evaluations, as well as to the recent Jefferson Lab measurements. By expanding the results in powers of the inverse nucleon  mass, we reproduce the known ``heavy-baryon'' expressions. This serves as a check of our calculation, as well as  demonstrates the differences between the manifestly Lorentz-invariant baryon $\chi$PT (B$\chi$PT)
and heavy-baryon (HB$\chi$PT) frameworks.
\end{abstract}
%\pacs{}
\date{\today}
\maketitle
\preprint{MITP/20-033}
\newpage
\tableofcontents

\newpage
\section{Introduction}

\begin{figure}[t]
\centering
       \includegraphics[width=5.5cm]{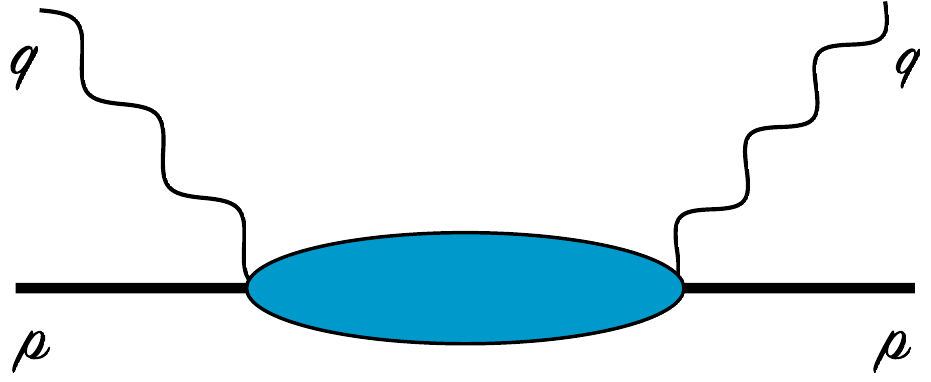}
\caption{The forward Compton scattering, or VVCS, in case of virtual photons, $q^2=-Q^2$. \label{fig:CSgeneric}}
\end{figure}

In the studies of nucleon structure,
the forward doubly-virtual Compton scattering (VVCS) amplitude,  Fig.~\ref{fig:CSgeneric}, is playing a central role (see, e.g., Refs.~\cite{Drechsel:2002ar,Kuhn:2008sy,Hagelstein:2015egb,Pasquini:2018wbl} for reviews). Traditionally, its general properties, such as unitarity, analyticity and crossing, are used
to establish various useful sum rules for the nucleon magnetic moment (Gerasimov--Drell--Hearn \cite{Gerasimov:1965et,Drell:1966jv} and Schwinger sum rules \cite{Schwinger:1975ti,Schwinger:1975tix,Schwinger:1975uq}) and polarizabilities (e.g., Baldin \cite{Baldin:1960} and Gell-Mann--Goldberger--Thirring sum rules \cite{GellMann:1954db}). More recently, the interest in nucleon VVCS has been renewed in connection
with precision atomic spectroscopy, where
this amplitude enters in the form of two-photon exchange (TPE) corrections. As the TPE corrections in atomic domain are dominated by low-energy VVCS, it makes sense to calculate them systematically using chiral perturbation theory ($\chi$PT), which is a low-energy
effective-field theory of the Standard Model.

In this paper, we present a state-of-the-art $\chi$PT calculation of the polarized nucleon VVCS, 
relevant to TPE corrections to hyperfine structure
of hydrogen and muonic hydrogen. This will extend 
the leading-order $\chi$PT  
evaluation of the TPE effects in hyperfine splittings~\cite{Pineda:2003,Peset:2014jxa,Peset:2016wjq,Hagelstein:2015lph,Hagelstein:2017cbl,Hagelstein:2018bdi}.
Here, we however, 
do not discuss the TPE evaluation, but rather focus
on testing the $\chi$PT framework against the
available empirical information on low-energy spin structure of the nucleon. 

It is
especially interesting to confront the $\chi$PT predictions with the recent measurements coming from 
the ongoing ``Spin Physics Program'' at Jefferson Laboratory~\cite{Prok:2008ev,Dutz:2003mm,Amarian:2004yf,Amarian:2002ar,Amarian:2003jy,Deur:2004ti,Slifer:2009ik,Solvignon:2013yun,Deur:2019pew,Sulkosky:2019zmn}, with the
exception of a recent measurement of the deuteron spin polarizability by the CLAS Collaboration \cite{Adhikari:2017wox}, which does not
treat correctly complications
due to deuteron spin \cite{Lensky:2018vdq}.

Our present calculation is extending Ref.~\cite{Alarcon:2020wjg} to the case of polarized VVCS.
We therefore use a manifestly-covariant extension of SU(2) $\chi$PT  to the baryon sector called Baryon $\chi$PT (B$\chi$PT). 
First attempts to calculate
VVCS in the straightforward B$\chi$PT framework (rather than the heavy-baryon expansion or the ``infrared regularization'') were done by Bernard {\it et~al.}~\cite{Bernard:2012hb} and our group~\cite{Lensky:2014dda}.  The two works obtained somewhat different results, 
most notably for the proton spin polarizability $\de_{LT}$. Here we improve on \cite{Lensky:2014dda}
in three important aspects appreciable at finite $Q^2$: 1) inclusion of the
Coulomb-quadrupole $(C2)$ $N\to \Delta$ transition \cite{Pascalutsa:2005vq,Pascalutsa:2005ts}, 2)
correct inclusion of the elastic form-factor contributions to 
the integrals $\Gamma_1(Q^2)$, $I_1(Q^2)$ and $I_A(Q^2)$ (see Sections \ref{IAsec} and \ref{I1sec} for details), and 3) cancellations between different orders in the chiral prediction and their effect on the convergence of the effective-field-theory calculation, and thus, the error estimate. These improvements, however, do
not bring us closer to the results of~\cite{Bernard:2012hb}, and the source of discrepancies most likely lies in the different counting and  renormalization of the $\pi \De$-loop contributions.
Bernard {\it et~al.}~\cite{Bernard:2012hb} use the
so-called small-scale expansion \cite{Hemmert:1996xg} for the $\De(1232)$ contributions, whereas we are using the  $\de$-counting scheme
\cite{Pascalutsa:2003aa} (see also Ref.~\cite[Sec.~4]{Pascalutsa:2006up} for review).

This paper is organized as follows. In Sec.~\ref{Sec:VVCS_SF_relation}, we introduce the polarized VVCS amplitudes and their relations to spin structure functions. In Sec.~\ref{Sec:VVCS_LEX}, we introduce the spin polarizabilities appearing in the low-energy expansion (LEX) of the polarized VVCS amplitudes. In Sec.~\ref{Sec:VVCS_calc}, we briefly describe our $\chi$PT calculation, focusing mainly on the uncertainty estimate. In Sec.~\ref{Sec:Pol}, we show our predictions for the proton and neutron polarizabilities, as well as some interesting moments of their structure functions. In Sec.~\ref{Sec:Summary}, we summarize the results obtained herein, comment on the improvements done with respect to previous calculations, and give an outlook to future applications. In App.~\ref{CrossSections}, we discuss the structure functions, in particular, we define the longitudinal-transverse response function, discuss the $\Delta$-pole contribution, and give analytical results for the tree-level $\pi N$- and $\De$-production channels of the photoabsorption cross sections. In App.~\ref{App:PolarizabilitiesAll}, we give analytical expressions for the $\pi N$-loop and $\Delta$-exchange contributions to the central values and slopes of the polarizabilities and moments of structure functions at $Q^2=0$. The complete expressions, also for the $\pi \Delta$-loop contributions, can be found in the {\it Supplemented material}.

\begin{table}[b]
\caption{Parameters (fundamental and low-energy constants) \cite{Agashe:2014kda} at the order they appear first. The $\pi N\Delta$ coupling constant $h_A$ is fit to the experimental Delta width and the $\gamma^* N \Delta$ coupling constants $g_M$, $g_E$ and $g_C$ are taken from the pion photoproduction study of Ref.~\cite{Pascalutsa:2005vq}.\label{tab:constants}} 
\begin{tabular}{lp{0.3cm}l}
\hline\\
$\mathcal{O}(p^2)$&&$\alpha=1/(137.04)$, $M_N=M_p=938.27$ MeV\\
$\mathcal{O}(p^3)$&&$g_A=1.270$, $f_\pi=92.21$ MeV, $m_\pi=139.57$ MeV \\
$\mathcal{O}(p^4/\varDelta)$&&$M_\Delta=1232$ MeV, $h_A\equiv 2g_{\pi N \Delta}=2.85$, $g_M=2.97$, $g_E=-1.0$, $g_C=-2.6$\\\\
\hline
\end{tabular}
\end{table}

\section{Calculation of unpolarized VVCS at NLO} \label{Sec:ChEFT_for_VVCS}

\subsection{VVCS amplitudes and relations to structure functions} \label{Sec:VVCS_SF_relation}

The polarized part of forward VVCS can be described in terms of two
independent Lorentz-covariant and gauge-invariant tensor structures and two scalar amplitudes
\cite{Hagelstein:2015egb}:
\beq
\label{Eq:T-Rel}
T^{\mu\nu}(p,q) = -   \frac{1}{M_N}\gamma^{\mu \nu \al} q_\al \,S_1(\nu, Q^2)  -  
\frac{1}{M_N^2} \Big( \gamma^{\mu\nu} q^2 + q^\mu \gamma^{\nu\al} q_\al  -  q^\nu \gamma^{\mu\al}
q_\al \Big) S_2(\nu, Q^2).
\eeq
Here, $q$ and $p$ are the photon and nucleon four-momenta (cf.\ Fig.~\ref{fig:CSgeneric}), $\nu$ is the photon lab-frame energy, $Q^2=-q^2$ is the photon virtuality, and $\gamma^{\mu \nu}=\frac{1}{2}\left[\gamma^\mu,\gamma^\nu\right]$ and 
$\gamma^{\mu \nu \al}=\frac{1}{2} \left(\gamma^\mu\gamma^\nu \ga^\al-\ga^\al \ga^\nu \ga^\mu \right)$ are the usual Dirac matrices. Alternatively, one can use the following
laboratory-frame amplitudes:
\begin{subequations}
\bea
 g_{TT}(\nu, Q^2) &=& \frac{\nu}{M_N}\Big[ S_1(\nu,Q^2) - \frac{Q^2}{M_N\,\nu} S_2(\nu,Q^2)\Big],\label{Eq:GGT-S1S2}\\
  g_{LT}(\nu, Q^2) &=& \frac{Q}{M_N}\Big[ S_1(\nu,Q^2) + \frac{\nu}{M_N} S_2(\nu,Q^2)\Big],\label{Eq:gLT-S1S2}
\eea
\end{subequations}
introduced in Eq.~(\ref{Eq:T-Compt-definition}).
The optical theorem relates the absorptive parts of the forward VVCS amplitudes to the nucleon structure functions or the cross sections of virtual photoabsorption:
\begin{subequations}
\eqlab{VVCSunitarity}
\bea
\im S_1(\nu,Q^2) &=& \frac{4\pi^2 \alpha}{\nu} \, g_1(x,Q^2)\eqlab{ImS1}=
\frac{M_N  \nu K(\nu,Q^2) }{\nu^2+Q^2}\left[\frac{Q}{\nu}\sigma_{LT}(\nu,Q^2)  + \sigma_{TT}(\nu,Q^2)\right],\\
\im S_2(\nu,Q^2) & =&  \frac{4\pi^2 \alpha M_N}{\nu^2} \, g_2(x, Q^2)\eqlab{ImS2}= \frac{M_N^2  K(\nu,Q^2) }{\nu^2+Q^2}\left[\frac{\nu}{Q}\sigma_{LT}(\nu,Q^2)  - \sigma_{TT}(\nu,Q^2)\right],
\eea
\end{subequations}
with $\alpha$ the fine structure constant, and $K(\nu,Q^2)$ the photon flux factor.
Note that the photon flux factor in the optical theorem and the cross sections, measured in electroproduction processes, is a matter of convention and has to be chosen in both quantities consistently. In what follows, we use Gilman's flux factor: 
\beq
K(\nu,Q^2)\equiv \vert \vec{q}\, \vert=\sqrt{\nu^2+Q^2}.
\eeq
The helicity-difference photoabsorption cross section is defined as $\sigma_{TT}=\nicefrac12\, (\sigma_{1/2}-\sigma_{3/2})$, where the photons are transversely polarized, and the subscripts on the right-hand side indicate the total helicities of the $\gamma^\ast N$ states. The cross section $\sigma_{LT}$ corresponds to a simultaneous helicity
change of the photon and nucleon spin flip, such that the
total helicity is conserved. A detailed derivation of the connection between this response function and the photoabsorption cross sections can be found in App.~\ref{CrossSections}. The forward VVCS amplitudes satisfy dispersion relations derived from the general principles of analyticity and causality:\footnote{The dispersion relation for $\nu S_2(\nu,Q^2)$ is used because it is pole-free in the limit $Q^2 \rightarrow 0$, and then $\nu \rightarrow 0$, cf.\ \Eqref{S2Born}.}
\begin{subequations}
\eqlab{genDRs}
\bea 
S_1 ( \nu, Q^2) &=&  \frac{16\pi\al M_N}{Q^2} \int_{0}^1 
\!\dd x\, 
\frac{g_1 (x, Q^2)}{1 - x^2 (\nu/\nu_{\mathrm{el}})^2  - i 0^+} \eqlab{S1DR}\\
&=&\frac{2M_N}{\pi} \int_{\nu_{\mathrm{el}}}^\infty \!\dd \nu' \, 
\frac{\nu^{\prime\,2}\big[ \frac{Q}{\nu'}\si_{LT}+\si_{TT}\big] ( \nu', Q^2)}{\sqrt{\nu^{\prime\,2}+Q^2}(\nu^{\prime\,2} -\nu^2- i 0^+)}\nn,\\
\nu S_2 ( \nu, Q^2) &=& \frac{16\pi\al M_N^2}{Q^2} \int_{0}^1 \!\dd x\, 
\frac{g_2 (x, Q^2)}{1 - x^2 (\nu/\nu_{\mathrm{el}})^2  - i 0^+}\eqlab{nuS2}  \\
&=& \frac{2M_N^2}{\pi} \int_{\nu_{\mathrm{el}}}^\infty \!\dd \nu' \, 
\frac{ \nu^{\prime\,2}\big[ \frac{\nu'}{Q}\si_{LT}-\si_{TT}\big] ( \nu', Q^2)}{\sqrt{\nu^{\prime\,2}+Q^2}(\nu^{\prime\,2} -\nu^2- i 0^+)},\nn
\eea 
\end{subequations}
with $\nu_{\mathrm{el}}=Q^2/2M_N$ the elastic threshold. 

\subsection{Low-energy expansions and relations to polarizabilities}
\label{Sec:VVCS_LEX}

The VVCS amplitudes naturally split into nucleon-pole ($S_i^{\mathrm{pole}}$) and non-pole ($S_i^{\mathrm{nonpole}}$)
parts, or Born ($S_i^{\mathrm{Born}}$) and non-Born ($\ol S_i$) parts:
\beq
S_i = S_i^{\mathrm{pole}} + S_i^{\mathrm{nonpole}}
= S_i^{\mathrm{Born}} + \ol S_i.
\eeq 
The Born amplitudes are given uniquely in terms
of the nucleon  form factors \cite{Drechsel:2002ar}:
\begin{subequations}
\eqlab{T12Born}
\bea
S_1^{\mathrm{Born}}(\nu, Q^2) &=& \frac{2\pi \alpha}{M_N}
\bigg\{\frac{ Q^2G_M(Q^2)F_1(Q^2)}{\nu_\mathrm{el}^2-\nu^2}-F_2^2(Q^2)\bigg\}\,, \eqlab{S1Born}\\
S_2^{\mathrm{Born}}(\nu, Q^2)&=&
-\, \frac{2 \pi \alpha  \nu}{\nu_\mathrm{el}^2-\nu^2}G_M(Q^2) F_2(Q^2) \,.
\eqlab{S2Born}
\eea
\end{subequations}
The same is true for the nucleon-pole amplitudes, which are related to the Born amplitudes in the following way:
\begin{subequations}
\bea
S_1^{\mathrm{pole}}(\nu, Q^2) &=& S_1^{\mathrm{Born}}(\nu, Q^2) + \frac{2\pi\al}{M_N}F_2^2(Q^2),\label{eq:S1pole}\\
S_2^{\mathrm{pole}}(\nu, Q^2)&=& 
S_2^{\mathrm{Born}}(\nu, Q^2) . \eqlab{S2subpole}
\eea
\end{subequations}
Here, we used the elastic Dirac and Pauli form factors $F_1(Q^2)$ and $F_2(Q^2)$, related to the electric and magnetic Sachs form factors $G_E(Q^2)$ and $G_M(Q^2)$ through:
\begin{subequations}
\bea
F_1(Q^2) &=&\frac{G_E(Q^2)+\tau G_M(Q^2)}{1+\tau},\\
F_2(Q^2)&=&\frac{ G_M(Q^2)-G_E(Q^2)}{1+\tau},
\eea
\end{subequations}
where $\tau=Q^2/4M_N^2$. 

A low-energy expansion (LEX) of \Eqref{genDRs}, in combination with the unitarity relations given in \Eqref{VVCSunitarity}, establishes various sum rules relating the nucleon properties (electromagnetic moments, polarizabilities) to experimentally observable response functions \cite{Drechsel:2002ar,Hagelstein:2015egb}.
The leading terms in the LEX of the real Compton scattering (RCS) amplitudes are determined uniquely by charge, mass and anomalous magnetic moment, as the global properties of the nucleon. These lowest-order terms represent the celebrated low-energy theorem (LET) of Low, Gell-Mann and Goldberger \cite{Low:1954kd, GellMann:1954kc}.
The polarizabilities, related to the internal structure of the nucleon,
enter the LEX at higher orders. They make up the non-Born amplitudes, and can be related to the moments of inelastic structure functions.

The process of VVCS can be realized experimentally in electron-nucleon scattering, where a virtual photon is exchanged between the electron and the nucleon. This virtual photon acts as a probe whose resolution depends on its virtuality $Q^2$.
In this way, one can access the so-called generalized polarizabilities, which extend the notion of polarizabilities to the case of response to finite momentum transfer. The generalized forward spin polarizability $\gamma_0(Q^2)$ and the longitudinal-transverse polarizability $\delta_{LT}(Q^2)$ are
most naturally defined via the LEX of the non-Born part of the lab-frame VVCS amplitudes \cite{Drechsel:2002ar}:
\begin{subequations}
\eqlab{fgFunctions}
\bea
  \frac{1}{4\pi}\, g_{TT}^\mathrm{nonpole}(\nu, Q^2)&=& \frac{2 \al}{M_N^2}I_A(Q^2) \,\nu+\gamma_0(Q^2)  \nu^3 + \bar \gamma_0(Q^2)  \nu^5 +\dots \label{Eq:LEX-GGT}\\
  \frac{1}{4\pi}\, g_{LT}^\mathrm{nonpole}(\nu, Q^2)&=& \frac{2 \al}{M_N^2} I_3(Q^2)Q+ \delta_{LT}(Q^2) \nu^2 Q + \dots \label{Eq:LEX-gLT}.
\eea
\end{subequations}
Their definitions in terms of integrals over structure functions are postponed to Eqs.~(\ref{Eq:gamma0Q2}) and (\ref{Eq:deltaLTQ2}). Here, we only give the definition of the moment $I_3(Q^2)$:
\beq
I_3(Q^2)=\frac{M_N^2}{4\pi^2 \al} \int_{\nu_0}^\infty \dd \nu\, \frac{K(\nu,Q^2)}{\nu Q} \sigma_{LT}(\nu,Q^2) =\frac{2M_N^2}{Q^2}\int_0^{x_0} \dd x \, \left[g_1(x,Q^2) + g_2(x,Q^2) \right],
\eeq
which is related to the Schwinger sum rule in the real photon limit \cite{Schwinger:1975ti,Schwinger:1975tix,Schwinger:1975uq,HarunarRashid:1976qz,Hagelstein:2017obr}.
The LEX of the non-pole part of the covariant VVCS amplitudes can be described entirely in terms of moments of inelastic spin structure functions (up to $\mathcal{O}(\nu^4)$ \cite{Lensky:2017dlc}):
\begin{subequations}
\eqlab{LEXS12}
\bea
\frac{1}{4\pi}S_1^\mathrm{nonpole}(\nu,Q^2)&=&\frac{2\al}{M_N}I_1(Q^2)+\left\{\frac{2\al}{M_N Q^2}\left[I_A(Q^2)-I_1(Q^2)\right]+M_N \delta_{LT}(Q^2)\right\}\nu^2,\qquad\eqlab{LEXS1}\\
\frac{1}{4\pi}\nu S_2^\mathrm{nonpole}(\nu,Q^2)&=&2 \al I_2(Q^2)+\frac{2\al}{Q^2}\left[I_1(Q^2)-I_A(Q^2)\right] \nu^2.
\eea
\end{subequations}
 $I_1(Q^2)$ and $I_A(Q^2)$ are generalizations of the famous Gerasimov--Drell--Hearn (GDH) sum rule~\cite{Gerasimov:1965et,Drell:1966jv} from RCS to the case of virtual photons \cite{Drechsel:2002ar}. Their definitions are given in Eqs.~(\ref{Eq:IA-SumRule}) and (\ref{Eq:I1-SumRule}).
 $I_2(Q^2)$ is the well-known Burkhardt-Cottingham (BC) sum rule~\cite{Burkhardt:1970ti}:
\bea
\label{BC2}
I_2(Q^2)& \equiv & \frac{2M_N^2}{Q^2}\int_0^{x_0}\dd x\,g_2(x,\,Q^2) =
\frac{1}{4} \, F_2(Q^2) G_M(Q^2),
\eea
which can be written as a ``superconvergence sum rule'':
\beq
\label{BC}
\frac{Q^2}{16 \pi \al M_N^2}\lim_{\nu \rightarrow 0} \nu S_2(\nu,Q^2)=\int_{0}^{1}\dd x \, g_{2}\,(x,\,Q^2)  = 0.
\eeq
The latter is valid for any value of $Q^2$
provided that the integral converges for $x \rightarrow 0$. 
Combining \Eqref{genDRs} with the above LEXs of the VVCS amplitudes, we can relate $I_A(Q^2)$, $I_1(Q^2)$, $\gamma_0(Q^2)$ and $\delta_{LT}(Q^2)$ to the moments of inelastic structure functions, see Sec.~\ref{Sec:Pol}. It is important to note that only $\gamma_0(Q^2)$ and $\delta_{LT}(Q^2)$ are generalized polarizabilities. The relation of the inelastic moments $I_A(Q^2)$ and $I_1(Q^2)$ to polarizabilities will be discussed in details in Secs.~\ref{IAsec} and \ref{I1sec}. The difference between $\ol S_1(\nu,Q
^2)$ and $S_1^\mathrm{nonpole}(\nu,Q
^2)$, cf.\ Eq.~(\ref{eq:S1pole}), will be important in this context.

\subsection{Details on \boldmath{$\chi$PT} calculation and uncertainty estimate}
\label{Sec:VVCS_calc}

In this work, we calculated the NLO prediction of B$\chi$PT for the polarized non-Born VVCS amplitudes. This includes the leading pion-nucleon ($\pi N$) loops, see Ref.~\cite[Fig.~1]{Lensky:2014dda}, as well as the subleading tree-level Delta-exchange ($\Delta$-exchange), see Ref.~\cite[Fig.~2]{Alarcon:2020wjg}, and the pion-Delta ($\pi \Delta$) loops, see Ref.~\cite[Fig.~2]{Lensky:2014dda}. In the $\delta$-power-counting scheme \cite{Pascalutsa:2003aa}, the LO and NLO non-Born VVCS amplitudes and polarizabilities are of $\mathcal{O}(p^3)$ and $\mathcal{O}(p^4/\varDelta)$, respectively.\footnote{In the full Compton amplitude, there is a lower order contribution coming from the Born terms, leading to a shift in nomenclature by one order: the LO contribution referred to as the NLO contribution, etc., see e.g.~Ref.~\cite{Lensky:2009uv}.} The low-energy constants (LECs) are listed in Table \ref{tab:constants}, sorted by the order at which they appear in our calculation. At the given orders, there are no 
``new'' LECs that would need to be fitted from Compton processes. For more details on the B$\chi$PT formalism, we refer to Ref.~\cite{Alarcon:2020wjg}, where power counting, predictive orders (Sec.\ III A), and the renormalization procedure (Sec.\ III B) are discussed.

A few remarks are in order for the inclusion of the $\Delta(1232)$ and the tree-level $\Delta$-exchange contribution. In contrast to Ref.~\cite{Lensky:2014dda}, we include the Coulomb-quadrupole $(C2)$ $N\to \Delta$ transition described by the LEC $g_C$. The relevant Lagrangian describing the non-minimal $\gamma^* N \Delta$ coupling~\cite{Pascalutsa:2005vq,Pascalutsa:2005ts}
(note that in these references the overall sign of $g_C$ is inconsistent between the Lagrangian and Feynman rules) reads:
\bea
\mathcal{L}^{(2)}_\Delta &=&  \frac{3e}{2M_N M_+}\,\overline N\, T_3\,\Big\{
i g_M  \tilde F^{\mu\nu} \,\partial_{\mu}\Delta_\nu- g_E \gamma_5 F^{\mu\nu}\,\partial_{\mu}\Delta_\nu\eqlab{gammaNDeltaLag}\\
&& +i \frac{g_C}{M_\Delta}\gamma_5 \gamma^\alpha (\partial_\alpha \Delta_\nu-\partial_\nu \Delta_\alpha)\partial_\mu F^{\mu \nu}\Big\}+\,\mbox{H.c.},
\nn
\eea
with $M_+=M_N+M_\Delta$ and the dual of the electromagnetic field strength tensor
 $\tilde F^{\mu\nu}=\frac{1}{2}\epsilon^{\mu\nu\rho\lambda}F_{\rho\lambda}$. 
Even though the Coulomb coupling is subleading
compared with the electric and magnetic couplings ($g_E$ and $g_M$), its  relatively large magnitude, cf.\ Table \ref{tab:constants}, makes it numerically important for instance in $\gamma_0(Q^2)$. Furthermore, we study the effect of modifying the magnetic coupling using a dipole form factor:
\begin{equation}
g_M\to \frac{g_M}{\big[1+\left(Q/\Lambda\right)^2\big]^2}\,,
\end{equation}
where $\Lambda^2=0.71$~GeV${^2}$. The inclusion of this $Q^2$ dependence mimics the form expected from 
vector-meson dominance. It is motivated by observing the importance of this form factor for the correct description of the electroproduction data \cite{Pascalutsa:2005vq}.

To estimate the uncertainties of our NLO predictions, we define
\begin{align}
 \tilde{\delta}(Q^2) = \sqrt{ \left(\frac{\varDelta}{M_N}\right)^2 + \left(\frac{Q^2}{2 M_N \varDelta}\right)^2 },\eqlab{dtilde}
\end{align}
such that the neglected next-to-next-to-leading order terms are expected to be of relative size $\tilde{\delta}^2$ \cite{Pascalutsa:2005vq}. The uncertainties in the values of the parameters in Table \ref{tab:constants} have a much smaller impact compared to the truncation
uncertainty and can be neglected. Unfortunately, $\Delta I_A(Q^2)$, $\ga_0(Q^2)$ and $\bar \ga_0(Q^2)$, i.e., the sum rules involving the cross section $\sigma_{TT}(\nu,Q^2)$, as well as the polarizability $\Delta I_1(Q^2)$, turn out to be numerically small. Their smallness suggests a cancellation of leading orders (which can indeed be confirmed by looking at separate contributions as shown below). Therefore, an error of $\tilde{\delta}^2(Q^2) P(Q^2)$, where $P(Q^2)$ is a generalized polarizability, might underestimate the theoretical uncertainty for some of the NLO predictions. To avoid this, we estimate the uncertainty of our NLO polarizability predictions by:
\bea
\sigma \!P(Q^2)&=&\Big\{ \mathrm{Max}\left[\tilde \delta^4(0)P(0)^2,\tilde\delta^4(0)P^\mathrm{LO}(0)^2,\tilde\delta^2(0)P^\mathrm{NLO}(0)^2\right]\nn \\
&&+ \mathrm{Max}\left[\tilde \delta^4(Q^2) \left[P(Q^2)-P(0)\right]^2,\right. \tilde \delta^4(Q^2) \left[P^\mathrm{LO}(Q^2)-P^\mathrm{LO}(0)\right]^2,\nn\\
&&\left.\tilde\delta^2(Q^2) \left[P^\mathrm{NLO}(Q^2)-P^\mathrm{NLO}(0)\right]^2 \right]\!\Big\}^{1/2},\eqlab{errorestimate}
\eea
where $P^\mathrm{LO}(Q^2)$ is the $\pi N$-loop contribution, $P^\mathrm{NLO}(Q^2)$ are the $\Delta$-exchange and $\pi \Delta$-loop contributions, and $P(Q^2)=P^\mathrm{LO}(Q^2)+P^\mathrm{NLO}(Q^2)$. 
This error prescription is similar to the one used in, e.g.,  Refs.~\cite{Griesshammer:2012we,Griesshammer:2015ahu,Epelbaum:2014efa,Epelbaum:2014sza}. 
Here, since we are interested in the generalized polarizabilities, we added in quadrature the error due to the real-photon piece
$P(0)$ and the $Q^2$-dependent remainder $P(Q^2)-P(0)$.  Note that $I_A(0)$ and $I_1(0)$ are given by the elastic Pauli form factor, which is not part of our B$\chi$PT prediction and is considered to be exact.

Note that our result for the spin polarizabilities (and the unpolarized moments \cite{Alarcon:2020wjg}) are NLO predictions only at low momentum transfers $Q\simeq m_\pi$. At larger values of $Q\gtrsim\varDelta$, they become incomplete LO predictions. Indeed, in this regime the $\Delta$~propagators do not carry additional suppression compared to the nucleon propagators, and the $\pi\Delta$ loops are promoted to LO. In general, we only expect a rather small contribution from omitted $\pi \Delta$ loops to the $Q^2$ dependence of the polarizabilities, since $\pi\Delta$ loops show rather weak dependence on $Q^2$ compared with the $\Delta$ exchange or $\pi N$ loops. Nevertheless, this issue has to be reflected in the error estimate. Since the polarizabilities at the real-photon point $P(0)$ are not affected, it is natural to separate the error on the $Q^2$-dependent remainder $P(Q^2)-P(0)$, as done in \Eqref{errorestimate}. To accommodate for the potential loss of precision above $Q\gtrsim\varDelta$, we define the relative error $\tilde{\delta}(Q^2)$ as growing with increasing $Q^2$, see \Eqref{dtilde}.

Upon expanding our results in powers of the inverse nucleon mass, $M_N^{-1}$, we are able to reproduce existing results
of heavy-baryon $\chi$PT (HB$\chi$PT) at LO. 
We, however, do not see a rationale to drop the higher-order $M_N^{-1}$ terms when they are not 
negligible (i.e., when their actual size exceeds by far the natural estimate
for the size of higher-order terms). Comparing our B$\chi$PT predictions to HB$\chi$PT, we will also see a deficiency of HB$\chi$PT in the description of the $Q^2$ behaviour of the polarizabilities. Note that the $\mathcal{O}(p^4)$ HB$\chi$PT results from Ref.~\cite{Kao:2002cp,Kao:2003jd}, which we use here for comparison, do not include the $\Delta$. These references studied the leading effect of the latter in the HB$\chi$PT framework, using the small-scale expansion \cite{Hemmert:1996xg}, observing no qualitative improvement in the HB$\chi$PT description of the empirical data \cite{Kao:2002cp,Kao:2003jd} when including it. We therefore choose to use the $\mathcal{O}(p^4)$ results as the representative HB$\chi$PT curves.

Another approach used in the literature to calculate the polarizabilities in $\chi$PT is the infrared regularization (IR) scheme, introduced in Ref.~\cite{Becher:1999he}.
This covariant approach tries to solve the power counting violation observed in Ref.~\cite{Gasser:1987rb} by dropping the regular parts of the loop integrals that contain the power-counting-breaking terms. However, this subtraction scheme modifies the analytic structure of the loop contributions and may lead to unexpected problems, as was shown in Ref.~\cite{Geng:2008mf}. As we will see in the next section, the IR approach also fails to describe the $Q^2$ behaviour of the polarizabilities.

\section{Results and discussion} \label{Sec:Pol}

We now present the NLO B$\chi$PT predictions for the nucleon polarizabilities and selected moments of the nucleon spin structure functions. Our results are obtained from the calculated non-Born VVCS amplitudes and the LEXs in Eqs.~\eref{fgFunctions} and \eref{LEXS12}. For a cross-check, we used the photoabsorption cross sections  described in App.~\ref{CrossSections}. In addition to the full NLO results, we also analyse the individual contributions from the $\pi N$ loops, the $\Delta$ exchange, and the $\pi \Delta$ loops.

\subsection{\boldmath{$\ga_0(Q^2)$} --- generalized forward spin polarizability}
\label{Sec:ForwardSpinPolarizability}

\begin{figure}[h!]
\begin{center}
\includegraphics[width=0.49\textwidth]{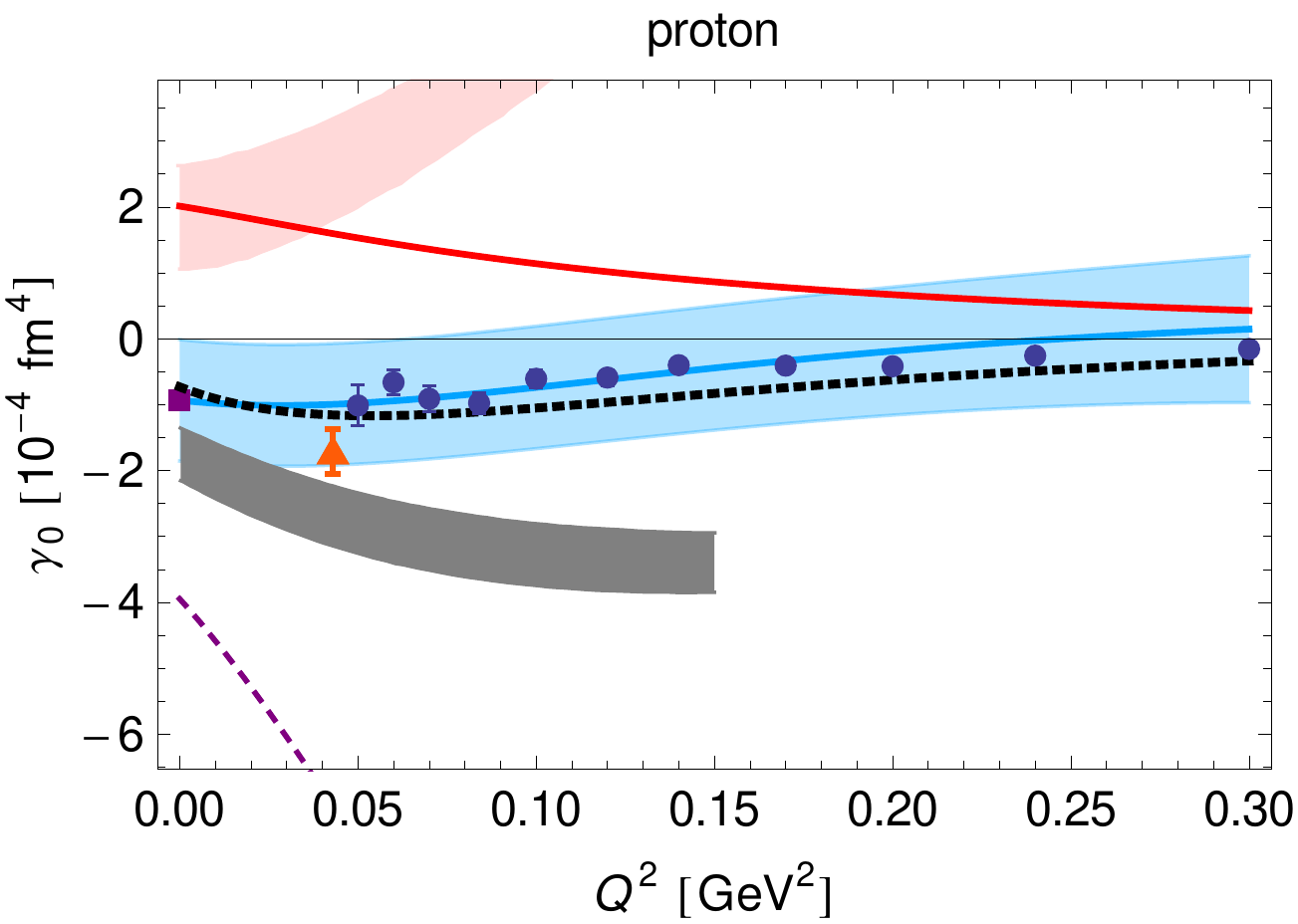}
\includegraphics[width=0.49\textwidth]{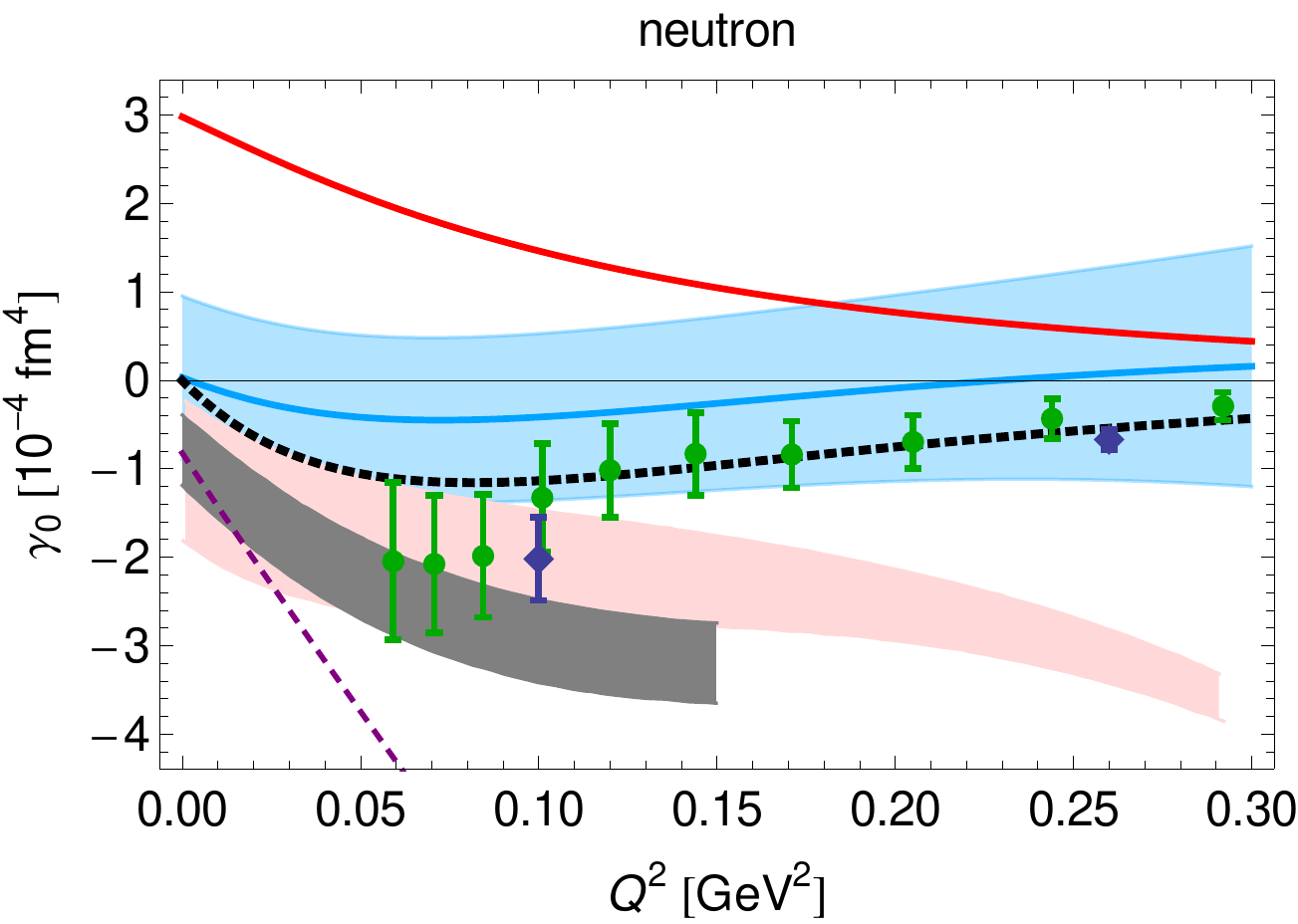}\\[0.2cm]
\includegraphics[width=0.49\textwidth]{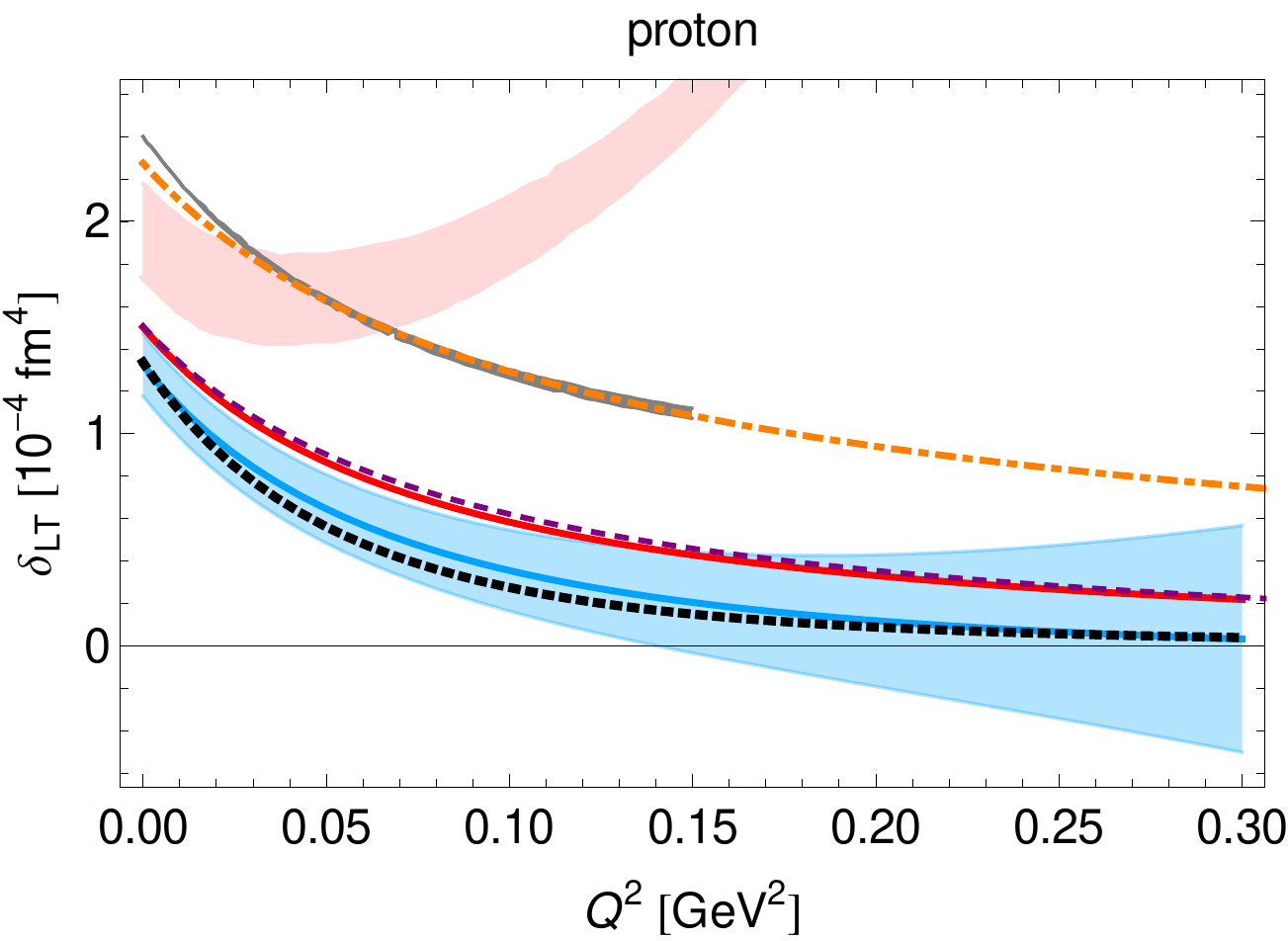}
\includegraphics[width=0.49\textwidth]{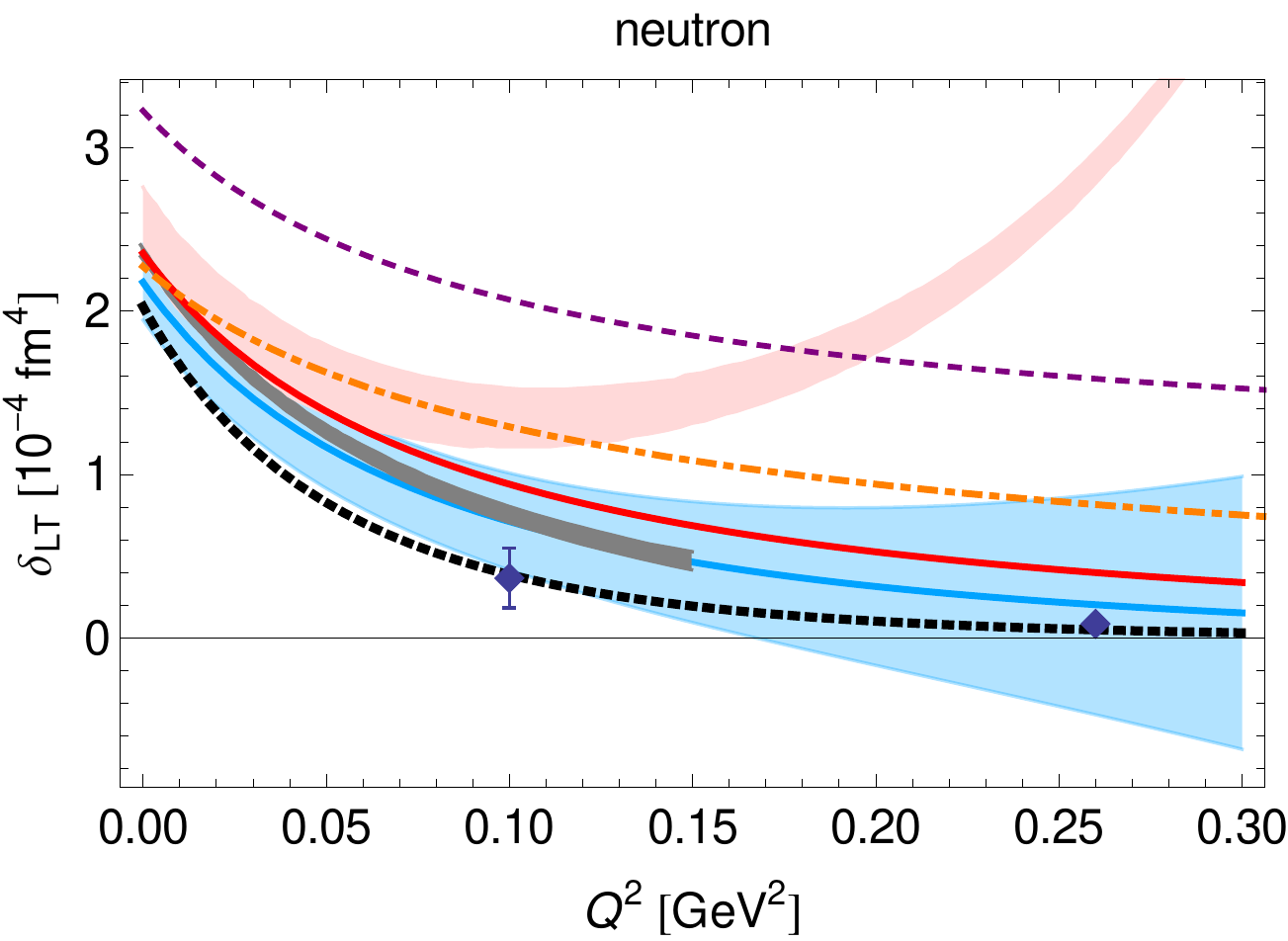}
\caption{Upper panel: Generalized forward spin polarizability for the proton (left) and neutron (right) as function of $Q^2$. The result of this work, the NLO B$\chi$PT prediction, is shown by the blue solid line and the blue band.
The red line represents the LO B$\chi$PT result. The purple short-dashed line is the $\mathcal{O}(p^4)$ HB result from Ref.~\cite{Kao:2002cp}.
The black dotted line is the MAID model prediction~\cite{Drechsel:2000ct,Drechsel:1998hk,private-Lothar}, which is taken from Ref.~\cite{Drechsel:2002ar} (proton)
and Ref.~\cite{Amarian:2004yf} (neutron). The pink band is the IR+$\Delta$ result from Ref.~\cite{Bernard:2002pw}, and the gray band is the  B$\chi$PT+$\Delta$ result from Ref.~\cite{Bernard:2012hb}.
Empirical extractions for the proton: Ref.~\cite{Prok:2008ev} (blue dots), Ref.~\cite{Gryniuk:2016gnm} (purple square) and Ref.~\cite{Zielinski:2017gwp} (orange triangle; uncertainties added in quadrature); and neutron: Ref.~\cite{Amarian:2004yf} (blue diamonds) and Ref.~\cite{Guler:2015hsw} (green dots; statistical and systematic uncertainties added in quadrature). Lower Panel: Longitudinal-transverse spin polarizability for the proton (left) and neutron (right).
The orange dot-dashed and purple short-dashed lines are the $\mathcal{O}(p^3)$ and $\mathcal{O}(p^4)$ HB results from Ref.~\cite{Kao:2002cp}. The pink band is the IR result from Ref.~\cite{Bernard:2002pw} and the gray band is the covariant B$\chi$PT+$\Delta$ result from Ref.~\cite{Bernard:2012hb}. The black dotted line is the MAID model prediction~\cite{Drechsel:2000ct,Drechsel:1998hk,private-Lothar}; note that for the proton we use the updated estimate from Ref.~\cite{Drechsel:2002ar} obtained using the $\pi,\eta,\pi\pi$ channels. \label{Fig:gamma0plot}}
\end{center}
\end{figure}

The forward spin polarizability, 
\bea
\gamma_0 (Q^2)&=& \frac{1}{2 \pi^2} \int_{\nu_0}^\infty \! \dd\nu\,\sqrt{1+\frac{Q^2}{\nu^2}} \,\frac{\sigma_{TT} (\nu,Q^2)}{\nu^3}\label{Eq:gamma0Q2}\\
&=&\frac{16 \al M_N^2}{Q^6}\int_0^{x_0}\!\dd x \, x^2 \! \left[g_1(x,Q^2)-\frac{4M_N^2 x^2}{Q^2}\,g_2(x,Q^2)\right]\!,\nn
\eea
provides information about the spin-dependent response of the nucleon to transversal photon probes. The RCS analogue of the above generalized forward spin polarizability sum rule is sometimes referred to as the Gell-Mann, Goldberger and Thirring (GGT) sum rule \cite{GellMann:1954db}.
At $Q^2=0$, the forward spin polarizability is expressed through the lowest-order spin polarizabilities of RCS as ${\gamma_0=-(\gamma_{E1E1} + \gamma_{M1M1} + \gamma_{E1M2}+\gamma_{M1E2})}$. 
The forward spin polarizability of the proton is relevant for an accurate knowledge of the (muonic-)hydrogen hyperfine splitting, as it controls the leading proton-polarizability correction \cite{Carlson:2008ke,Hagelstein:2017cbl}.

The $\pi N$-loop, $\Delta$-exchange, and $\pi \Delta$-loop contributions to the NLO B$\chi$PT prediction of the forward spin polarizability amount to, in units of $10^{-4}$~fm$^4$:
\begin{subequations}
\begin{align}
&\gamma_{0p} =-0.93(92)\approx 2.01 - 2.84 -0.10, \label{Eq:gamma0ProtonRealPoint}\\
&\gamma_{0n} =\hphantom{-} 0.03(92) \approx 2.98 - 2.84-0.10,
\end{align}
\end{subequations}
while the slope is composed as follows, in units of $10^{-4}$~fm$^6$:
\begin{subequations}
\bea
\left.\frac{\dd\gamma_{0p} (Q^2)}{\dd Q^2}\right|_{Q^2=0}&=& -0.22(4)\approx-0.33  +0.11 +0.01 , \\
\left.\frac{\dd\gamma_{0n} (Q^2)}{\dd Q^2}\right|_{Q^2=0}&=&-0.61(7) \approx -0.73 +0.11 +0.01.
\eea
\end{subequations}
Figure~\ref{Fig:gamma0plot} \{upper panel\} shows our NLO prediction, as well as the LO $\pi N$ loops,
compared to different experimental and theoretical results. For the proton, we have one determination at the real-photon point by the GDH collaboration \cite{Dutz:2003mm}, $\gamma_{0p}=-1.00(8)(12)  \times 10^{-4}$~fm$^4$, and further Jefferson Laboratory data \cite{Zielinski:2017gwp,Prok:2008ev} at very low $Q^2$.
For the neutron, only data at finite $Q^2$ are available \cite{Amarian:2004yf, Guler:2015hsw}. 
The experimental data for the proton are fairly well reproduced in the whole $Q^2$ range considered here, while for the neutron the agreement improves with increasing $Q^2$.
The HB limit of our $\pi N$-loop contribution reproduces the results published in Refs.~\cite{Kao:2002cp,Bernard:1995dp} for arbitrary $Q^2$. 
In addition, our prediction is compared to the MAID model \cite{Drechsel:2002ar,Amarian:2004yf}, the IR+$\Delta$ calculation of Ref.~\cite{Bernard:2002pw} and the B$\chi$PT+$\Delta$ result of Ref.~\cite{Bernard:2012hb}.

\begin{figure}
\begin{center}
\includegraphics[width=0.49\textwidth]{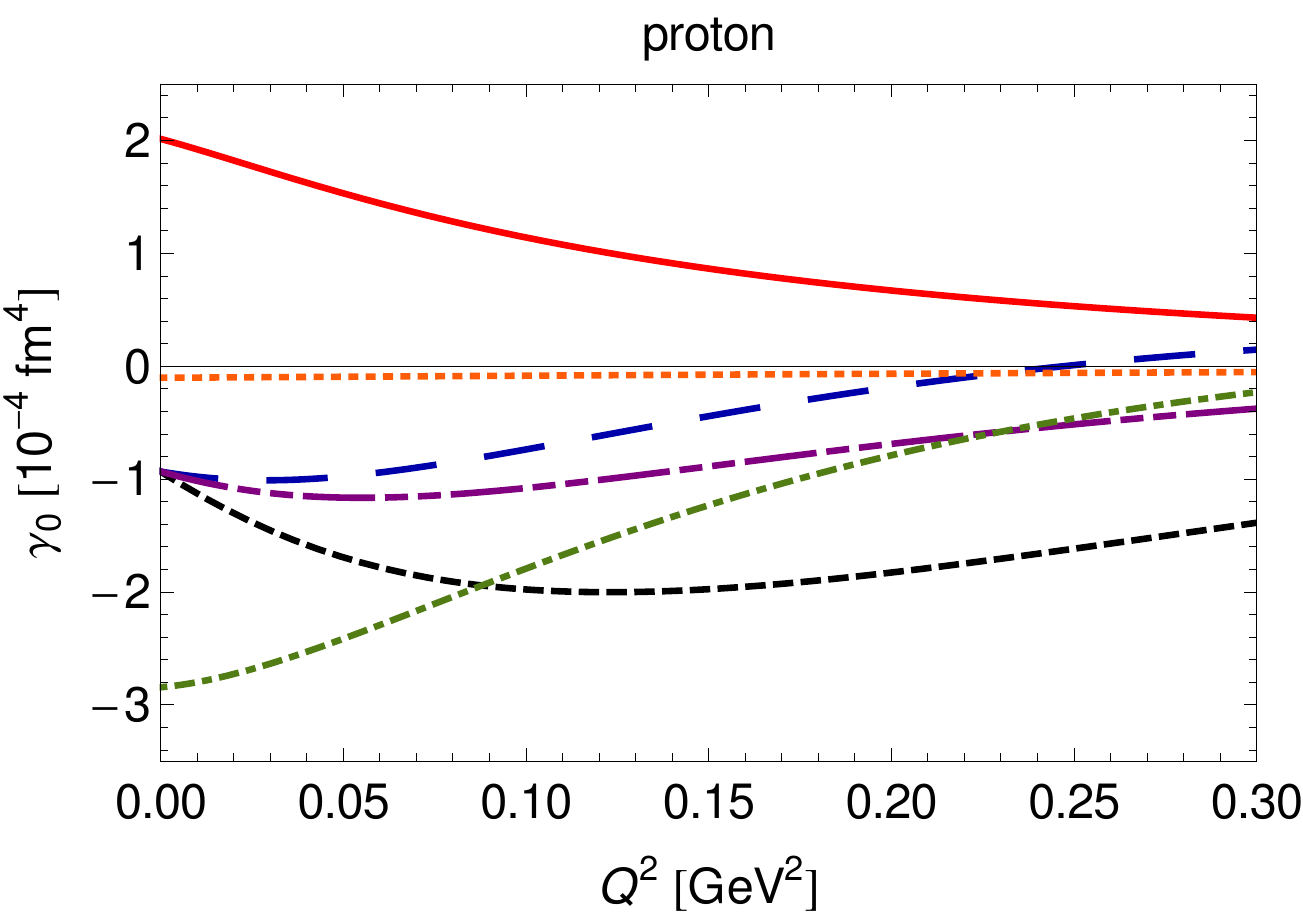}
\includegraphics[width=0.49\textwidth]{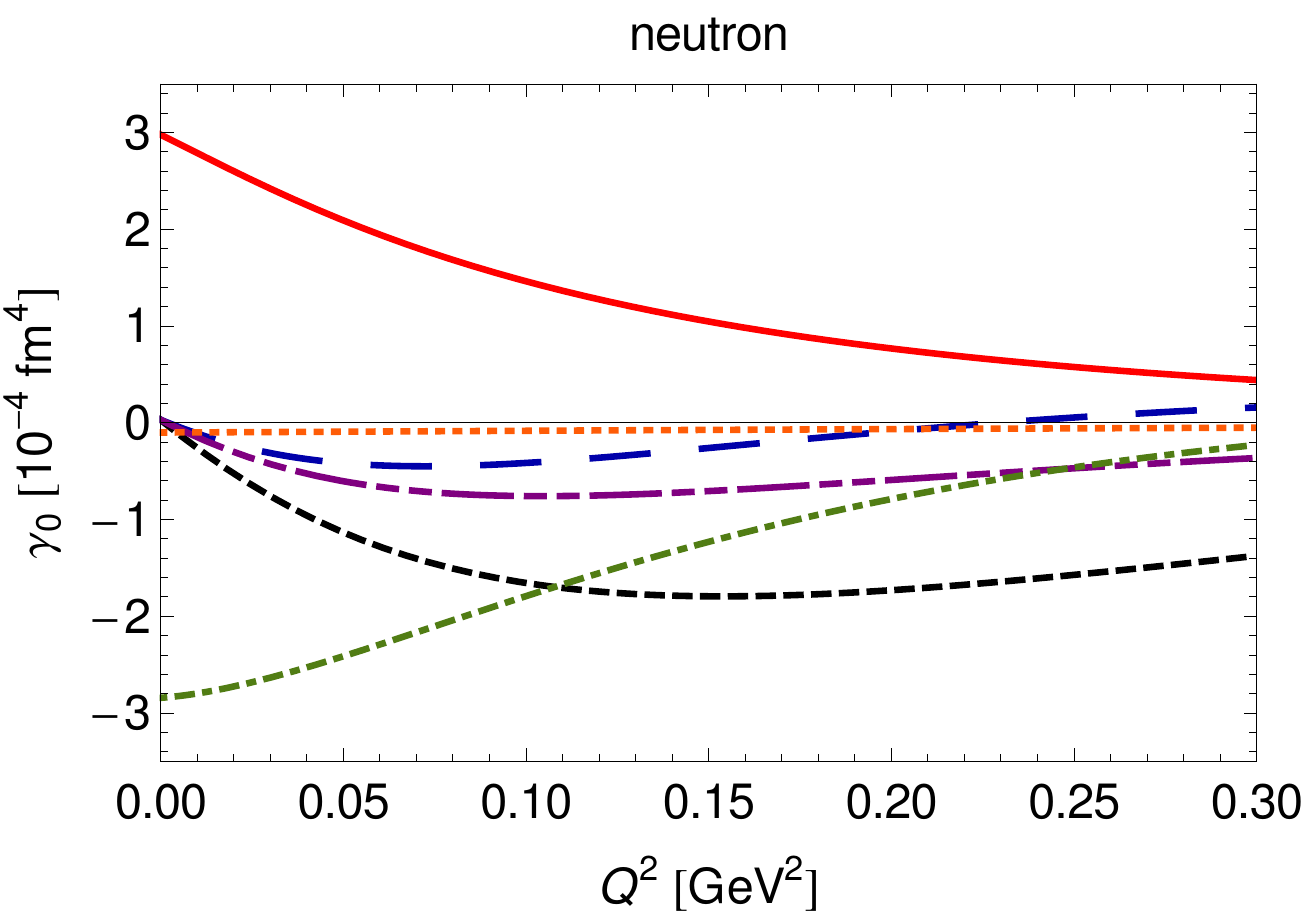} \\[0.2cm]
\includegraphics[width=0.49\textwidth]{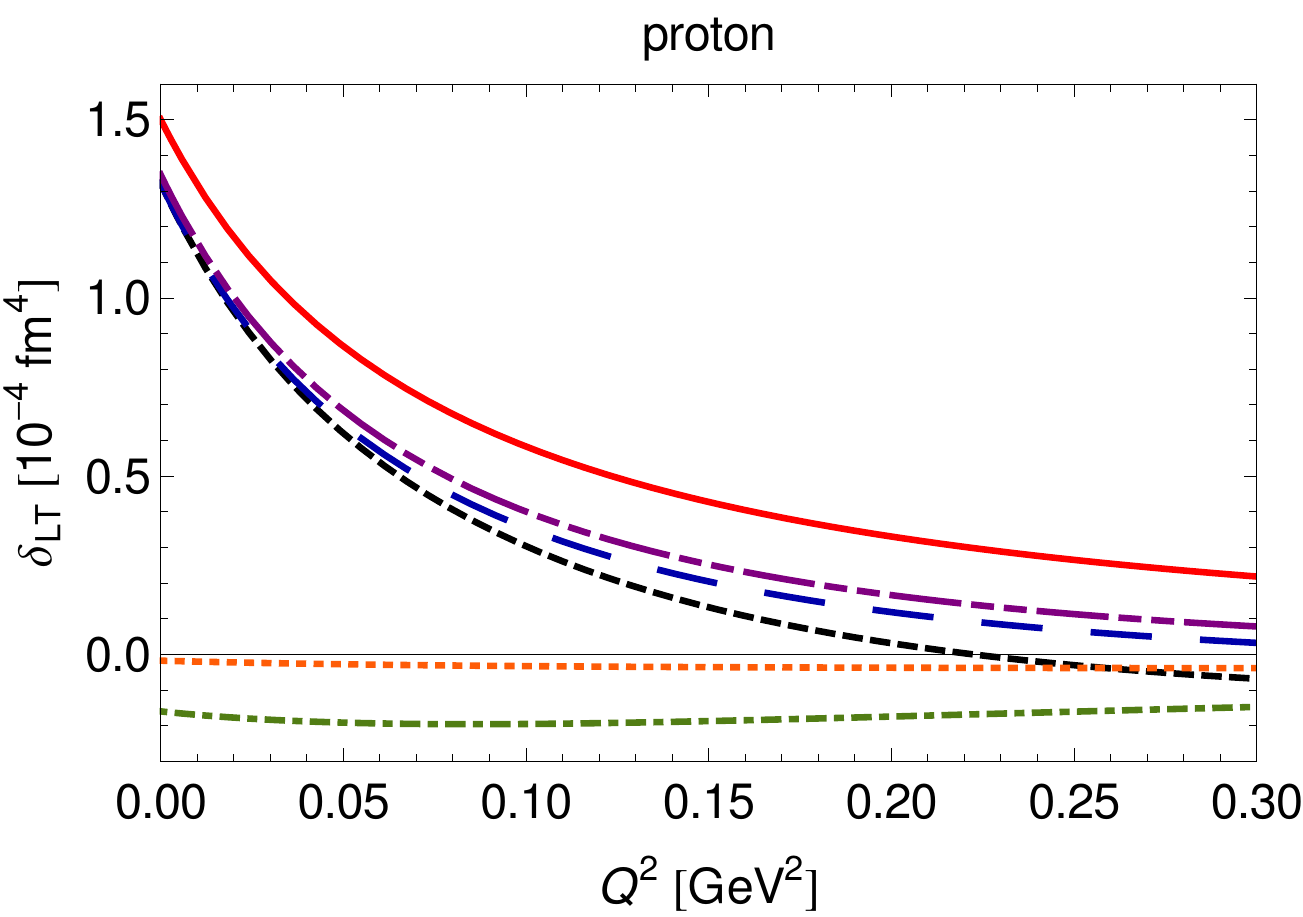}
\includegraphics[width=0.49\textwidth]{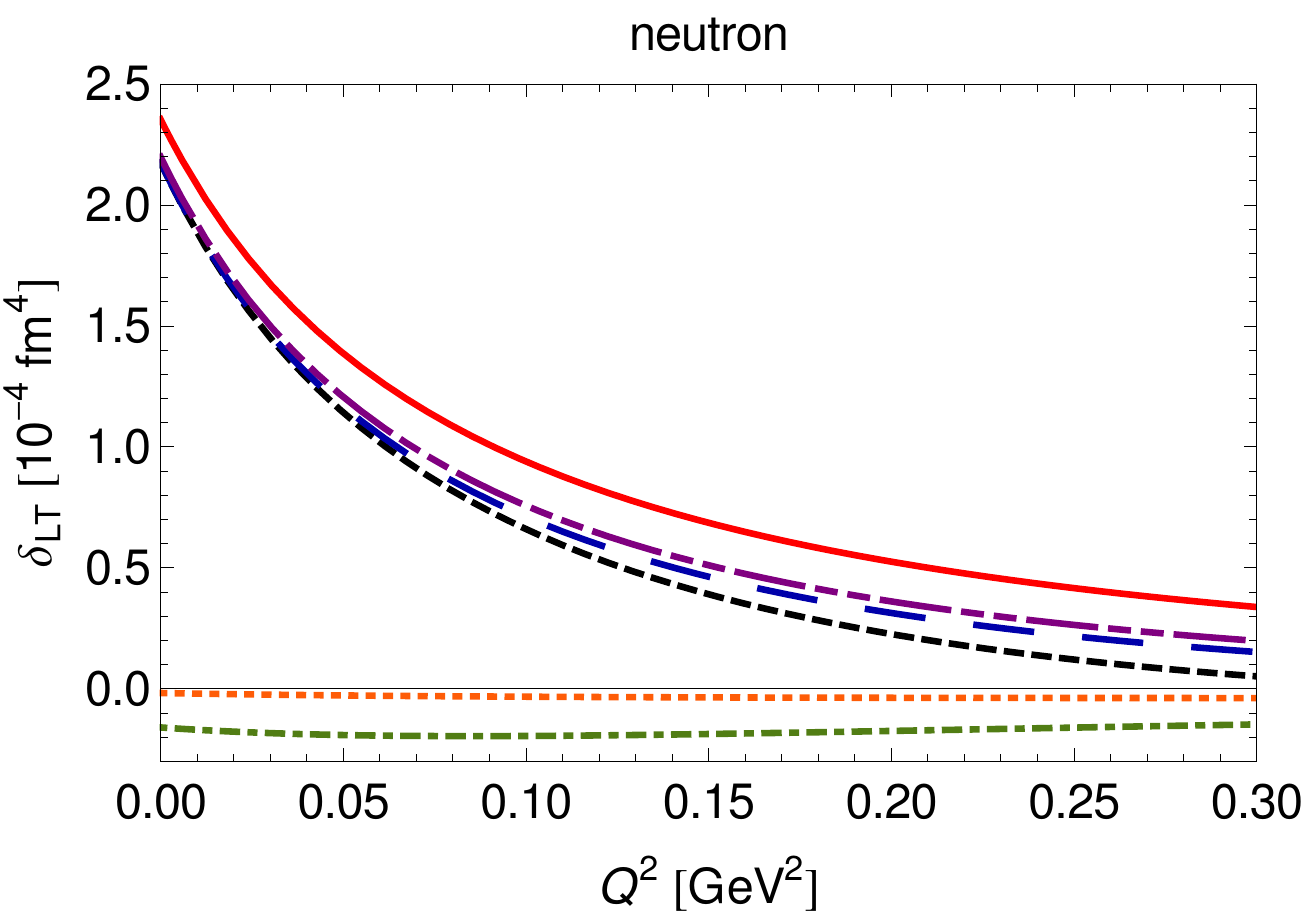}
\caption{Contributions of the different orders to the chiral predictions of $\gamma_0(Q^2)$ \{upper panel\} and $\delta_{LT}(Q^2)$ \{lower panel\} for the proton (left) and neutron (right). Red solid line: $\pi N$-loop contribution, green dot-dashed line: $\Delta$-exchange contribution, orange dotted line: $\pi \Delta$-loop contribution, blue long-dashed line: total result, purple dot-dot-dashed line: total result without $g_C$ contribution, black short-dashed line: total result without $g_M$ dipole.  \label{Fig:gamma0-orders}}
\end{center}
\end{figure}

The $\pi N$-production channel gives a positive contribution to the photoabsorption cross section $\sigma_{TT}(\nu,Q^2)$ at low $Q^2$, cf.\ Fig.~\ref{Fig:SummaryCrossSections}. Accordingly, one observes that the $\pi N$ loops give a sizeable positive contribution to $\ga_0(Q^2)$.
The Delta, on the other hand, has a very large effect by cancelling the $\pi N$ loops and bringing the result close to the empirical data. From Fig.~\ref{Fig:gamma0-orders} \{upper panel\}, one can see that it is the $\Delta$ exchange what dominates, while $\pi \Delta$ loops are negligible. 
This was expected, since the forward spin polarizability sum rule is an integral over the helicity-difference cross section, in which $\sigma_{3/2}$ is governed by the Delta at low energies (the relevant energy region for the sum rule). 

To elucidate the difference between the present calculation and the one from Ref.~\cite{Bernard:2012hb},
we note that the two calculations differ in the following important aspects.
Firstly, Ref.~\cite{Bernard:2012hb} uses the small-scale counting~\cite{Hemmert:1996rw} that considers $\varDelta$ and $m_\pi$ as being of the same size, $\varDelta\sim m_\pi$. In practice, this results in a set of $\pi\Delta$-loop graphs which contains graphs with one or two $\gamma\Delta\Delta$ couplings and hence two or three Delta propagators. Such graphs are suppressed in the $\delta$-counting and thus omitted from our calculation while present in that of Ref.~\cite{Bernard:2012hb}.
Secondly, the Lagrangians describing the interaction of the Delta are constructed differently and assume slightly different values for the coupling constants. In particular, we employ (where possible)
the so-called ``consistent'' couplings to the Delta field, i.e., those couplings that
project out the spurious degree of freedom, see Refs.~\cite{Pascalutsa:2006up,Pascalutsa:1999zz,Pascalutsa:1998pw}. The authors of Ref.~\cite{Bernard:2012hb}, on the other hand, use couplings where the consistency in this sense is not
enforced. The effects of these differences are of higher order in the $\delta$-counting expansion, and their contribution to the $Q^2$ dependence of the considered polarizabilities is expected
to be rather small; however, the differences at $Q^2=0$ could be noticeable \cite{Krebs:2019ddp}.

Finally, as mentioned in Sec.~\ref{Sec:VVCS_calc}, we are including a dipole form factor in the magnetic coupling $g_{M}$, a modification absent in Ref.~\cite{Bernard:2012hb}. This modification is expected to be needed in order to generate the correct $Q^2$ behaviour of the polarizabilities that receive a significant contribution from the magnetic $\gamma^* N\Delta$ transition, such as $\gamma_0(Q^2)$. Figure ~\ref{Fig:gamma0plot} \{upper panel\}
shows that our predictions for the $Q^2$ dependence of $\gamma_0(Q^2)$ differ quite significantly from those of Ref.~\cite{Bernard:2012hb}. The main reason for this is the dipole form factor that indeed drives the curves closer to the experimental data. Another polarizability that shows a similar importance of the dipole form factor is the closely related $I_A(Q^2)$, considered below in Sec.~\ref{IAsec}, and shown in Fig.~\ref{Fig:IAplot} \{upper panel\}.
In other polarizabilities, like $\delta_{LT}(Q^2)$ shown in Fig.~\ref{Fig:gamma0plot} \{lower panel\}, the magnetic transition is not so prominent, and so is the effect of the dipole form factor on the $Q^2$ dependence. The effect of the form factor on the polarizabilities is further illustrated in Figs.~\ref{Fig:gamma0-orders}, \ref{Fig:IA-orders-plot} and \ref{Fig:d2-orders-plot}, where one can see the total result with the $g_M$ dipole compared to the total result without it. 

Concerning the generalized forward spin polarizability, the experimental data for $\gamma_{0n}(Q^2)$ at very low $Q
^2$ slightly favor the B$\chi$PT prediction without inclusion of the dipole form factor \cite{Bernard:2012hb}. In general, our B$\chi$PT prediction is able to describe all experimental data within errors and shows perfect agreement for $\gamma_{0p}(Q^2)$ at $Q^2 <0.3$ GeV$^2$ and for $\gamma_{0n}(Q^2)$ in the region of $0.15 < Q^2 <0.3$ GeV$^2$. 
The $\pi \Delta$-loop contribution does not modify the $Q^2$ behavior of $\gamma_0(Q^2)$, and only differs from Ref.~\cite{Bernard:2012hb} by a small global shift. 
Note also the relatively large effect of $g_C$, which generates a sign change for virtualities above $\sim0.2$ GeV$^2$, see Fig.~\ref{Fig:gamma0-orders} \{upper panel\}.

\subsection{\boldmath{$\delta_{LT}(Q^2)$} --- longitudinal-transverse polarizability}
\label{Sec:LongitudinalTransversePolarizability}

The longitudinal-transverse spin polarizability,
\bea
\delta_{LT} (Q^2)&=& \frac{1}{2 \pi^2} \int_{\nu_0}^\infty \! \dd\nu\,\sqrt{1+\frac{Q^2}{\nu^{2}}}\, \frac{\sigma_{LT} (\nu,Q^2)}{Q\,\nu^2} \label{Eq:deltaLTQ2} \\
&=&\frac{16\al M_N^2}{Q^6}\!\int_0^{x_0}\!\dd x \, x^2\left[g_1(x,Q^2)+g_2(x,Q^2)\right]\!,\quad\nn
\eea
contains information about the spin structure of the nucleon, and is another important input in the determination of the (muonic-)hydrogen hyperfine splitting \cite{Carlson:2008ke,Hagelstein:2017cbl}. It is also relevant in studies of higher-twist corrections to the structure function $g_2(x,Q^2)$, given by the moment $d_2(Q^2)$ \cite{Kao:2003jd}, see Section \ref{d2Sec}.
The peculiarity of the response encoded in this polarizability is that it involves a spin flip of the nucleon and a polarization change of the photon, see App.~\ref{CrossSections} and  Fig.~\ref{Fig:sigmaLT}.

It is expected that the Delta isobar gives only a small contribution to $\delta_{LT}(Q^2)$, what makes this polarizability a potentially clean test case for chiral calculations. 
Consequently, there are relatively many different theoretical calculations of $\delta_{LT}(Q^2)$
coming from different versions of $\chi$PT with baryons (HB, IR and covariant). Ref.~\cite{Kao:2002cp} found a systematic deviation of the HB result for $\delta_{LTn}(Q^2)$ from the MAID model prediction. 
This disagreement was identified by the authors of Ref.~\cite{Kochelev:2011bh}
as a puzzle involving the neutron $\delta_{LT}$ polarizability---the $\delta_{LT}$ puzzle. The IR calculation in Ref.~\cite{Bernard:2002pw} also showed a deviation from the data and predicted a rapid rise of $\delta_{LT}(Q^2)$ with growing $Q^2$. 
The problem is solved by keeping the relativistic structure of the theory, as the B$\chi$PT+$\Delta$ result of Ref.~\cite{Bernard:2012hb} showed.

As expected, already the leading $\pi N$ loops provide a reasonable agreement with the experimental data, cf.\ Fig.~\ref{Fig:gamma0plot} \{lower panel\}. Since the $\Delta$-exchange contribution to $\delta_{LT}(Q^2)$ is small, the effect of the $g_M$ form
factor is negligible in this polarizability, as is that of the $g_C$ coupling, cf.\ Fig.~\ref{Fig:gamma0-orders} \{lower panel\}. In fact, we predict both the $\Delta$-exchange and the $\pi\Delta$-loop contributions to be small and negative. 
This is in agreement with the MAID model, which predicts a small and negative contribution of the $P_{33}$ wave to $\delta_{LT}(Q^2)$.
However, in the calculation of Ref.~\cite{Bernard:2012hb}, which is different from the one presented here only in the way the $\Delta(1232)$ is included, the contribution of this resonance to $\delta_{LTp}(Q^2)$ is sizeable and positive.
The authors of that work attributed this large contribution to diagrams where the photons couple directly to the Delta inside a loop. As mentioned in Sec.~\ref{Sec:ForwardSpinPolarizability},
the effect of such loop diagrams does not  change the $Q^2$ behaviour of the polarizabilities. On the other hand, it can produce a substantial shift of the $\delta_{LT}(Q^2)$ as a whole.  A higher-order calculation should resolve the discrepancy
between the two covariant approaches, however, it will partially lose the predictive power since
the LECs appearing at higher orders will have to be fitted to experimental data.

The $\pi N$-loop, $\Delta$-exchange, and $\pi \Delta$-loop contributions to the NLO B$\chi$PT prediction of the longitudinal-transverse polarizability are, in units of $10^{-4}$~fm$^4$:
\begin{subequations}
\bea\label{Eq:deltaLTRealPoint}
\delta_{LTp} 
&=& 1.32(15)\approx 1.50 -0.16 -0.02 , \\
\delta_{LTn}&=&2.18(23)\approx 2.35 -0.16 -0.02,
\eea
\end{subequations}
while the slopes are, in units of $10^{-4}$~fm$^6$:
\begin{subequations}
\bea
\left.\frac{\dd\delta_{LTp} (Q^2)}{\dd Q^2}\right|_{Q^2=0}&=& -0.85(8)\approx  -0.80 - 0.04 - 0.01 ,\\
\left.\frac{\dd\delta_{LTn} (Q^2)}{\dd Q^2}\right|_{Q^2=0}&=& -1.24(12)\approx  -1.19 - 0.04 -0.01 .
\eea
\end{subequations}

\subsection{\boldmath{$I_A(Q^2)$} --- a generalized GDH integral}
\label{IAsec}

\begin{figure}[t]
\begin{center}
\includegraphics[width=0.49\textwidth]{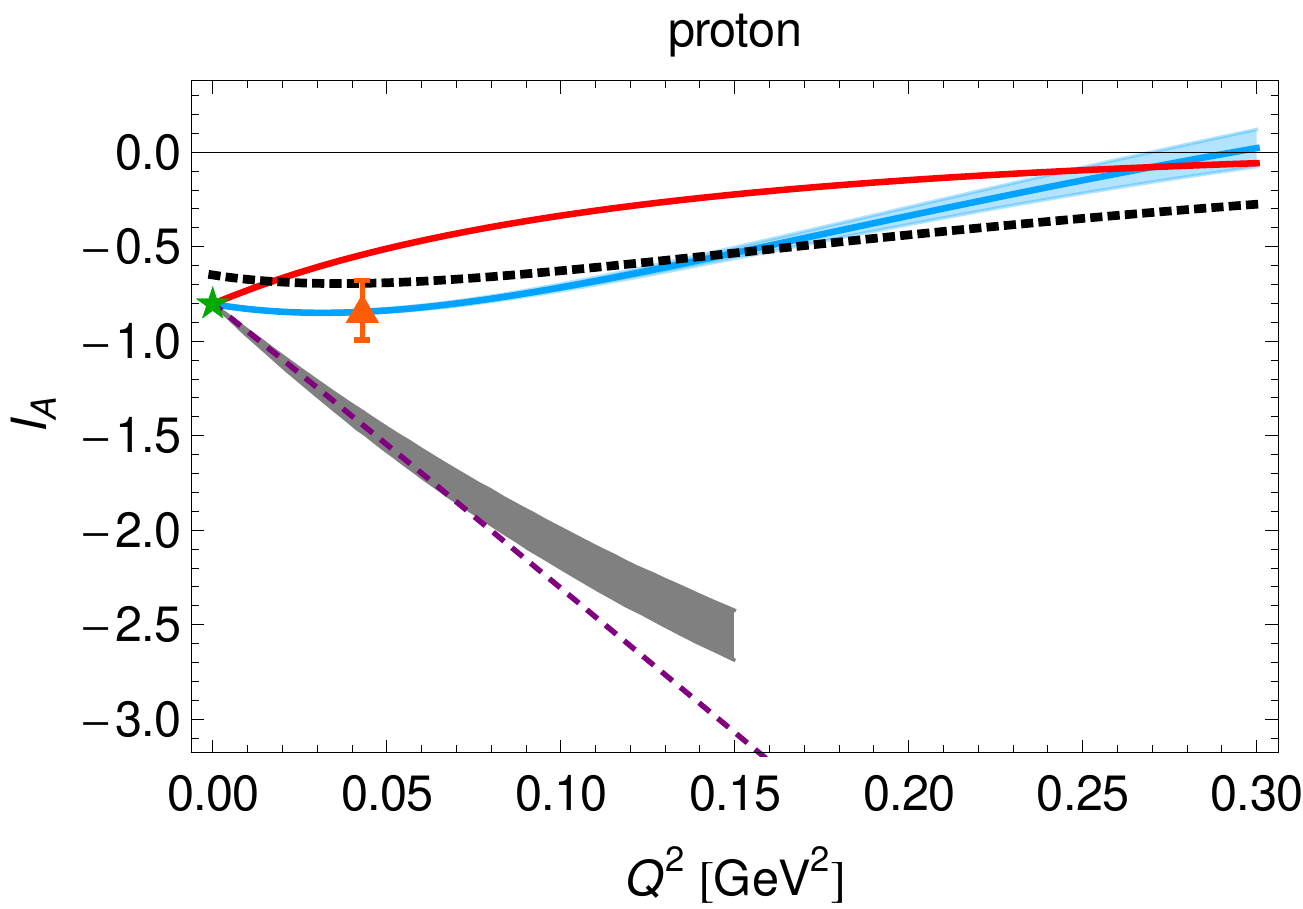}
\includegraphics[width=0.49\textwidth]{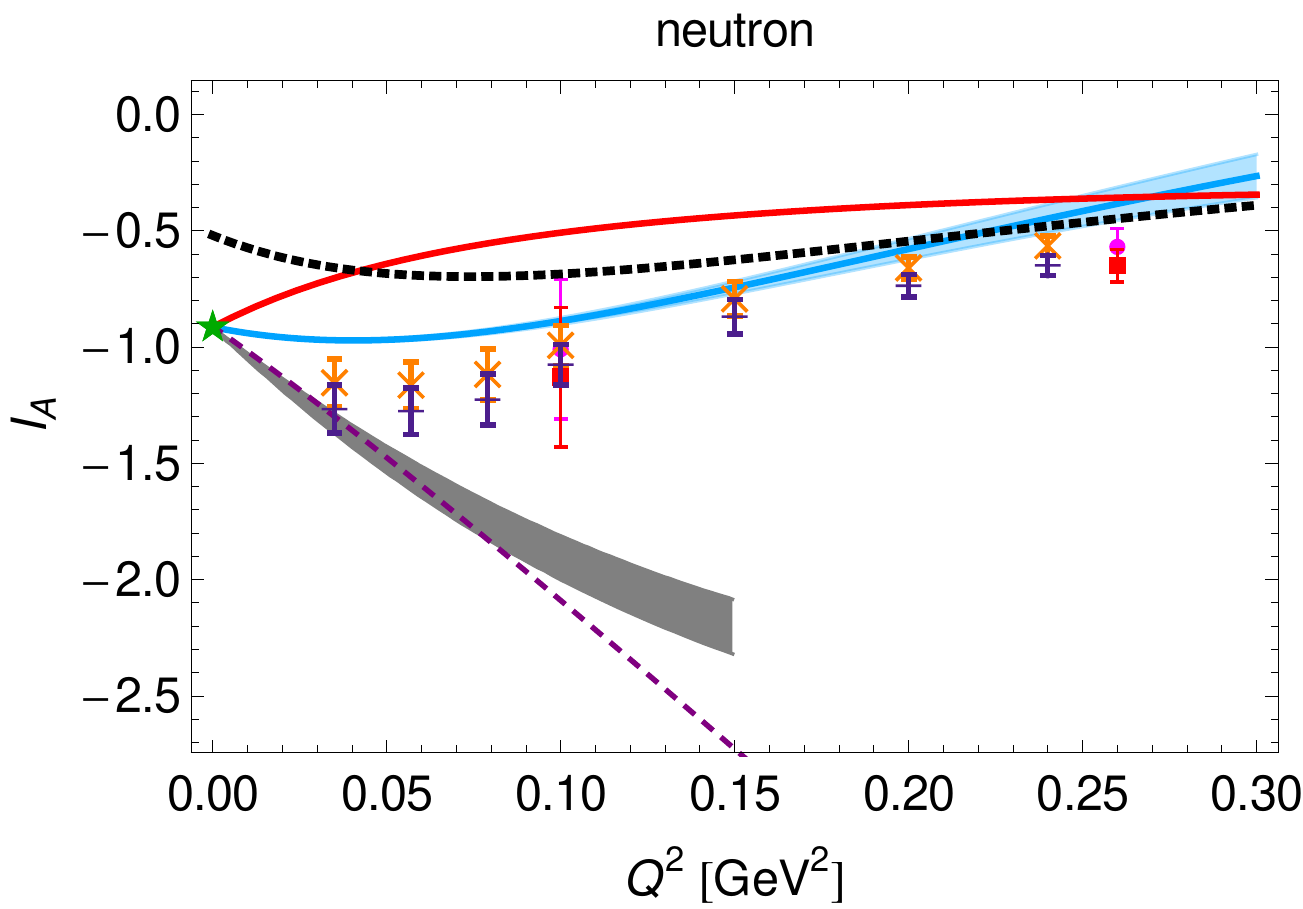}\\[0.2cm]
\includegraphics[width=0.49\textwidth]{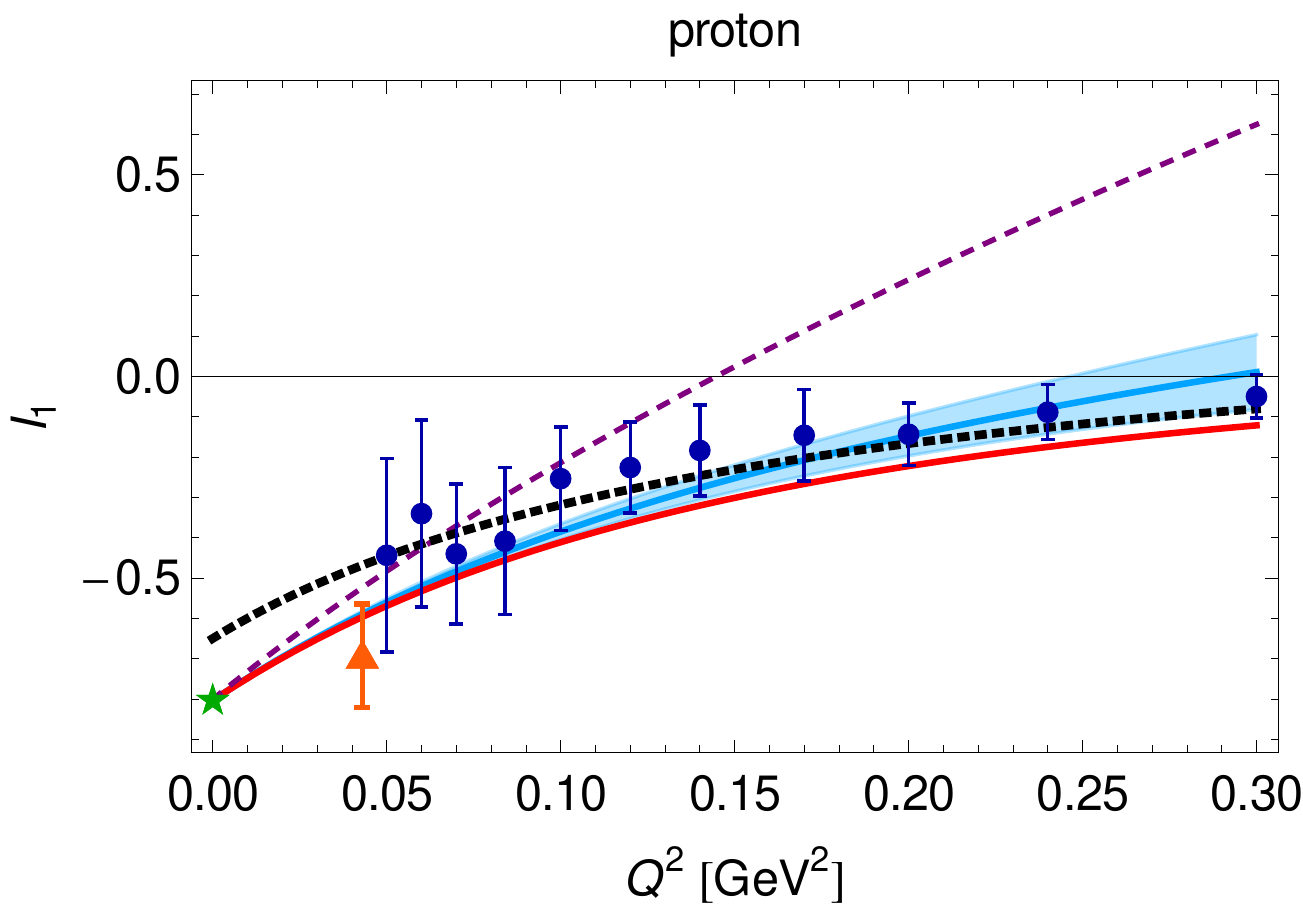}
\includegraphics[width=0.49\textwidth]{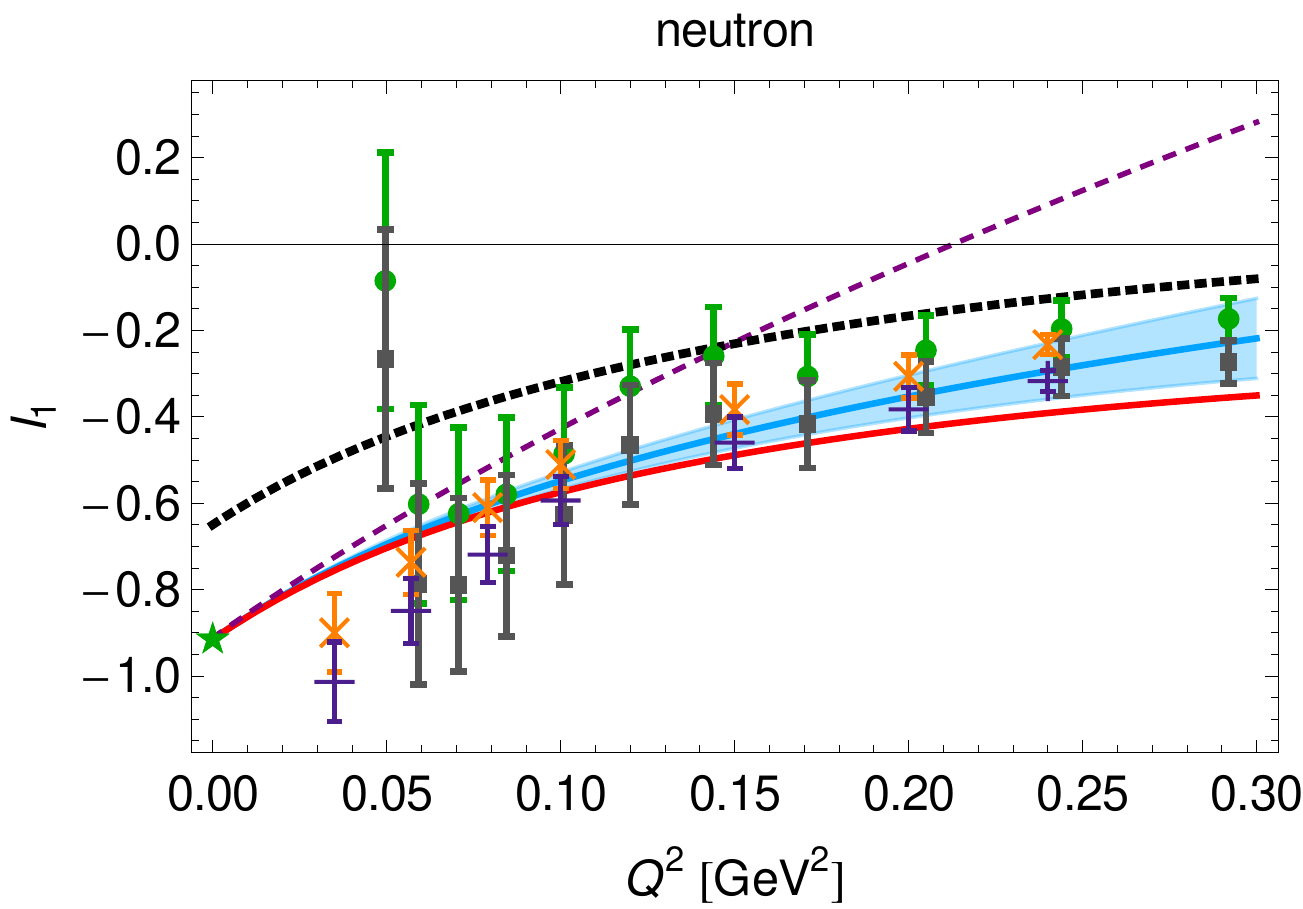}
\caption{Upper panel: The generalized GDH integral $I_A(Q^2)$ for the proton (left) and neutron (right) as function of $Q^2$. The result of this work, the NLO B$\chi$PT prediction, is shown by the blue solid line and the blue band.
The red line represents the LO B$\chi$PT result. The purple short-dashed line is the $\mathcal{O}(p^4)$ HB result from Ref.~\cite{Kao:2002cp,Kao:2003jd}. The gray band is the B$\chi$PT+$\Delta$ result from Ref.~\cite{Bernard:2012hb}. The black dotted line is the MAID model prediction \cite{MAID}. Experimental extractions for the proton: Ref.~\cite{Zielinski:2017gwp} (orange triangle; uncertainties added in quadrature); and neutron: Refs.~\cite{Amarian:2002ar}/\cite{Sulkosky:2019zmn}, where magenta dots/orange diagonal crosses correspond to data and red squares/lilac crosses correspond to data plus extrapolation to unmeasured energy regions. The green stars at the real-photon point are derived from the anomalous magnetic moments: $\varkappa_p \approx 1.793$ and $\varkappa_n\approx -1.913$ \cite{Mohr:2012aa}. Lower panel: The generalized GDH integral $I_1 (Q^2)$ for the proton (left) and neutron (right) as function of $Q^2$. The purple short-dashed line is the HB result from Ref.~\cite{Kao:2003jd}. Experimental extractions for the proton: Ref.~\cite{Prok:2008ev} (blue dots) and Ref.~\cite{Zielinski:2017gwp} (orange triangle; uncertainties added in quadrature); and neutron: Ref.~\cite{Guler:2015hsw}/\cite{Sulkosky:2019zmn} (uncertainties added in quadrature) where green dots/orange diagonal crosses correspond to data  and gray squares/lilac crosses correspond to data plus extrapolation to unmeasured energy regions. \label{Fig:IAplot}}
\end{center}
\end{figure}

The helicity-difference cross section $\sigma_{TT}$ exhibits a faster fall-off in $\nu$ than its spin-averaged counterpart $\sigma_{T}$. 
This is due to a cancellation between the leading (constant) terms of $\sigma_{1/2}$ and $\sigma_{3/2}$ at large $\nu$.\footnote{Notice that a constant term in $\sigma_{TT}$ at $\nu\to \infty$ is forbidden by crossing symmetry.}
The resulting $1/\nu$ fall-off of the helicity-difference cross section allows one to write an unsubtracted dispersion relation for the VVCS amplitude $g_{TT}^\mathrm{nonpole}(\nu,Q^2)$, cf.\ Eq.~(\ref{Eq:LEX-GGT}).
This is the origin of the GDH sum rule \cite{Gerasimov:1965et,Drell:1966jv},
\beq
-\frac{\alpha}{2 M_N^2}\varkappa^2 = \frac{1}{2 \pi^2} \int_{\nu_0}^\infty \! \dd\nu\,\frac{\sigma_{TT} (\nu)}{\nu}, \label{Eq:GDH-SumRule}
\eeq
which establishes a relation to the anomalous magnetic moment $\varkappa$. It is  experimentally verified for the nucleon by MAMI (Mainz) and ELSA (Bonn)~\cite{Ahrens:2001qt, Helbing:2002eg}. 

There are two extensions of the  GDH sum rule to finite $Q^2$: the generalized GDH integrals $I_A(Q^2)$ and $I_1(Q^2)$. The latter will be discussed in Sec.~\ref{I1sec}. The former is defined as:\footnote{Note that $I_A(Q^2)$ is sometimes called $I_{TT}(Q^2)$.}
\bea
-\frac{\alpha}{2 M_N^2}I_A(Q^2) &=& -\frac{1}{8 \pi^2} \int_{\nu_0}^\infty \! \dd\nu\, \sqrt{1+\frac{Q^2}{\nu^{2}}}\, \frac{\sigma_{TT} (\nu,Q^2)}{\nu} \label{Eq:IA-SumRule}\qquad\\
&=&\frac{\alpha}{Q^2}\int_0^{x_0}\!\dd x\,\Big[\frac{4M_N^2 x^2}{Q^2}g_2(x,Q^2)-g_1(x,Q^2)\Big].\nn
\eea
 Due to its energy weighting, the integral in Eq.~\eqref{Eq:IA-SumRule} converges slower than the one in the generalized forward spin polarizability sum rule~\eqref{Eq:gamma0Q2}. Therefore, knowledge of the cross  section at higher energies is required and the evaluation of the generalized GDH integral $I_A(Q^2)$ is not as simple as the evaluation of $\gamma_0(Q^2)$.

\begin{figure}
\begin{center}
\includegraphics[width=0.49\textwidth]{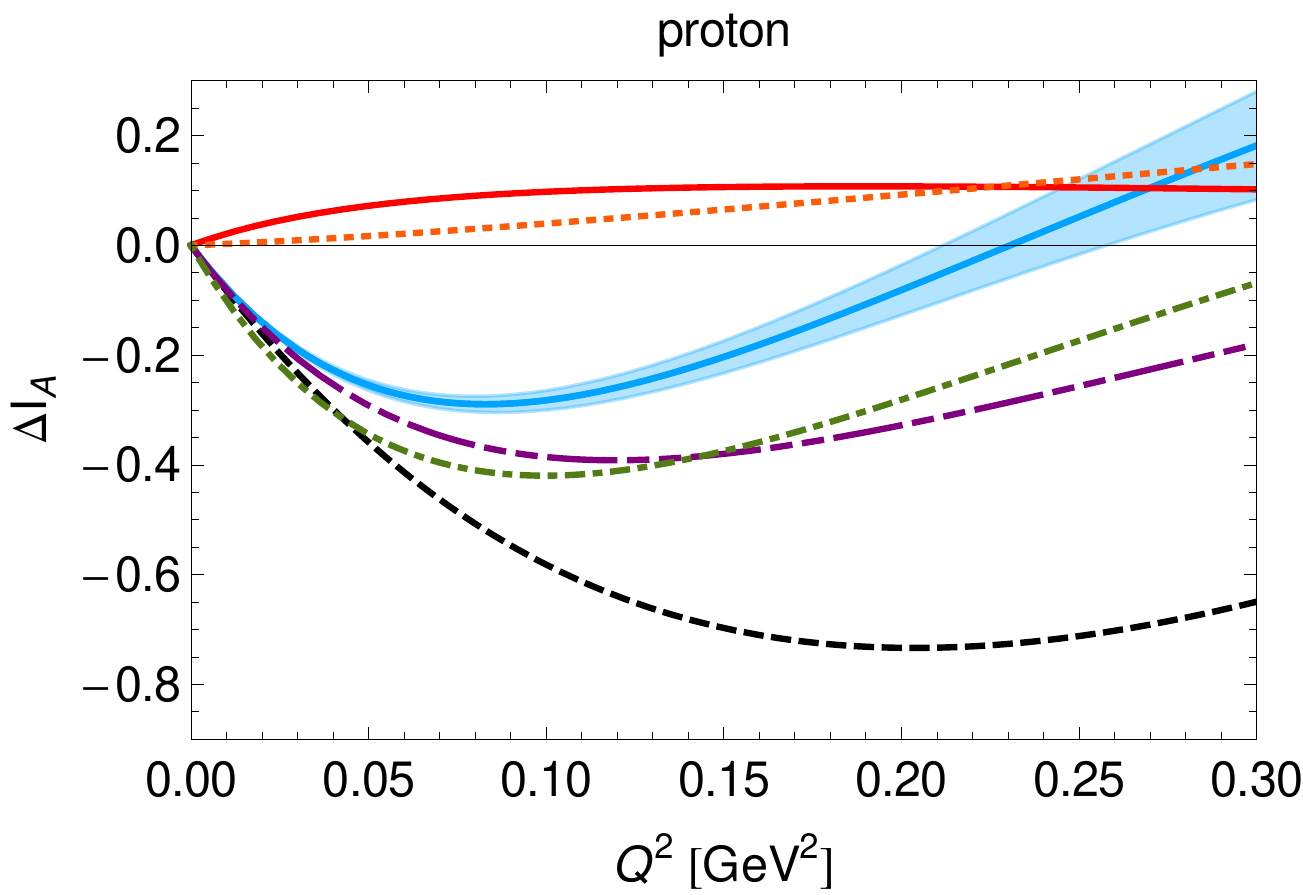}
\includegraphics[width=0.49\textwidth]{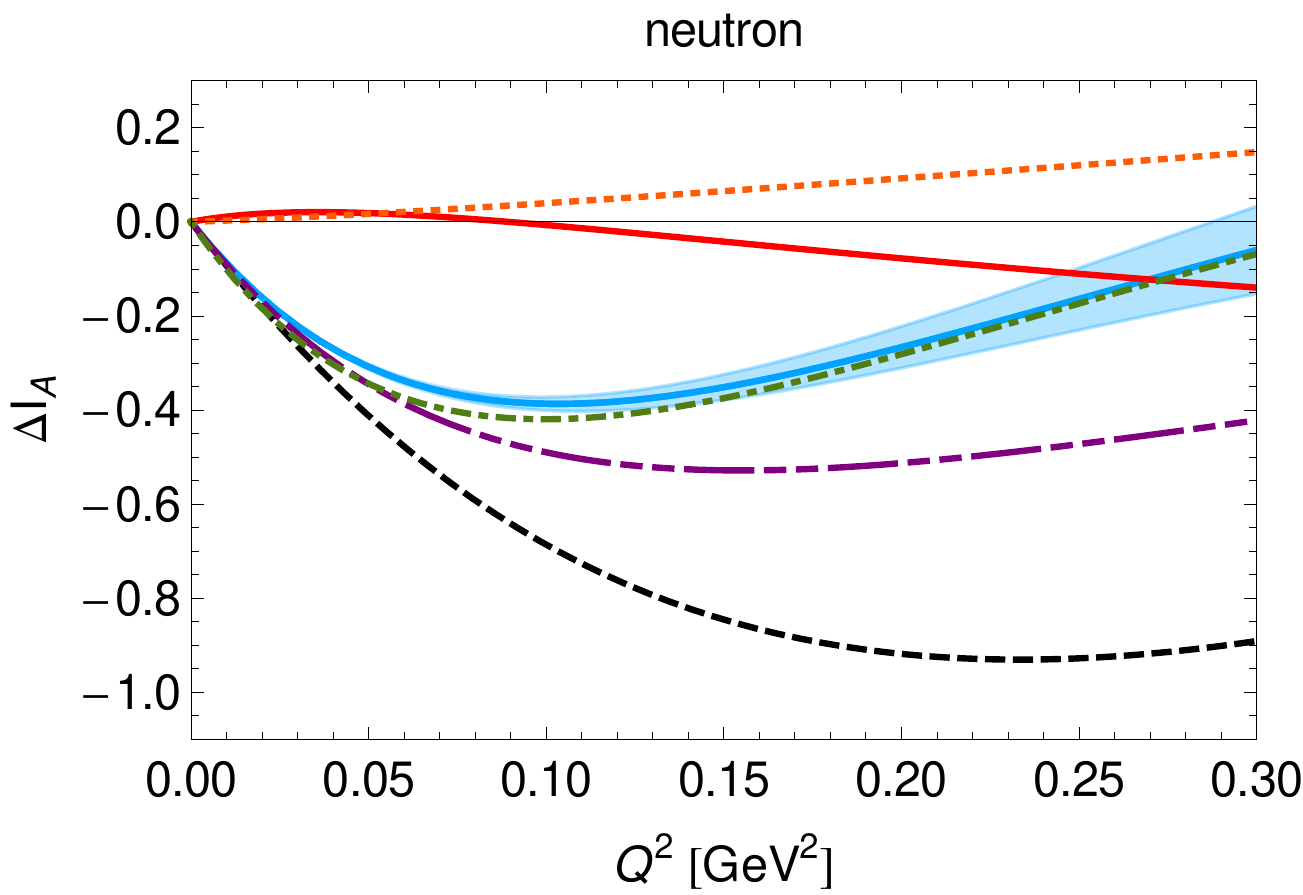}\\[0.2cm]
\includegraphics[width=0.49\textwidth]{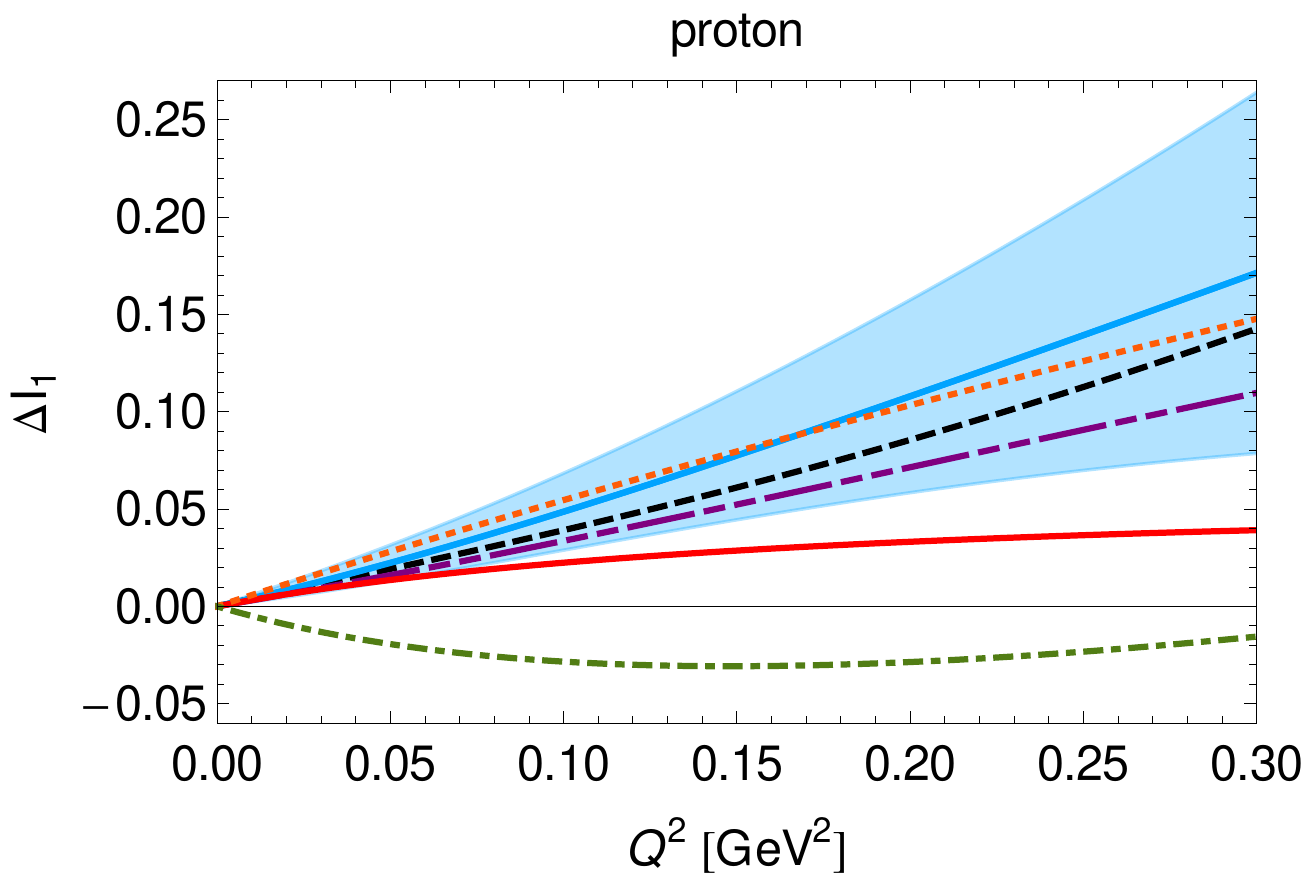}
\includegraphics[width=0.49\textwidth]{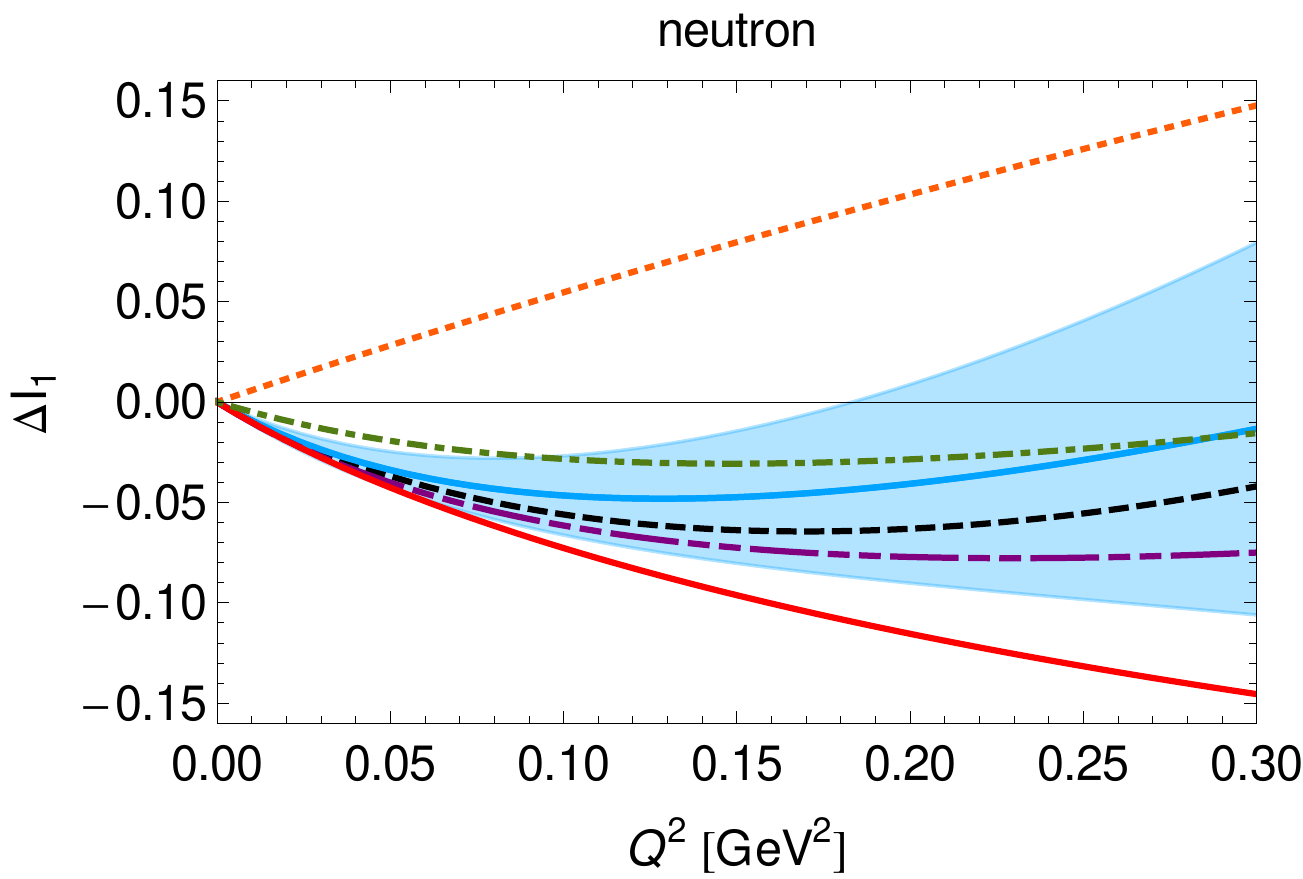}
\caption{Contributions of the different orders to the chiral predictions of $\Delta I_A(Q^2)$ \{upper panel\} and $\Delta I_1(Q^2)$ \{lower panel\} for the proton (left) and neutron (right). Red solid line: $\pi N$-loop contribution, green dot-dashed line: $\Delta$-exchange contribution, orange dotted line: $\pi \Delta$-loop contribution, blue solid line and blue band: total result, purple dot-dot-dashed line: total result without $g_C$ contribution, black short-dashed line: total result without $g_M$ dipole. \label{Fig:IA-orders-plot}}
\end{center}
\end{figure}

The generalized GDH integral $I_A(Q^2)$ is directly related to the non-pole amplitude $g_{TT}^\mathrm{nonpole}(\nu,Q^2)$, which differs from non-Born amplitude $\ol g_{TT}(\nu,Q^2)$ by a term involving the elastic Pauli form factor:
\beq
g_{TT}^\mathrm{nonpole}(\nu,Q^2)=\ol g_{TT}(\nu,Q^2)-\frac{2\pi\al \nu}{M_N^2}F_2^2(Q^2),
\eeq
cf.\ Eqs.~(\ref{Eq:GGT-S1S2}) and (\ref{eq:S1pole}). Consequently, $I_A(Q^2)$ is not a pure polarizability, but also contains an elastic contribution. 
The ``non-polarizability'' or the Born part of $I_A(Q^2)$ is given by:
\beq
I_A^\mathrm{Born}(Q^2)=I_A(Q^2)-\Delta I_A(Q^2)=-\frac{1}{4}F_2^2(Q^2),\eqlab{genGDHnonpoleA}
\eeq
where we refer to the polarizability part as $\Delta I_A(Q^2)$. The same is true for the generalized GDH integral $I_1(Q^2)$, which is directly related to $S_1^\mathrm{nonpole}(\nu,Q^2)$:
\beq
I_1^\mathrm{Born}(Q^2)=I_1(Q^2)-\Delta I_1(Q^2)=-\frac{1}{4}F_2^2(Q^2).\eqlab{genGDHnonpole}
\eeq
In the following, we will add the Born parts to our LO and NLO B$\chi$PT predictions for the polarizabilities $\Delta I_A(Q^2)$ and $\Delta I_1(Q^2)$, employing an empirical parametrization for the  elastic Pauli form factor \cite{Bradford:2006yz}. This allows us to compare to the experimental results for $I_A(Q^2)$ and $I_1(Q^2)$, cf.\ Fig.~\ref{Fig:IAplot}. Note that the blue error bands only describe the uncertainties of our B$\chi$PT predictions of the polarizabilities, while the elastic contributions are considered to be exact, as explained in Sec.~\ref{Sec:VVCS_calc}. The uncertainties of the polarizability predictions are therefore better reflected in Fig.~\ref{Fig:IA-orders-plot}, where we show the contributions of the different orders to the  B$\chi$PT predictions of $\Delta I_A(Q^2)$ and $\Delta I_1(Q^2)$, as well as the total results with error bands.

The E97-110 experiment at Jefferson Lab has recently published their data for $I_{An}(Q^2)$ in the region of $0.035\, \mathrm{GeV}^2<Q^2<0.24\, \mathrm{GeV}^2$ \cite{Sulkosky:2019zmn}. In addition, there are results for $I_{An}(Q^2)$ from the earlier E94-010 experiment \cite{Amarian:2002ar}, and for $I_{Ap}(Q^2)$ from the E08-027 experiment \cite{Zielinski:2017gwp}.  
The $\mathcal{O}(p^4)$ HB calculation gives a large negative effect \cite{Kao:2003jd}, which does not describe the data. The B$\chi$PT+$\Delta$ result from Ref.~\cite{Bernard:2012hb}, which mainly differs from our work by the absence of the dipole form factor in $g_M$,  looks similar to this HB result and only describes the data points at lowest $Q^2$.  Our NLO prediction, however, follows closely the $Q^2$ evolution of the data. In Fig.~\ref{Fig:IA-orders-plot} \{upper panel\}, we show the polarizability $\Delta I_A(Q^2)$, whose $Q^2$ evolution is clearly dominated by the $\Delta$ exchange.
Similar to the case of $\gamma_{0p}(Q^2)$, inclusion of the dipole in $g_M$ and the Coulomb coupling $g_C$ is very important in order to describe the experimental data. The LO prediction, on the other hand, slightly overestimates the data, cf.\ Fig.~\ref{Fig:IAplot} \{upper panel\}.

At the real-photon point: $I_A (0) = - \frac{\varkappa^2}{4}$ and $\Delta I_{A}(0)=0$. Therefore, we give only the slope of the polarizability $\Delta I_{A}(Q^2)$ [showing also the separate contributions from $\pi N$ loops,  $\Delta$ exchange and $\pi\Delta$ loops] in units of GeV$^{-2}$:
\begin{subequations}
\bea
\left.\frac{\dd \Delta I_{Ap} (Q^2)}{\dd Q^2}\right|_{Q^2=0}&=&-8.58(3.43)\approx 2.38 -11.21 + 0.25  ,\\
\left.\frac{\dd \Delta I_{An} (Q^2)}{\dd Q^2}\right|_{Q^2=0}&=& -9.55(3.43)\approx1.41 -11.21 + 0.25 .
\eea
\end{subequations}
Including the empirical Pauli form factor \cite{Bradford:2006yz}, we find, in units of GeV$^{-2}$:
\beq
\left.\frac{\dd I_{Ap} (Q^2)}{\dd Q^2}\right|_{Q^2=0}=-3.18, \qquad \left.\frac{\dd I_{An} (Q^2)}{\dd Q^2}\right|_{Q^2=0}= -3.00.
\eeq

\subsection{\boldmath{$\Gamma_1(Q^2)$} and \boldmath{$I_1(Q^2)$} --- the first moment of the structure function \boldmath{$g_1(x,Q^2)$}} 
\label{I1sec}

The second variant for a generalization of the GDH sum rule to finite $Q^2$ is defined as:
\bea
-\frac{\alpha}{2 M_N^2}I_1(Q^2) &=& -\frac{1}{8 \pi^2} \int_{\nu_0}^\infty \! \dd\nu\, \frac{1}{\sqrt{\nu^{2}+Q^2}} \Big[ \sigma_{TT} (\nu,Q^2) + \frac{Q}{\nu} \sigma_{LT} (\nu,Q^2)\Big] \label{Eq:I1-SumRule}\\
&=&-\frac{\alpha}{Q^2}\int_0^{x_0}\!\dd x\,g_1(x,Q^2),\nn
\eea
where $I_1 (0) = - \frac{\varkappa^2}{4}$. This generalized GDH integral directly stems from the amplitude $S_1^\mathrm{nonpole}(\nu,Q^2)$ with the LEX from \Eqref{LEXS1}.
It is given by the first moment of the structure function $g_1(x,Q^2)$,
$\Gamma_1(Q^2) = \int^{x_0}_0\! \dd x\, g_1(x,Q^2)$, 
as follows:
$
I_1(Q^2) = \frac{2M_N^2}{Q^2}\,\Gamma_1(Q^2).
$
 The isovector combination:
 \begin{align}
 \Gamma_{1(p-n)}(Q^2) = \int^{x_0}_0\! \dd x\, \left[g_{1p}(x,Q^2) - g_{1n}(x,Q^2) \right], 
\end{align}
is related to the axial coupling of the nucleon through the Bjorken sum rule \cite{Bjorken:1966jh,Bjorken:1969mm}:
 \begin{align}
\lim_{Q^2 \to \infty} \Gamma_{1(p-n)}(Q^2) = \frac{g_A}{6}.
\end{align}

\begin{figure}[t]
\begin{center}
\includegraphics[width=0.49\textwidth]{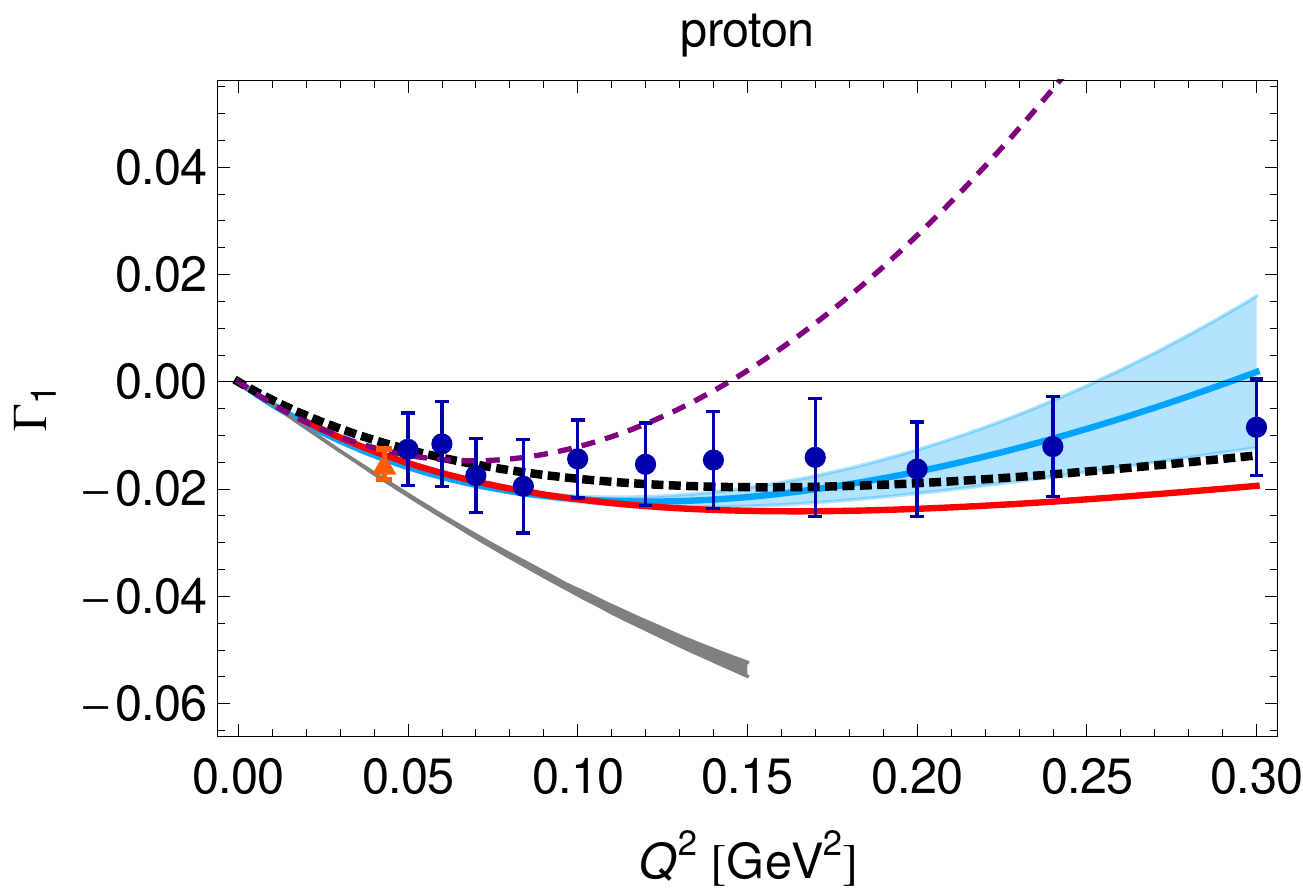}
\includegraphics[width=0.49\textwidth]{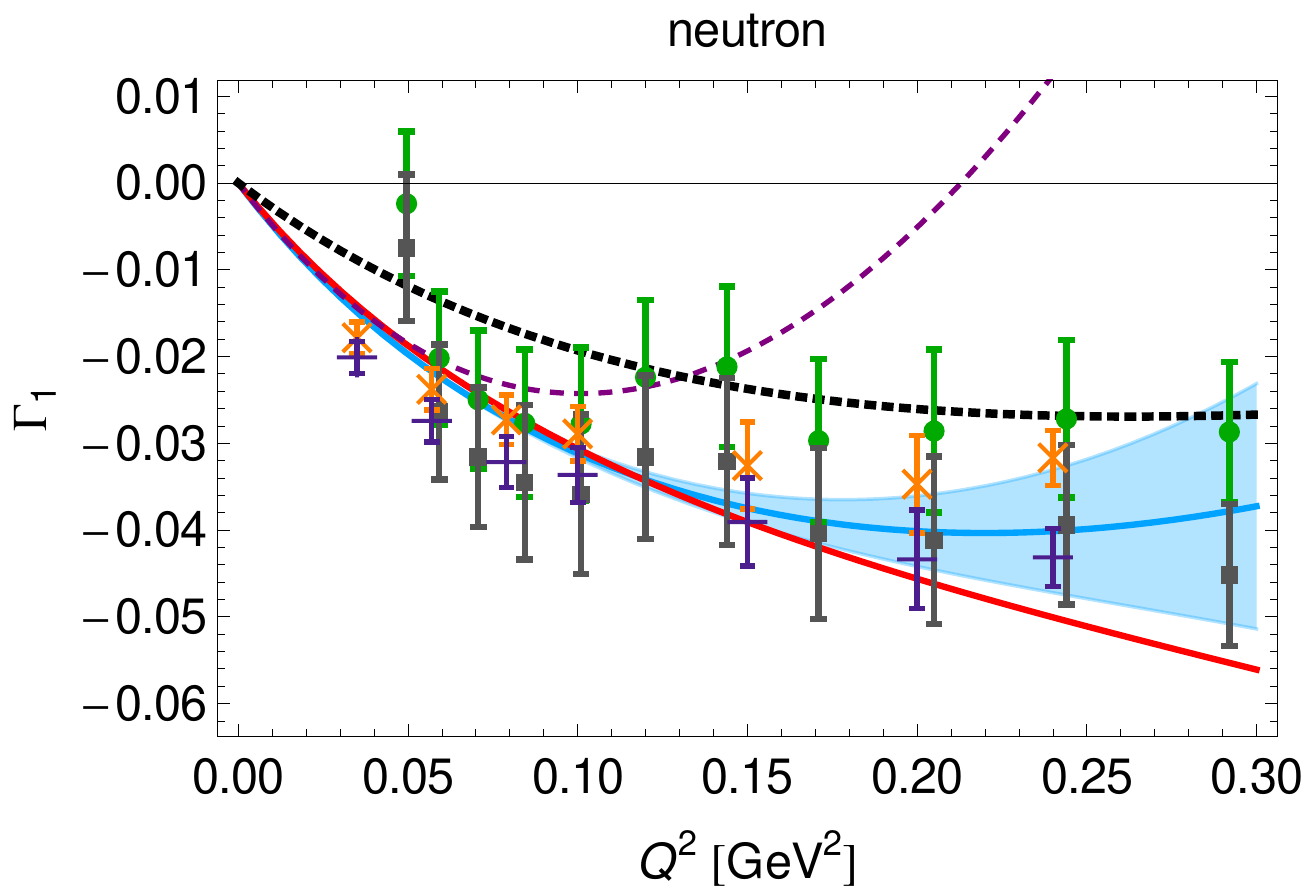}
\caption{First moment of the structure function $g_1(x,Q^2)$ for the proton (left) and neutron (right) as function of $Q^2$. The legend is the same as in Fig.~\ref{Fig:IAplot}.\label{Fig:Gamma1plot}} 
\end{center}
\end{figure}

As explained in \Eqref{genGDHnonpoleA}, the moment $I_1(Q^2)$ splits into a polarizability part $\Delta I_1(Q^2)$ and a Born part $I_1^\mathrm{Born}(Q^2)$. Figure~\ref{Fig:IAplot} \{lower panel\} shows the $Q^2$ dependence of $I_1(Q^2)$ which, in contrast to $I_A(Q^2)$ shown in Figure~\ref{Fig:IAplot} \{upper panel\}, is clearly dominated by its Born part and the elastic Pauli form factor. 
 The $\pi N$-loop, $\Delta$-exchange and $\pi\Delta$-loop contributions to the polarizability $\Delta I_1(Q^2)$ are  shown in Fig.~\ref{Fig:IA-orders-plot} \{lower panel\}. Comparing to Fig.~\ref{Fig:IA-orders-plot} \{upper panel\}, one sees that $\Delta I_1(Q^2)$ is less sensitive to $g_C$ and the dipole form factor in $g_M$ than $\Delta I_A(Q^2)$. 

For the proton, our NLO B$\chi$PT prediction gives a very good description of the experimental data \cite{Prok:2008ev,Zielinski:2017gwp} and is in reasonable agreement with the MAID prediction \cite{MAID}. For the neutron, one observes good agreement with the empirical evaluations including extrapolations to unmeasured energy regions starting from $Q^2>0.1$ GeV$^2$ \cite{Sulkosky:2019zmn,Guler:2015hsw}. In the region of $Q^2<0.05$ GeV$^2$, one observes an interesting tension between the recent E97-110 experiment \cite{Sulkosky:2019zmn} and the data from CLAS \cite{Guler:2015hsw}. While the newest measurement finds $I_{1n}(0.035\,\mathrm{GeV}^2)<\varkappa^2_n/4$, thus suggesting a negative slope at low $Q$, the older measurement found a rather large value for $I_{1n}(0.0496\,\mathrm{GeV}^2)$. A similar but milder behaviour is seen in the E97-110 \cite{Sulkosky:2019zmn} and E94-010 \cite{Amarian:2002ar} data for $I_{An}$. The MAID predictions do not agree with the CODATA recommended values for the anomalous magnetic moments of the proton and neutron \cite{Mohr:2012aa}, which in our work are imposed by using empirical parametrizations for the elastic Pauli form factors \cite{Bradford:2006yz}.
The slope of the HB result from Ref.~\cite{Kao:2003jd} is too large and therefore only reproduces the data at very low $Q^2$.

\begin{figure}[t]
\begin{center}
\includegraphics[width=0.49\textwidth]{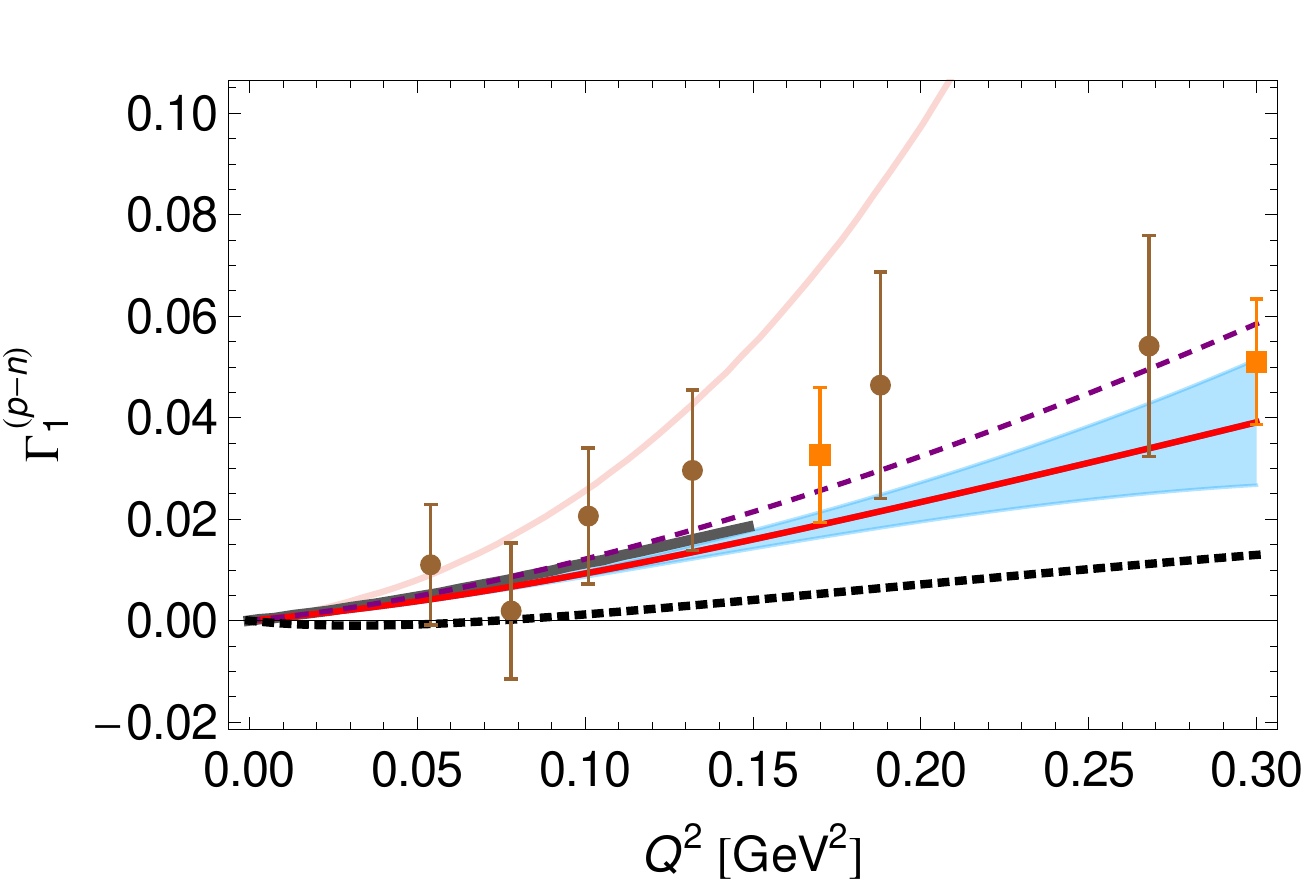}
\caption{Isovector combination of $\Gamma_1(Q^2)$ as function of $Q^2$. The legend is the same as in Fig.~\ref{Fig:Gamma1plot}. 
The pink curve is the IR result from Ref.~\cite{Bernard:2002pw,Bernard:2002bs}. 
The experimental points are from Ref.~\cite{Deur:2008ej} (brown dots) and Ref.~\cite{Deur:2004ti} (orange squares).  \label{Fig:Gamma1-isovector-plot}}
\end{center}
\end{figure}

Figure~\ref{Fig:Gamma1plot} shows the moment $\Gamma_1(Q^2)$ for the proton and neutron, while Fig.~\ref{Fig:Gamma1-isovector-plot} shows the isovector combination $\Gamma_{1,\,p-n}(Q^2)$. 
The LO and NLO B$\chi$PT predictions are identical, because our calculation
produces the same Delta contributions for the proton and the neutron.
For the isovector combination, the MAID model only agrees with the data at very low $Q^2 < 0.10$~GeV$^2$. The same is true for the IR result \cite{Bernard:2002pw,Bernard:2002bs}, while all other chiral results describe the data: NLO B$\chi$PT (this work), B$\chi$PT+$\Delta$ \cite{Bernard:2012hb} and HB$\chi$PT \cite{Kao:2003jd}.

At the real-photon point: $I_1 (0) = - \frac{\varkappa^2}{4}$ and $\Delta I_{1}(0)=0$. Therefore, we give only the slope of the polarizability $\Delta I_{1}(Q^2)$ [showing also the separate contributions from $\pi N$ loops,  $\Delta$ exchange and $\pi\Delta$ loops] in units of GeV$^{-2}$:
\begin{subequations}
\bea
\left.\frac{\dd \Delta I_{1p} (Q^2)}{\dd Q^2}\right|_{Q^2=0}&=&\hphantom{-} 0.39(4)\approx 0.34 - 0.53 + 0.58   ,\\
\left.\frac{\dd \Delta I_{1n} (Q^2)}{\dd Q^2}\right|_{Q^2=0}&=&-1.01(10)\approx -1.07 - 0.53 + 0.58  .
\eea
\end{subequations}
Including the empirical Pauli form factor \cite{Bradford:2006yz}, we find, in units of GeV$^{-2}$:
\beq
\left.\frac{\dd  I_{1p} (Q^2)}{\dd Q^2}\right|_{Q^2=0}=5.80, \qquad 
\left.\frac{\dd  I_{1n} (Q^2)}{\dd Q^2}\right|_{Q^2=0}=5.53 .
\eeq

\begin{figure}[tbh]
\begin{center}
\includegraphics[width=0.49\textwidth]{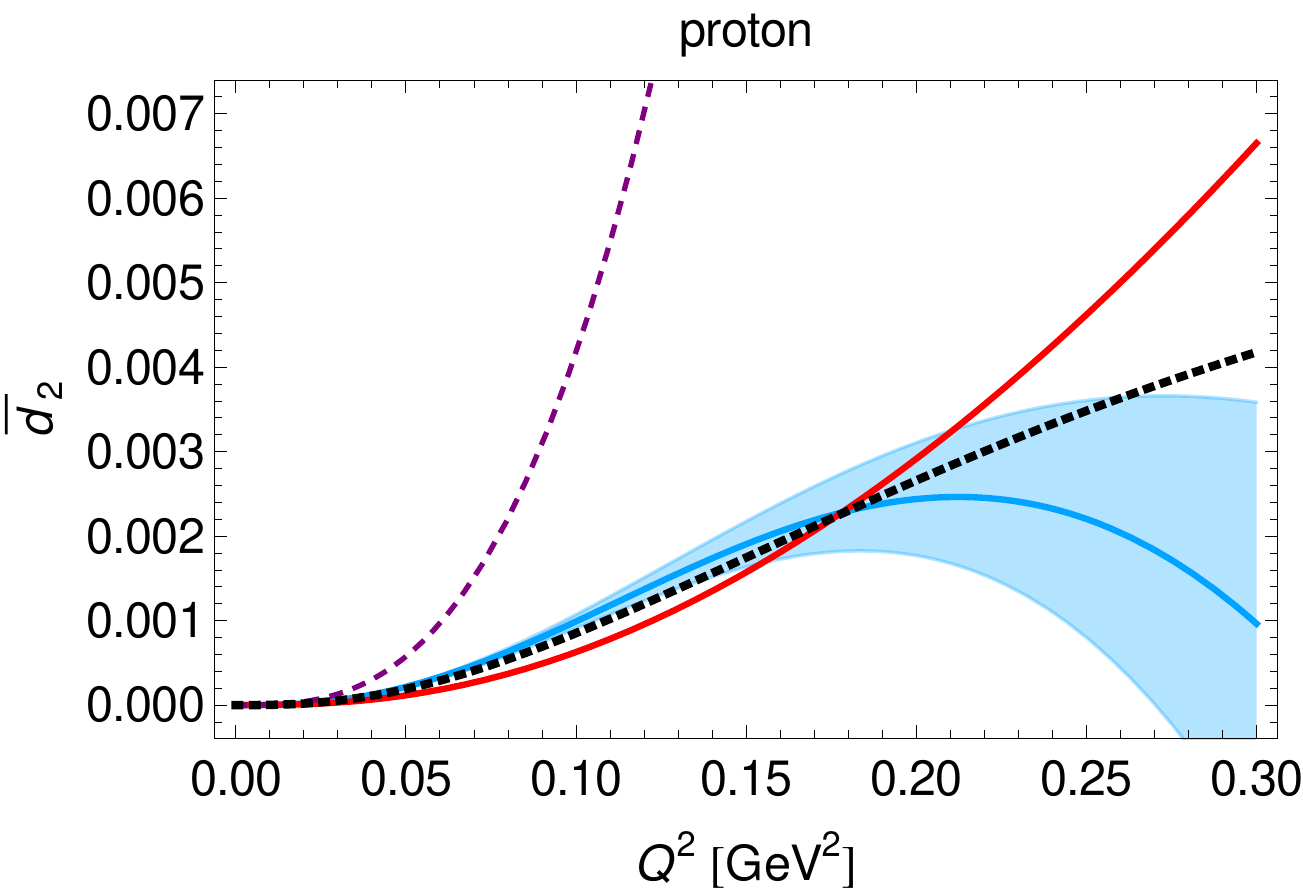}
\includegraphics[width=0.49\textwidth]{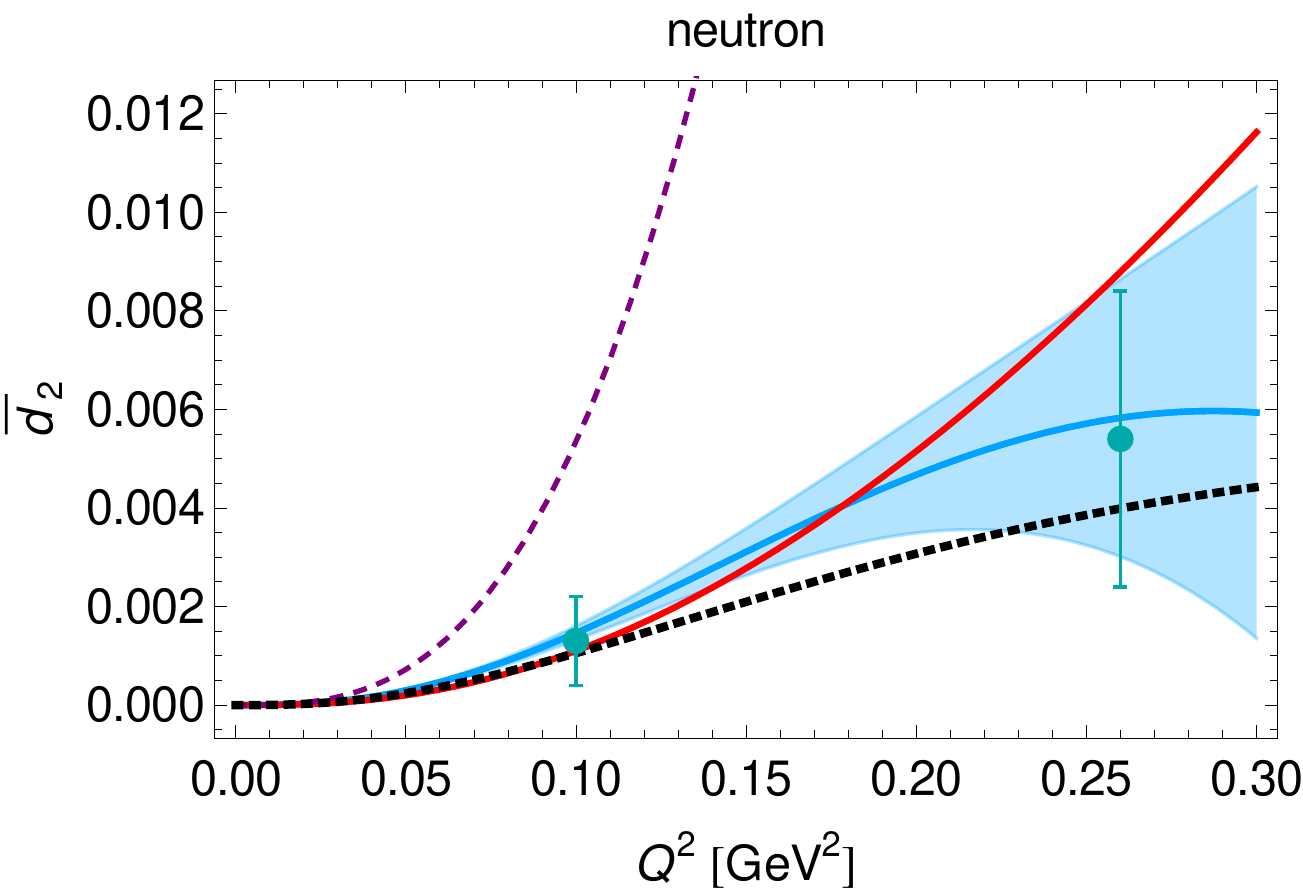}\\[0.2cm]
\includegraphics[width=0.49\textwidth]{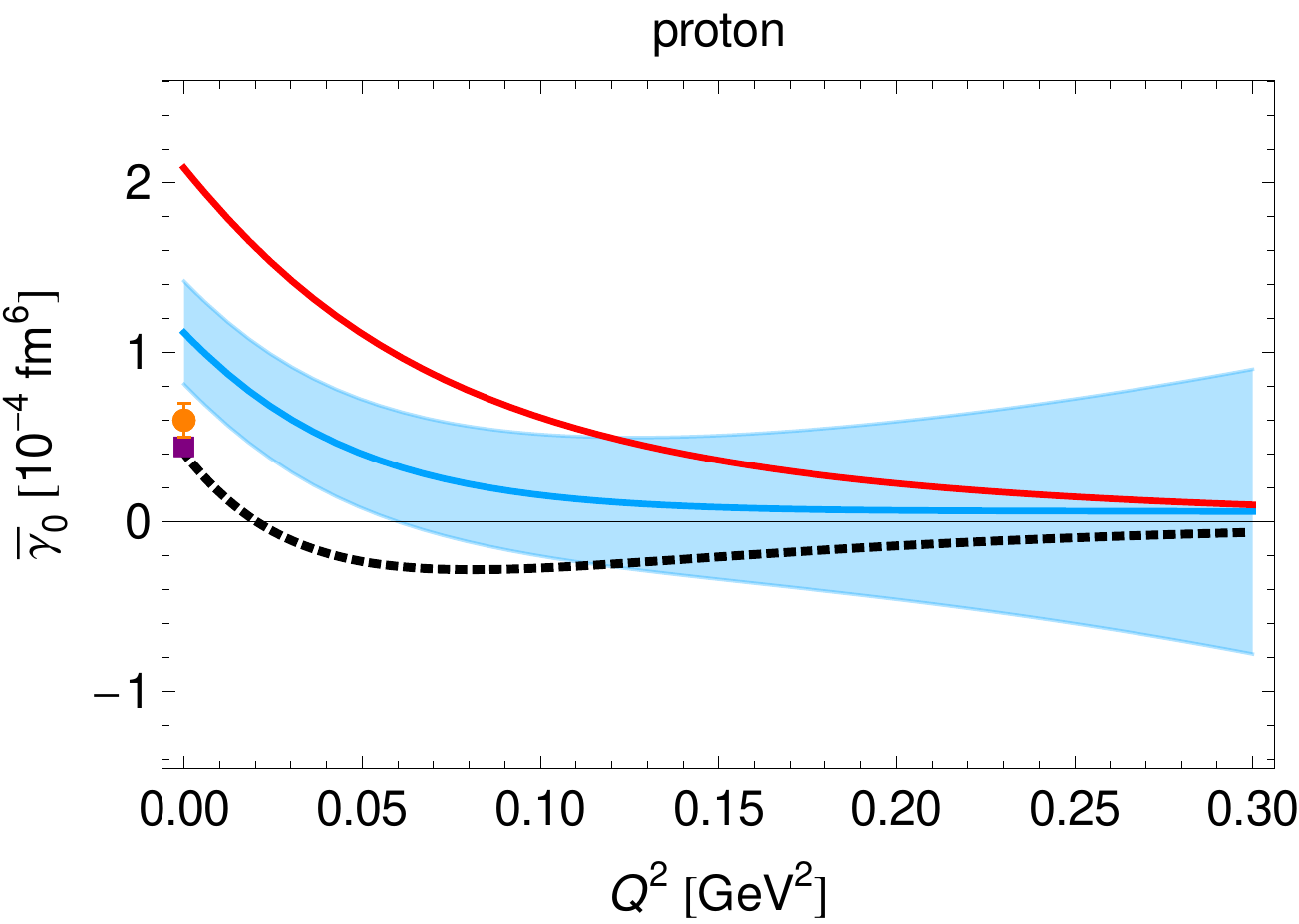}
\includegraphics[width=0.49\textwidth]{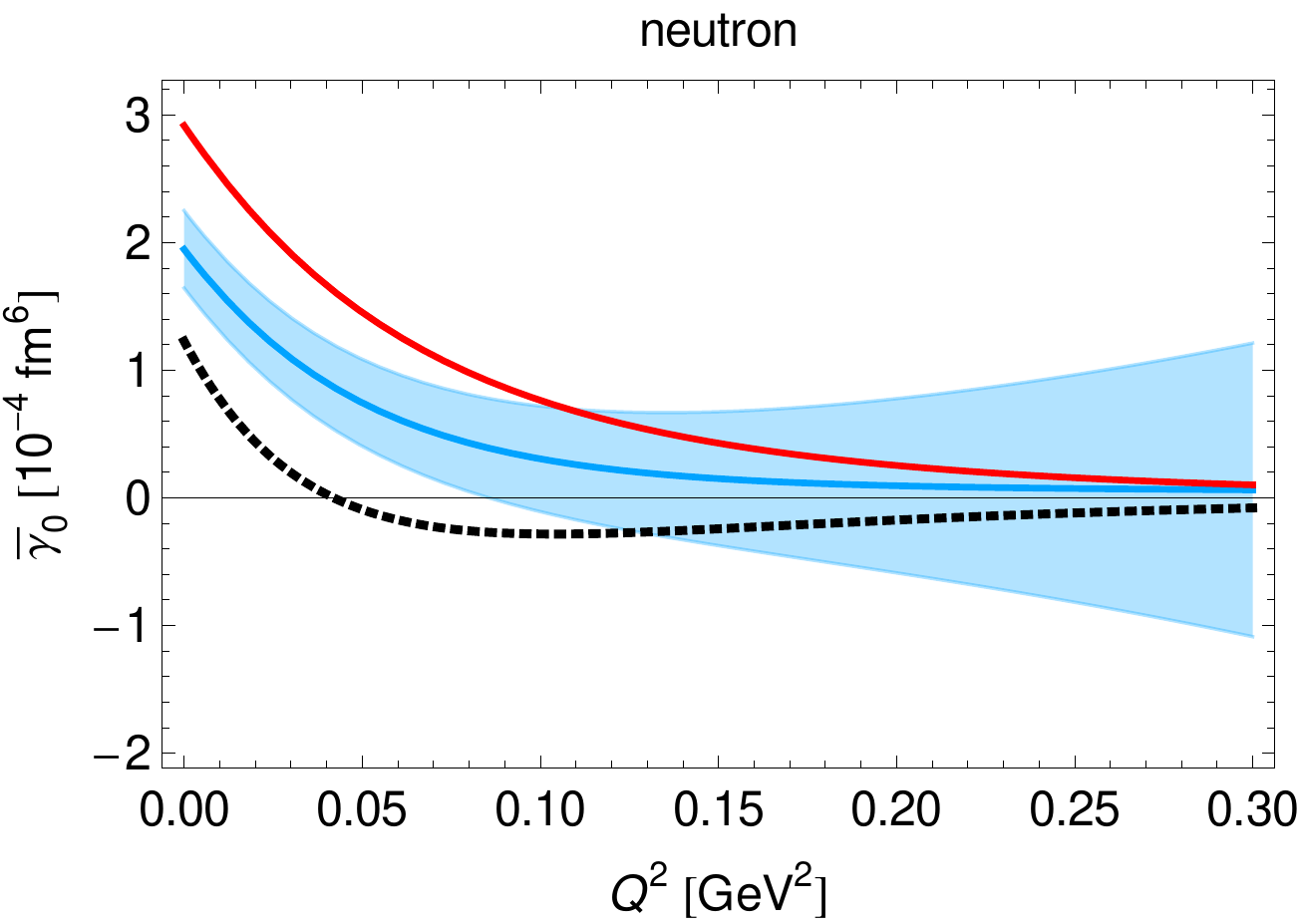}
\caption{Upper panel: The inelastic moment $\bar{d}_2(Q^2)$ for the proton (left) and neutron (right) as function of $Q^2$. The result of this work, the NLO B$\chi$PT prediction, is shown by the blue solid line and the blue band.
The red line represents the LO B$\chi$PT result. The purple short-dashed line is the $\mathcal{O}(p^4)$ HB result from Ref.~\cite{Kao:2002cp,Kao:2003jd}. The black dotted line is the MAID model prediction \cite{MAID}.  The experimental points for the neutron (cyan dots) are from Ref.~\cite{Amarian:2003jy}. Lower panel: Fifth-order generalized forward spin polarizability $\ol \ga_0(Q^2)$, for the proton (left) and neutron (right) as function of $Q^2$. The experimental points for the proton are from Ref.~\cite{Gryniuk:2016gnm} (purple square) and Ref.~\cite{Pasquini:2010zr} (orange dot).
\label{Fig:d2plot}}
\end{center}
\end{figure}
\subsection{\boldmath{$\bar{d}_2(Q^2)$} --- a measure of color polarizability}\label{d2Sec}

Another interesting moment to consider is  $d_2(Q^2)$, which is related to the twist-3 part of the spin structure function $g_2(x,Q^2)$ \cite{Jaffe:1989xx,Shuryak:1981pi}:
\begin{align}\label{Eq:d2Def}
 d_2(Q^2) \equiv 3 \int^1_0 \!\! \dd x\ x^2 [g_2(x,Q^2) - g_2^{WW}(x,Q^2)],
\end{align}
where $g_2^{WW}(x,Q^2)$ is the twist-2 part of $g_2(x,Q^2)$. Using the Wandzura-Wilczek relation \cite{Wandzura:1977qf}, one can relate $d_2(Q^2)$ to the moments of the spin structure functions $g_1(x,Q^2)$ and $g_2(x,Q^2)$:
\begin{align}\label{Eq:d2WW}
 d_2(Q^2) =  \int^1_0 \!\! \dd x\ x^2 \,[3 g_2(x,Q^2) + 2 g_1(x,Q^2)].
\end{align}
This relation, however, only holds for asymptotically large $Q^2$. It is also in the high-$Q^2$ region, where $d_2(Q^2)$ is a measure of color polarizability \cite{Filippone:2001ux,Burkardt:2009rf}, through its relation to the gluon field strength tensor \cite{Shuryak:1981pi}. We refer to Ref.~\cite{Deur:2018roz} for a recent review on the spin structure of the nucleon, including a discussion of sum rules for deep inelastic scattering and color polarizabilities.

What we consider in the following is the inelastic part of $d_2(Q^2)$, defined as the moment of $g_1(x,Q^2)$ and $g_2(x,Q^2)$ spin structure functions, cf.\ Eq.~(\ref{Eq:d2WW}): 
\begin{align}
 \bar{d}_2(Q^2) =  \int^{x_0}_0 \!\! \dd x\ x^2 \,[3 g_2(x,Q^2) + 2 g_1(x,Q^2)].
\end{align}
This moment provides another testing ground for our B$\chi$PT predictions through comparison with experiments on the neutron \cite{Amarian:2003jy}. Going towards the low-$Q^2$ region, the interpretation of $\bar{d}_2(Q^2)$ in terms of color polarizabilities will fade out. The above definition, however, implies it is related to other VVCS polarizabilities:
\beq
\label{Eq:d2inelDef}
 \bar{d}_2(Q^2) = \frac{Q^4}{8 M_N^4}\left[ \frac{M_N^2 Q^2}{\alpha}\delta_{LT}(Q^2) +I_1(Q^2) - I_A(Q^2) \right].
\eeq
Note that $\bar{d}_2(Q^2)$ and its first two derivatives with respect to $Q^2$ vanish at $Q^2=0$. The considerations in Eqs.~\eref{genGDHnonpoleA} and \eref{genGDHnonpole} have no effect on $\bar{d}_2(Q^2)$, since the Born contribution from $I_A(Q^2)$ and $I_1(Q^2)$ cancel out. Therefore, $\bar{d}_2(Q^2)$ is a pure polarizability.

In Fig.~\ref{Fig:d2plot} \{upper panel\}, we show our NLO B$\chi$PT prediction and other results for $\bar{d}_2(Q^2)$. While MAID \cite{MAID} and B$\chi$PT describe the experimental data for the neutron \cite{Amarian:2003jy} very well, the HB limit \cite{Kao:2002cp,Kao:2003jd} is showing a fast growth with $Q^2$. This illustrates the importance of keeping the relativistic result. Note also that, even though the $\pi N$-loop contribution is dominant, both $g_C$ and the form factor in $g_M$ are essential to obtain a curvature that reproduces the data, cf.\ Fig.~\ref{Fig:d2-orders-plot} \{upper panel\}. For the proton there are, to our knowledge, no experimental results to compare with. However, the agreement between the NLO B$\chi$PT prediction and the MAID prediction at low energies is reasonable.

\subsection{\boldmath{$\ol \ga_0(Q^2)$} --- fifth-order generalized forward spin polarizability}
\label{Sec:ForwardSpinPolarizability5thOrder}

It is interesting to compare the generalized fifth-order forward spin polarizability sum rule,
\bea
\bar\gamma_0 (Q^2)&=& \frac{1}{2 \pi^2} \int_{\nu_0}^\infty \! \dd\nu\,\sqrt{1+\frac{Q^2}{\nu^2}} \,\frac{\sigma_{TT} (\nu,Q^2)}{\nu^5}\label{Eq:bargamma0Q2}\\
&=&\frac{64 \al M_N^4}{Q^{10}}\int_0^{x_0}\!\dd x \, x^4 \! \left[g_1(x,Q^2)-\frac{4M_N^2 x^2}{Q^2}\,g_2(x,Q^2)\right]\!,\nn
\eea
to the sum rule integrals for $I_A(Q^2)$ and $\ga_0(Q^2)$, since they differ merely by their energy weighting of $\sigma_{TT}(\nu,Q
^2)$ and a constant prefactor, cf.\ Eqs.~\eqref{Eq:gamma0Q2}, \eqref{Eq:IA-SumRule} and \eqref{Eq:bargamma0Q2}. From $I_A(Q^2)$ to $\ga_0(Q^2)$ to $\ol \ga_0(Q^2)$, the energy suppression is increasing by a factor of $\nu^{-2}$, respectively. Therefore, the description of $\ol \ga_0(Q^2)$ should be easiest in a low-energy effective-field theory such as $\chi$PT, whereas $\ga_0(Q^2)$ and $I_A(Q^2)$ receive larger contributions from higher energies.  

\begin{figure}
\begin{center} 
\includegraphics[width=0.49\textwidth]{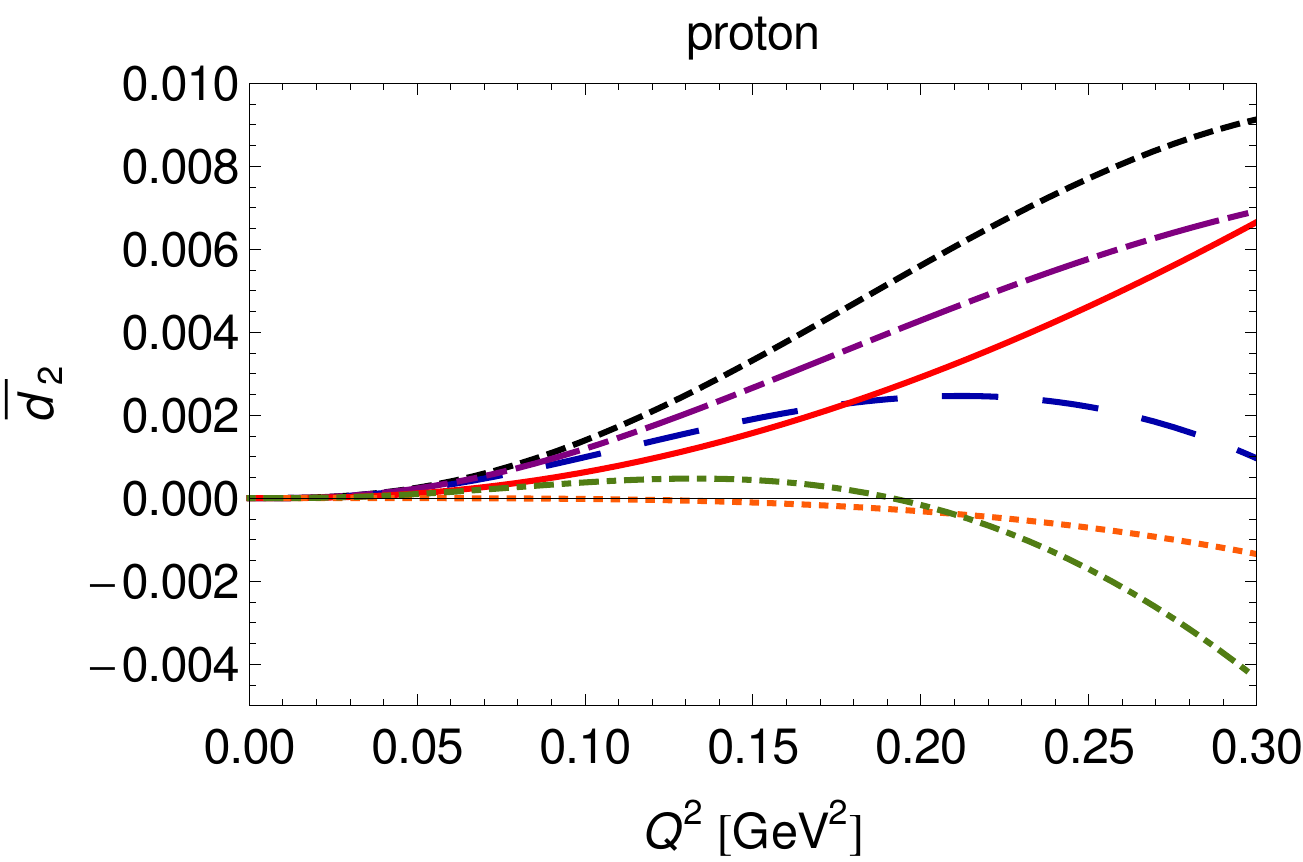}
\includegraphics[width=0.49\textwidth]{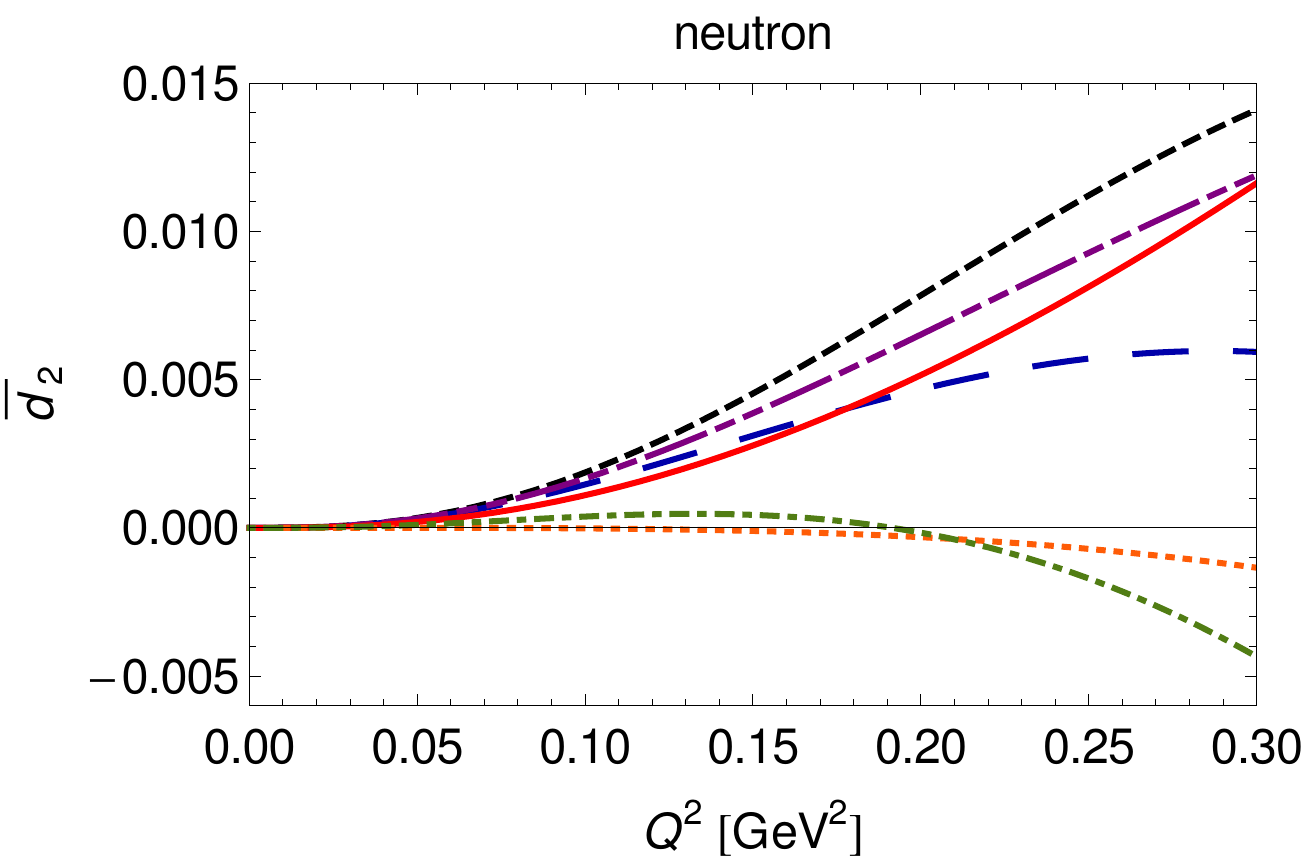}\\[0.2cm]
\includegraphics[width=0.49\textwidth]{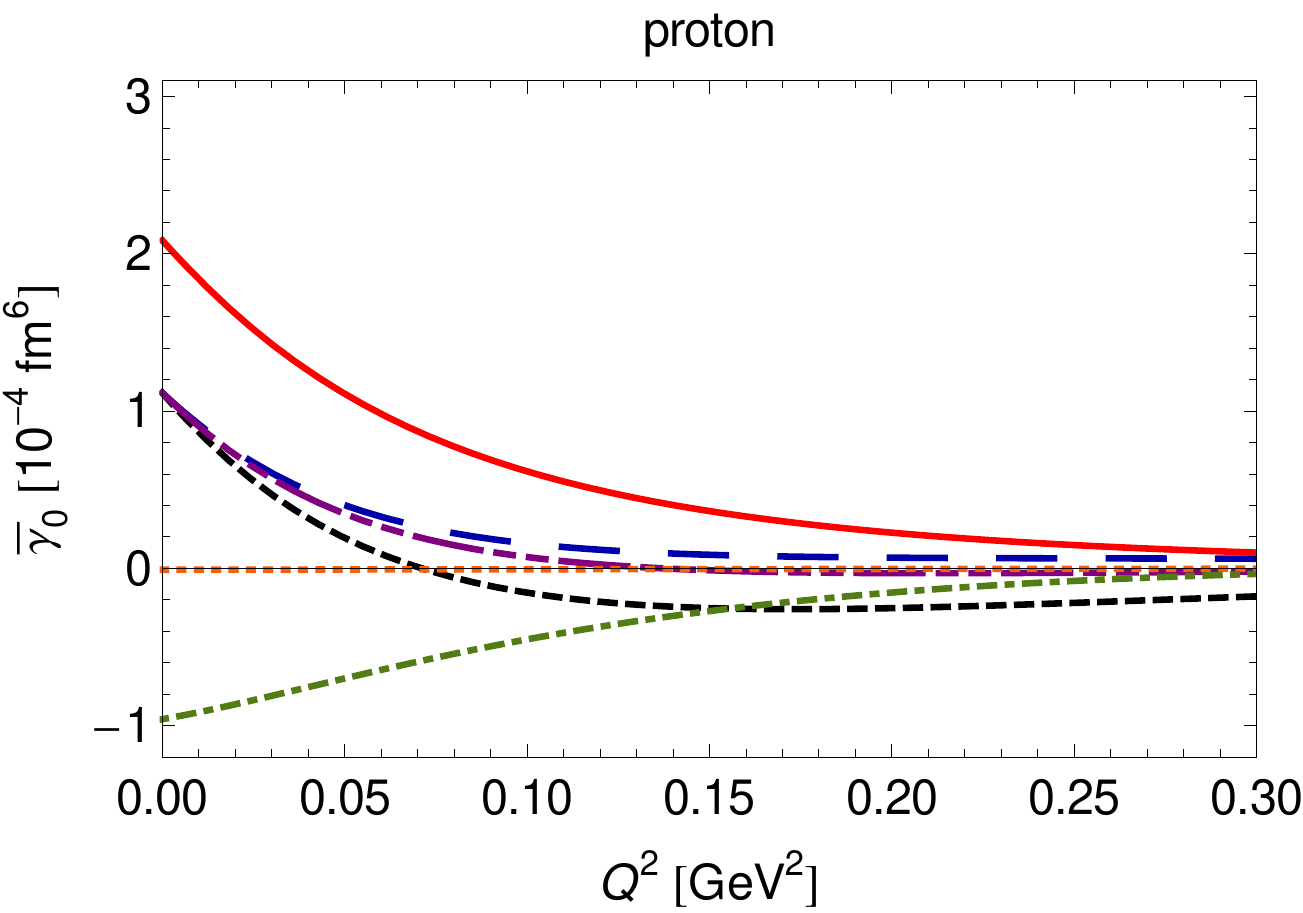}
\includegraphics[width=0.49\textwidth]{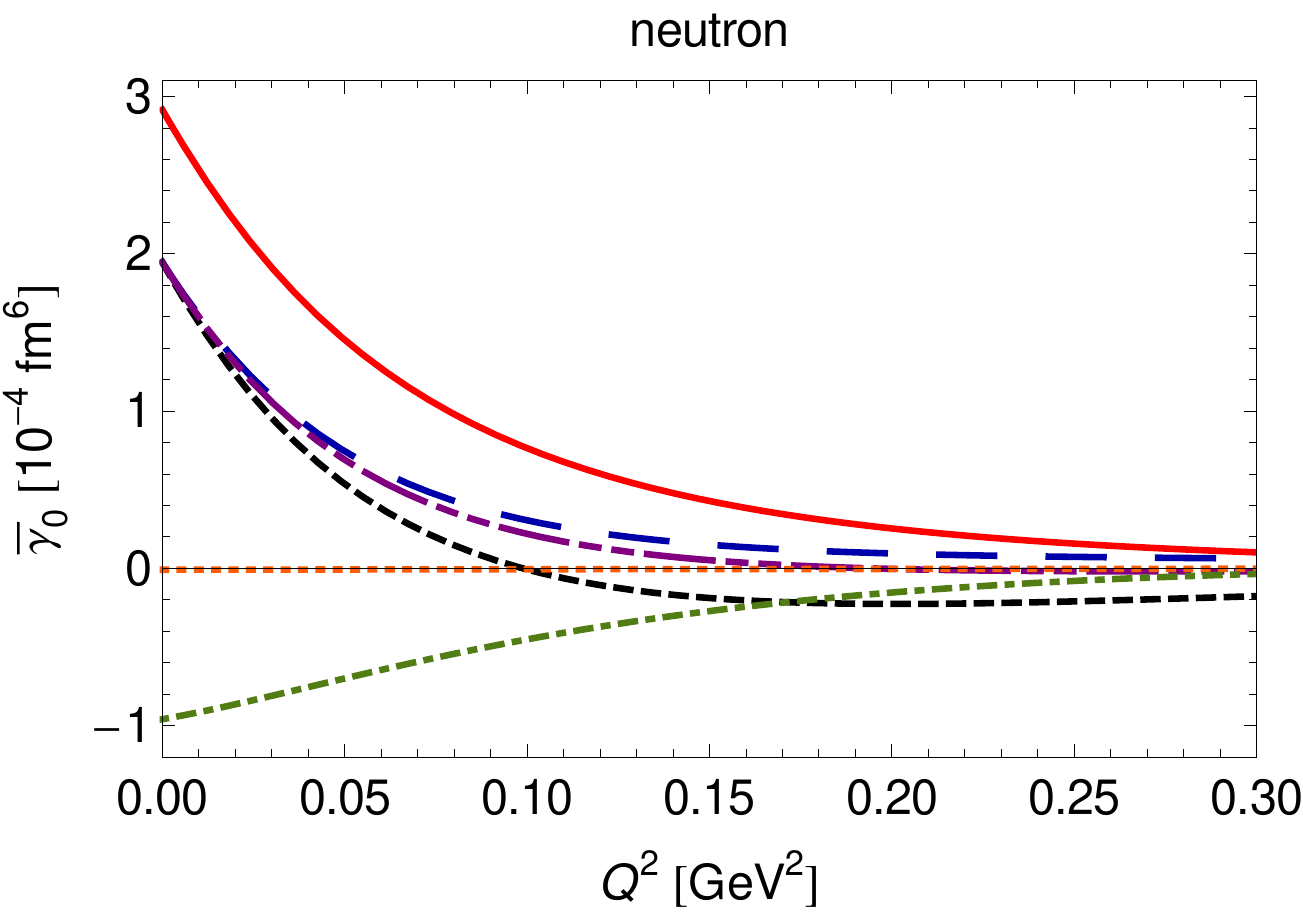}
\caption{Contributions of the different orders to the chiral predictions of $\bar{d}_2(Q^2)$
\{upper panel\} and $\bar\gamma_0(Q^2)$ \{lower panel\}
for the proton (left) and neutron (right). Red solid line: $\pi N$-loop contribution, green dot-dashed line: $\Delta$-exchange contribution, orange dotted line: $\pi \Delta$-loop contribution, blue long-dashed line: total result, purple dot-dot-dashed line: total result without $g_C$ contribution, black short-dashed line: total result without $g_M$ dipole. \label{Fig:d2-orders-plot}}
\end{center}
\end{figure}

In Fig.~\ref{Fig:d2plot} \{lower panel\}, we show our LO and NLO B$\chi$PT predictions for $\ol \ga_0(Q^2)$. One can see that the $\pi N$-loop contribution is positive (in accordance to what we see for the cross section $\sigma_{TT}$, see Fig.\ \ref{Fig:SummaryCrossSections}). The Delta shifts it substantially, especially in the low $Q^2$ region, bringing it into a better agreement with data.
In general, the B$\chi$PT curves start above the empirical data points at the real-photon point, and then decrease asymptotically to zero above $Q^2>0.1$ GeV$^2$. On the other hand, the MAID prediction reproduces the empirical data at the real-photon point, then decreases to negative values until about $Q^2>0.06$ GeV$^2$, from where it also starts to asymptotically approach zero. Consequently, our NLO B$\chi$PT prediction of $\ol \ga_0(Q^2)$ is consistently above the MAID prediction. This is very different to what we saw for $I_A(Q^2)$ in Fig.~\ref{Fig:IAplot} \{upper panel\}, where the MAID prediction at the real-photon point is above the experimental value. While the agreement of our predictions with the empirical data is in general quite good for all moments of $\sigma_{TT}(\nu,Q
^2)$, one should point out that both for $\ga_{0n}(Q^2)$ and $\ol \ga_{0p}(Q^2)$ we overestimate the data at low $Q^2$. For $I_A(Q^2)$ such observation cannot be made because $\Delta I_A(0)=0$, and thus, $I_A(0)$ is given by the empirical Pauli form factor only. From $I_A(Q^2)$, $\ga_0(Q^2)$ and $\ol \ga_0(Q^2)$, the latter has the smallest, however, non-negligible dependence on $g_C$ and the dipole in $g_M$, cf.\ Fig.~\ref{Fig:d2-orders-plot} \{lower panel\}.

The $\pi N$-loop, $\Delta$-exchange, and $\pi \Delta$-loop contributions to the NLO B$\chi$PT prediction of the fifth-order forward spin polarizability amount to, in units of $10^{-4}$~fm$^6$:
\begin{subequations}
\begin{align}
&\bar\gamma_{0p} =1.12(30)\approx 2.08 -0.96 -0.01, \label{Eq:gamma0BarProtonRealPoint}\\
&\bar\gamma_{0n} =1.95 (30) \approx 2.92-0.96 -0.01,
\end{align}
\end{subequations}
while the slope is composed as follows, in units of $10^{-4}$~fm$^8$:
\begin{subequations}
\bea
\left.\frac{\dd\bar\gamma_{0p} (Q^2)}{\dd Q^2}\right|_{Q^2=0}&=&-0.84 (10)\approx -1.00+ 0.16 +0.00, \\
\left.\frac{\dd\bar\gamma_{0n} (Q^2)}{\dd Q^2}\right|_{Q^2=0}&=&-1.42(15) \approx -1.58+ 0.16+0.00.
\eea
\end{subequations}
Note that the HB prediction of $\bar \ga_{0p}$ ($4.23$ at $\mathcal{O}(p^3)$ and $3.65 $ at $\mathcal{O}(\epsilon^3)$ \cite{Holstein:1999uu,Pasquini:2010zr}) is almost one order of magnitude larger than the empirical value, and therefore not shown in Fig.~\ref{Fig:d2plot}.

\subsection{Summary}\label{Sec:Summary}

\begin{table}[tbh]
\caption{The NLO B$\chi$PT predictions for the forward VVCS polarizabilities and their slopes at $Q^2=0$. The contributions of the $\pi N$ loops, the $\Delta$ exchange and the $\pi\Delta$ loops are shown, together with the combined total result. Note that $I_A(0)=I_1(0)=\bar d_2(0)=0$ and $(\bar d_2)'=0$.
\label{Table:Individual-Results-Pol}}

\begin{tabular}{cc|c|c|c|l|}
\cline{3-6} 
&& $\pi N$ loops & $\Delta$ exchange   & $\pi\Delta$ loops & \multicolumn{1}{|c|}{Total} \\
\hline
\multicolumn{1}{|c||}{$\,\gamma_0 \,$}&$p$ &$\hpm2.01$&\multirow{2}{*}{$-2.84$}&\multirow{2}{*}{$-0.10$}&$-0.93(92)$ \\
\multicolumn{1}{|c||}{$(10^{-4}$~fm$^4)$} &$n$&$\hpm2.98$&&&$\hpm 0.03(92)$\\
\hline
\multicolumn{1}{|c||}{$\delta_{LT}$}&$p$&$\hpm1.50$&\multirow{2}{*}{$-0.16$}&\multirow{2}{*}{$-0.02$}&$\hpm1.32(15)$\\
\multicolumn{1}{|c||}{$(10^{-4}$~fm$^4)$}&$n$ &$\hpm2.35$&&&$\hpm2.18(23)$\\
\hline
\multicolumn{1}{|c||}{$\ol \ga_0$}&$p$&$\hpm2.08$&\multirow{2}{*}{$-0.96$}&\multirow{2}{*}{$-0.01$}&$\hpm 1.12(30)$\\
\multicolumn{1}{|c||}{$(10^{-4}$~fm$^6)$}&$n$ &$\hpm2.92$&&&$\hpm1.95(30)$\\
\hline
\multicolumn{1}{|c||}{$(\gamma_0 )'$}&$p$ &$-0.33$&\multirow{2}{*}{$\hpm0.11$}&\multirow{2}{*}{$\hpm0.01$}&$-0.22(4)$\\
\multicolumn{1}{|c||}{$(10^{-4}$~fm$^6)$} &$n$&$-0.73$&&&$-0.61(7)$\\
\hline
\multicolumn{1}{|c||}{$(\delta_{LT})'$}&$p$&$-0.80$&\multirow{2}{*}{$-0.04$}&\multirow{2}{*}{$-0.01$}&$-0.85(8)$\\
\multicolumn{1}{|c||}{$(10^{-4}$~fm$^6)$}&$n$ &$-1.19$&&&$-1.24(12)$\\
\hline
\multicolumn{1}{|c||}{$(\bar\gamma_0 )'$}&$p$ &$-1.00$&\multirow{2}{*}{$\hpm 0.16$}&\multirow{2}{*}{$\hpm  0.00$}&$-0.84(10)$\\
\multicolumn{1}{|c||}{$(10^{-4}$~fm$^8)$} &$n$&$-1.58$&&&$-1.42(15)$\\
\hline
\multicolumn{1}{|c||}{$(\Delta I_A)'$}&$p$ &$\hpm2.38$&\multirow{2}{*}{$-11.21$}&\multirow{2}{*}{$\hpm0.25$}&$-8.58(3.43)$\\
\multicolumn{1}{|c||}{$($GeV$^{-2})$} &$n$&$\hpm1.41$&&&$-9.55(3.43)$\\
\hline
\multicolumn{1}{|c||}{$(\Delta I_1)'$}&$p$&$\hpm0.34$&\multirow{2}{*}{$-0.53$}&\multirow{2}{*}{$\hpm0.58$}&$\hpm0.39(4)$\\
\multicolumn{1}{|c||}{$($GeV$^{-2})$}&$n$ &$-1.07$&&&$-1.01(10)$\\
\hline
\end{tabular}
\end{table}

\begin{table}[bht]
\caption{Our NLO B$\chi$PT predictions for the spin polarizabilities at $Q^2=0$,
compared with the B$\chi$PT+$\Delta$ predictions from Bernard {\it et~al.}~\cite{Bernard:2012hb}, and the available empirical information. Where the reference is
not given, the empirical number is provided by the MAID analysis \cite{Drechsel:2000ct,MAID} with unspecified uncertainty.
\label{Table:Results-Pol}}
\begin{tabular}{c|c|c|c||c|c|c|}
\cline{2-7} 
&  \multicolumn{3}{|c||}{ Proton} & 
\multicolumn{3}{|c|}{Neutron} \\
\cline{2-7} 
  & \, This work \, & B$\chi$PT+$\Delta$& Empirical   & \, This work\,& B$\chi$PT+$\Delta$ & \,\,Empirical\,\, \\
\hline
\multicolumn{1}{|c|}{$\,\gamma_0 \,$} & $-0.93(92)$&$-1.74(40)$ &$-1.00(8)(12)$ \cite{Dutz:2003mm} &$0.03(92)$ &$-0.77(40)$ & $-0.005$ \\
\multicolumn{1}{|c|}{$(10^{-4}$~fm$^4)$} & &  &$-0.90(8)(11)$ \cite{Pasquini:2010zr}&&& [MAID]\\
\multicolumn{1}{|c|}{}&&&$-0.929(105)$ \cite{Gryniuk:2016gnm}&&&\\
\hline
\multicolumn{1}{|c|}{$\delta_{LT}$} & $\hpm1.32(15)$  &$2.40(1)$ & $1.34$& $2.18(23)$ &$2.38(3)$& $2.03$\\
\multicolumn{1}{|c|}{$(10^{-4}$~fm$^4)$} &&&[MAID]&&&[MAID]\\
\hline
\multicolumn{1}{|c|}{$\ol \ga_0$} & $\hpm1.12(30)$ & & $0.60(7)(7)$ \cite{Pasquini:2010zr}& $1.95(30)$ && $1.23$\\
\multicolumn{1}{|c|}{$(10^{-4}$~fm$^6)$} &&&$0.484(82)$ \cite{Gryniuk:2016gnm}&&&[MAID]\\
\hline
\end{tabular}
\end{table}

 Our results are summarized in Table \ref{Table:Individual-Results-Pol}, where we give the contributions of the different orders to the chiral predictions of the polarizabilities and their slopes at the real-photon point. A quantitative comparison of our predictions for the spin polarizabilities to the work of Bernard {\it et~al.}~\cite{Bernard:2012hb} and different empirical evaluations is shown in Table \ref{Table:Results-Pol}. We can see that the inclusion of the Delta turns out to be very important for all moments of the helicity-difference cross section. To describe the $Q^2$ behavior of the polarizabilities, the magnetic coupling of the $N \rightarrow \Delta$ transition should be modified by a dipole form factor, as has been observed previously in the description of electroproduction data \cite{Pascalutsa:2005vq}. This dipole form factor effectively takes account of vector-meson exchanges.
 The Coulomb-quadrupole $N \rightarrow \Delta$ transition, despite its subleading order, is important in the description of some moments of spin structure functions. This is contrary to what we saw for the moments of unpolarized structure functions \cite{Alarcon:2020wjg}, where the Coulomb coupling had a negligible effect. The $\pi\Delta$ loops are mainly relevant for the generalized GDH integrals.

\section{Conclusions}\label{Sec:Concl}

We have presented a complete NLO 
calculation of the polarized non-Born VVCS amplitudes in covariant B$\chi$PT, with pion, nucleon, and $\Delta(1232)$ fields. The dispersion relations between the VVCS amplitudes and the tree-level photoabsorption cross sections served as a cross-check of these calculations. 

The obtained moments of the proton and neutron spin structure functions, related to generalized polarizabilities and GDH-type integrals, agree well with the available experimental data. 
The description of their $Q^2$ evolution is improved compared to the previous $\chi$PT predictions.
In particular, the NLO B$\chi$PT predictions obtained here give a better description of the empirical data (e.g., from the Jefferson Laboratory ``Spin Physics Program'') than the HB \cite{Kao:2002cp,Kao:2003jd} and IR \cite{Bernard:2002pw} calculations. 

The demonstrated predictive power of the $\chi$PT framework  amplitudes makes it well suited for
extending the $\chi$PT evaluation of the TPE effect in the hyperfine structure of (muonic-)hydrogen  ~\cite{Hagelstein:2015lph,Hagelstein:2017cbl,Hagelstein:2018bdi} to next-to-leading order.

\section*{Acknowledgements}

We thank Lothar Tiator and Marc Vanderhaeghen for helpful discussions. This work is supported by the Deutsche Forschungsgemeinschaft (DFG) through the
Collaborative Research Center [The Low-Energy Frontier of the Standard Model (SFB 1044)]. JMA acknowledges support from the Community of Madrid through the ``Programa de atracci\'on de talento investigador 2017 (Modalidad 1)'', and the Spanish MECD grants FPA2016-77313-P. FH gratefully acknowledges financial support from the Swiss National Science Foundation.

\appendix
\small

\section{Tensor decompositions of the VVCS amplitudes}\label{labframeamplitudes}
In this appendix, we review the decomposition of the forward VVCS process into tensor structures and scalar amplitudes. In particular, we consider the connection between the covariant and the semi-relativistic decomposition in the lab frame that is defined in terms of the conventional
transverse, longitudinal, transverse-transverse, and transverse-longitudinal amplitudes. 

As explained in Sec.~\ref{Sec:VVCS_SF_relation}, the process of forward VVCS off the nucleon can be described in terms of four explicitly covariant amplitudes $S_{1,\,2}$ and $T_{1,\,2}$
\cite{Hagelstein:2015egb}:
\bea
\label{Eq:T-Rel_Spin}
\hspace{-0.5cm}T(\nu,Q^2) & = &  \Bigg\{ \!\!
\left( -g^{\mu\nu}+\frac{q^{\mu}q^{\nu}}{q^2}\right)
T_1(\nu, Q^2) +\frac{1}{M_N^2} \left(p^{\mu}-\frac{p\cdot
q}{q^2}\,q^{\mu}\right) \left(p^{\nu}-\frac{p\cdot
q}{q^2}\, q^{\nu} \right) T_2(\nu, Q^2)\eqlab{fVVCS}\\
&&-   \frac{1}{M_N}\gamma^{\mu \nu \al} q_\al \,S_1(\nu, Q^2)  -  
\frac{1}{M_N^2} \Big( \gamma^{\mu\nu} q^2 + q^\mu \gamma^{\nu\al} q_\al  -  q^\nu \gamma^{\mu\al}
q_\al \Big) S_2(\nu, Q^2)\Bigg\}\,\epsilon_{\mu}^{\prime *} \epsilon_\nu\,,
\nn
\eea
where $\epsilon_\mu$ ($\epsilon_\mu^{\prime *}$) are the incoming (outgoing) photon polarization vectors, $\nu$ is the photon lab-frame energy and $Q^2$ is the photon virtuality. Alternatively, the decomposition in the laboratory frame (which in the forward case coincides with the Breit frame) is parametrized
in terms of the nucleon Pauli matrices $\vec{\sigma}$ and the four scalar functions $f_L$, $f_T$,
$g_{TT}$, and $g_{LT}$:
\bea\label{Eq:T-Compt-definition}
T(\nu,Q^2)&=&\varepsilon_0\,\varepsilon_0^{\,\prime*}\,f_{L}(\nu,Q^2) +(\vec{\varepsilon}^{\,\, \prime *} \cdot \vec{\varepsilon}\,) \,f_{T}(\nu,Q^2)  +  i \vec{\sigma}\cdot (\vec{\varepsilon}^{\,\, \prime *} \times \vec{\varepsilon}\,)\, g_{TT}(\nu,Q^2) \\
&&-  i \vec{\sigma}\cdot [(\varepsilon_0\vec{\varepsilon}^{\,\, \prime *} - \vec{\varepsilon}\,\varepsilon^{\,\prime *}_0)\times \hat{q}]\, g_{LT}(\nu,Q^2)\,.\nn 
\eea
Here, $\vec q$ and $\hat{q}=\vec{q}/\vert \vec{q}\, \vert$ are the photon three-momentum in the lab system and its unit vector. The modified polarization vector components are given by:
\begin{align}
\varepsilon_0&=\left[\epsilon_0-\frac{\nu}{\left|\vec q\,\right|}\, (\vec\epsilon\cdot\hat q\,)\right]\frac{\left|\vec q\,\right|}{Q}\,,
&\vec\varepsilon = \vec\epsilon-\hat q\,(\vec\epsilon\cdot\hat q\,)\,,\eqlab{modified_polarization_vectors}
\end{align}
 where $\epsilon=(\epsilon_0,\vec\epsilon\,)$ is the usual incoming photon polarization vector, and $\epsilon^{\prime *}$ the outgoing polarization vector. The LEX of the lab frame amplitudes
 [\Eqref{fgFunctions}]
 can serve, in particular, as the definition of the generalized polarizabilities.
 The lab frame amplitudes are also conveniently used for the definition of the response functions,
 see the example of the scalar amplitude $g_{LT}(\nu,Q^2)$ and the corresponding response function $\sigma_{LT}(\nu,Q^2)$ below in App.~\ref{CrossSections}.

\section{Photoabsorption cross sections}\label{CrossSections}

\begin{figure*}
\begin{center} 
\includegraphics[width=0.9\textwidth]{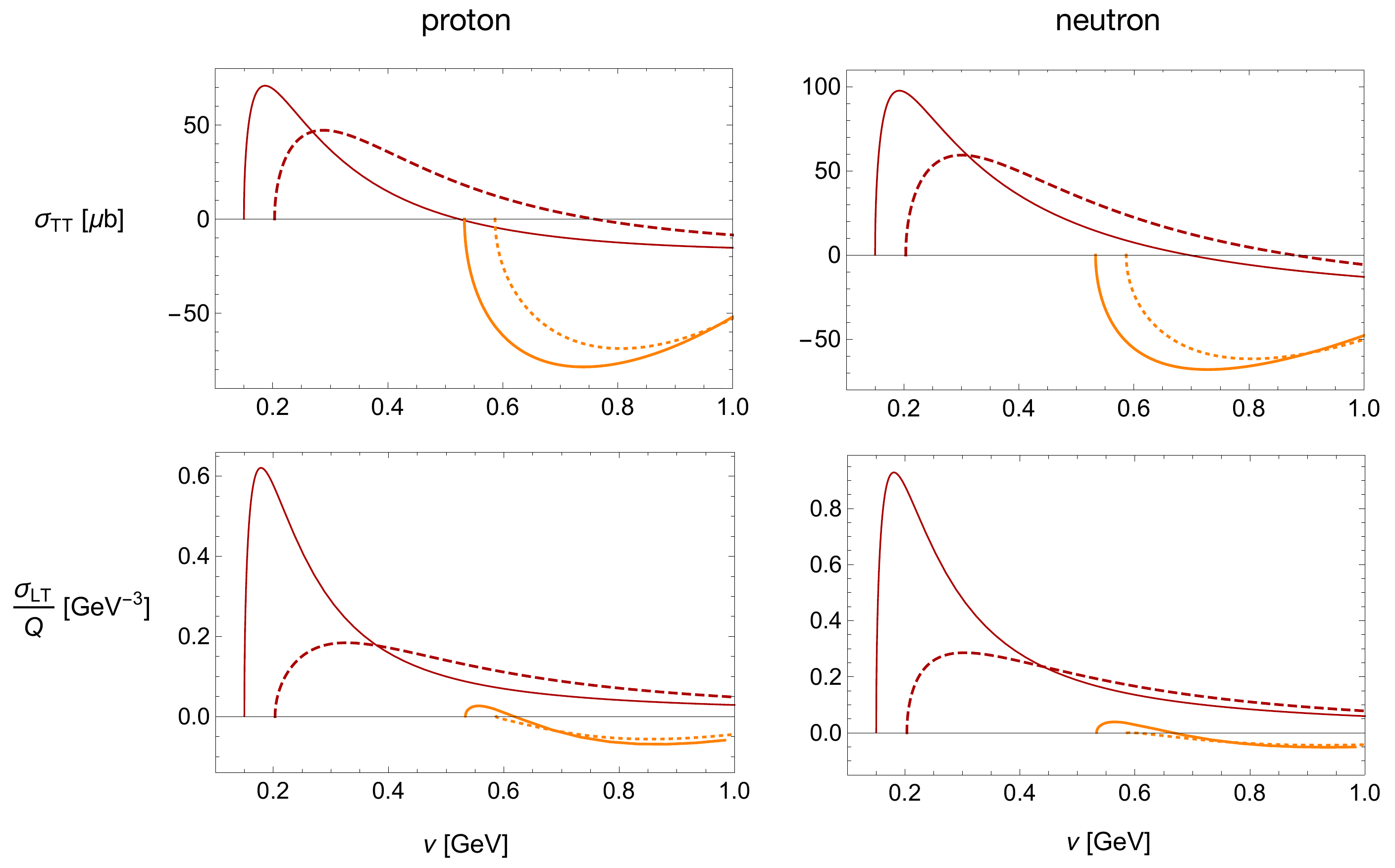}
\caption{Photoabsorption cross sections for $\pi N$ (red) and $\pi \Delta$ production (orange) with $Q^2=0$ (solid) and $Q^2=0.1$ GeV$^2$ (dashed for $\pi N$ and dotted for $\pi \Delta$ channel). \label{Fig:SummaryCrossSections}}
\end{center}
\end{figure*}

In the forward kinematics, the spin-dependent VVCS amplitudes and the spin polarizabilities can be described in terms of the polarized structure functions $g_1(x,Q^2)$ and $g_2(x,Q^2)$, or equivalently, the helicity-difference cross section $\sigma_{TT}(\nu,Q^2)$ and the longitudinal-transverse response function $\sigma_{LT}(\nu,Q^2)$, with the help of dispersion relations \eref{genDRs} and the optical theorem \eref{VVCSunitarity}. In this way, the photoabsorption cross sections, measured in electroproduction processes, form the basis for most empirical evaluations shown throughout Sec.~\ref{Sec:Pol}. In the following, we present the B$\chi$PT predictions for the tree-level cross sections of $\pi N$-, $\pi \Delta$- and $\Delta$-production through photoabsorption on the nucleon, cf.\ Figs.~8, 9 and 10 in Ref.~\cite{Alarcon:2020wjg}. In Secs.~\ref{piNXS} and \ref{DeltaXS}, we will discuss the leading $\pi N$-production channel and the $\Delta$-production channel, respectively. We used these cross sections to verify the polarizability predictions obtained otherwise from the calculated non-Born VVCS amplitudes. Due to the bad high-energy behavior of the $\pi \Delta$-production cross sections in B$\chi$PT, cf.\  Fig.~\ref{Fig:SummaryCrossSections}, the dispersion relations in \Eqref{genDRs} require further subtractions for a reconstruction of the $\pi \Delta$-loop contribution to the spin-dependent VVCS amplitudes. Therefore, not all polarizabilities could be verified, but only those appearing as higher-order terms in the LEX of the VVCS amplitudes, such as $\bar \ga_0$ \cite{Hagelstein:2017cbl}.

\subsection{$\pi N$-production channel}\label{piNXS}

\begin{figure}[t]
\begin{center}
\epsfig{file=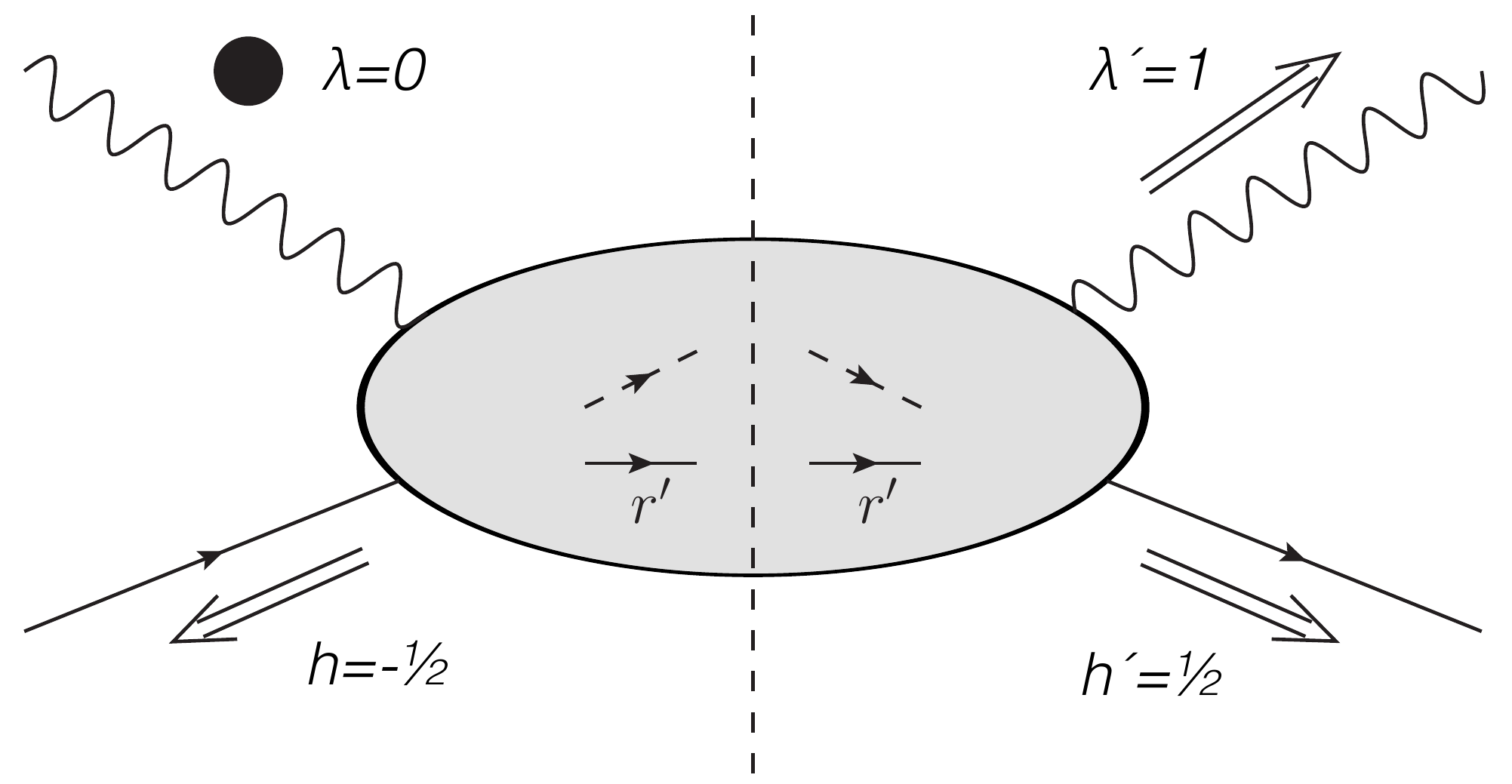,width=7cm,angle=0} 
\caption{Relation between the forward Compton process and the photoabsorption process given by the optical theorem. In particular, we show the longitudinal-transverse contribution. The double-line arrows represent the spin of the external particles, while the dot represents the scalar (longitudinal) polarization of the incoming photon. Inside the blob the intermediate states are represented: e.g., nucleons with spins $r'$ (which are averaged in the calculation of the cross section) and pions. \label{Fig:sigmaLT}}
\end{center}
\end{figure}

In order to extract the response function $\sigma_{LT}(\nu,Q^2)$, we have developed a method similar to the one used to calculate $\sigma_{TT}(\nu,Q^2)$, see, for example, Ref.~\cite{Holstein:2005db}. 
For $\sigma_{LT}(\nu,Q^2)$, however, the calculation is more complicated because one has to take into account that the associated Compton process  involves a spin-flip of the nucleon, as illustrated in Fig.~\ref{Fig:sigmaLT}. When calculating the cross section, the product of the incoming nucleon spinors 
has to reflect this flip. 

The forward VVCS amplitude related to $\sigma_{LT}(\nu,Q^2)$ --- and $\delta_{LT}(Q^2)$ --- is $g_{LT}(\nu,Q^2)$. 
It can be extracted from Eq.~\eqref{Eq:T-Compt-definition} if one takes the modified polarization vector components in \Eqref{modified_polarization_vectors} with $\epsilon=\epsilon_L$ and $\epsilon^{\prime *}=\epsilon^*_\pm$ as input, where $\epsilon_L=\frac{1}{Q}(|\vec{q}\,|,0,0,q^0)$ and $\epsilon_{\pm}=\mp\frac{1}{\sqrt{2}}(0,1,\pm i , 0)$ are the standard longitudinal and transverse polarization vectors, respectively. 
For $\epsilon_L$ and $\epsilon^*_\pm$, only the choice of helicities $h'=\pm1/2$ and $h=\mp 1/2$  gives a non-zero contribution, and one obtains: 
\beq\label{Eq:Derivation-LT1}
\chi^\dagger_{h'}\,T(\nu,Q^2)\,\chi_h= \chi^\dagger_{h'} \{-  i \vec{\sigma}\cdot [(\varepsilon_0\vec{\varepsilon}^{\,\, \prime *} - \vec{\varepsilon}\,\varepsilon^{\,\prime *}_0)\times \hat{q}]\, g_{LT}(\nu,Q^2)\} \chi_h
= \sqrt{2}\, g_{LT}(\nu,Q^2),
\eeq
 where $\chi_h$ and $\chi_{h'}^\dagger$ are two-component Pauli spinors with opposite helicities, or here, spins.

Let us now consider the related photoabsorption process and, in particular, the tree-level $\gamma^* N \rightarrow  \pi N$ channel, see diagrams in Fig.~8 of Ref.~\cite{Alarcon:2020wjg}. We define the $\pi N$-production amplitude as:
\begin{align}\label{Eq:T-decomposition}
 \mathcal{T}= \ol{u}_{h_B}(P_B) \sum_{i}  \mathcal{A}_i (s,t) \Gamma_i \,u_{h_A}(P_A),
\end{align}
with the
Dirac structures:
\begin{subequations}
\begin{align}\label{Eq:DefGammas}
  \Gamma_1&= \gamma_5, \\
  \Gamma_2&= \frac{1}{2}\left[\slashed{q}_A,\slashed{\epsilon}\right]\gamma_5, 
\end{align}
\end{subequations}
where $u_{h_A}(P_A)$ and ${u}_{h_B}^\dagger(P_B)$ are the Dirac spinors, and $P_A$ and $P_B$ are the four-momenta of the incoming and outgoing nucleons, respectively.  When calculating the photoabsorption cross section, related to the VVCS amplitude in Eq.~(\ref{Eq:Derivation-LT1}), the nucleon spin flip should be implemented by $\ol u_{h'}(P_A)$ in $\mathcal{T}^\dagger$ and $u_{h}(P_A)$ in $\mathcal{T}$, together with the appropriate transverse and longitudinal  photon polarization vectors $\eps
^*_\pm$ and $\eps_L$. 

However, if one wants to use the properties of the Dirac matrices, it is more useful to construct an operator to produce this spin flip in the external nucleons of Fig.~\ref{Fig:sigmaLT}. 
This is accomplished by introducing the projector $\Gamma_{LT}\equiv \frac{1}{2\sqrt{2}}(\gamma^1+ i \gamma^2)\gamma_5$, which also takes into account the extra factor $\sqrt{2}$ in Eq.~\eqref{Eq:Derivation-LT1}. 
We checked that with this projector one correctly extracts $\delta_{LT}$ by comparing the HB limit of our result to the HB result of Ref.~\cite{Kao:2002cp}, where the authors calculate this polarizability from the Compton amplitude directly.
With all those ingredients, the longitudinal-transverse cross section is calculated in the following way:
\begin{align}\label{Eq:Derivation-LT2}
&\sigma_{LT}(\nu,Q^2)=\frac{1}{32\pi\,s}\frac{|\vec{p}_f|_\mathrm{cm}}{|\vec{p}_i|_\mathrm{cm}} \int_{-1}^{1}\! \dd\!\cos\theta \sum_{i,j} \mathcal{A}_i  \mathcal{A}^\dagger_j \mathcal{X}_{ij}, 
\end{align}
with 
\begin{align}
\mathcal{X}_{ij}=\text{Tr}[(\slashed{P}_B+M_N)\Gamma_i(\slashed{P}_A+M_N)\Gamma_{LT}\gamma^0\Gamma_j^\dagger\gamma^0 ],
\end{align}
where $\theta$ is the scattering angle in the center-of-mass (cm) frame, and $\vert \vec p_i \vert_\mathrm{cm}$ ($\vert \vec p_f \vert_\mathrm{cm}$) is the three-momentum of an incoming (outgoing) particle in the cm frame. An explicit calculation of the matrix $\mathcal{X}_{ij}$ leads to:
\begin{align}
 \mathcal{X}=M_N Q \left(\begin{array}{cc}
  0   & 2  (P_B-P_A)\cdot \epsilon_L \\
 -\sqrt{2} |\vec{q}_f|_\mathrm{cm} \sin\theta &  (s-u) \\
 \end{array}
 \right),
 \end{align}
where $\vert \vec q_i \vert_\mathrm{cm}$ ($\vert \vec q_f \vert_\mathrm{cm}$) is the relative three-momentum of the incoming (outgoing) particles in the cm frame. Here, $s$, $t$ and $u$ are the usual Mandelstam variables. For the different $\ga^* N \rightarrow \pi N$ channels, we obtain the following amplitudes $\mathcal{A}_i$, where we introduce $q_A$ as the four-momentum of the incoming photon and $q_B$ as the four-momentum of the outgoing pion:
\begin{itemize}
 \item $\gamma^* p \to \pi^0 p$
 \begin{subequations}
\begin{align}
  &\mathcal{A}_1= \frac{e\, g_A M_N}{f_\pi}\left[\frac{2 P_A\cdot \epsilon + q_A\cdot \epsilon}{s-M_N^2} + \frac{2 P_B\cdot \epsilon - q_A\cdot \epsilon}{u-M_N^2}\right], \\
   &\mathcal{A}_2=\frac{e\, g_A M_N}{f_\pi}\left[\frac{1}{s-M_N^2} + \frac{1}{u-M_N^2}\right];
\end{align}
 \end{subequations}
 \item $\gamma^* p \to \pi^+ n$
  \begin{subequations}
\begin{align}
  &\mathcal{A}_1= \frac{\sqrt{2}\, e\, g_A M_N}{f_\pi}\left[\frac{2 P_A\cdot \epsilon + q_A\cdot \epsilon}{s-M_N^2} + \frac{2 (P_A - P_B)\cdot \epsilon + q_A\cdot \epsilon}{t-m_\pi^2}\right],\\
   &\mathcal{A}_2=\frac{\sqrt{2}\, e\, g_A M_N}{f_\pi(s-M_N^2)};
\end{align}
 \end{subequations}
 \item $\gamma^* n \to \pi^0 n$
  \begin{subequations}
\begin{align}
  &\mathcal{A}_1= 0,\\
  & \mathcal{A}_2=0;
\end{align}
 \end{subequations}
 \item $\gamma^* n \to \pi^- p$
  \begin{subequations}
\begin{align}
  &\mathcal{A}_1= \frac{\sqrt{2}\, e\, g_A M_N}{f_\pi}\left[\frac{2 P_B\cdot \epsilon - q_A\cdot \epsilon}{u-M_N^2}  - \frac{2 (P_A - P_B)\cdot \epsilon + q_A\cdot \epsilon}{t-m_\pi^2}\right],\\
   &\mathcal{A}_2=\frac{\sqrt{2}\, e\, g_A M_N}{f_\pi(u-M_N^2)};
\end{align}
 \end{subequations}
\end{itemize}
The analytical expressions shown above were checked with the amplitudes given in Ref.~\cite{Pasquini:2007fw}. Analytical expressions for the  tree-level $\ga^* N \rightarrow \pi N$ channel of the $\sigma_{LT}(\nu,Q^2)$ and $\sigma_{TT}(\nu,Q^2)$ cross sections are given below (proton channels: $\pi^+n$ and $\pi^0 p$; neutron channel $\pi^-p$).
We  checked that they reproduce the known results in the real-photon limit \cite{Lensky:2009uv,Holstein:2005db}.
To shorten the final expressions for the cross sections, which are considerably longer for finite $Q^2$ than in the real-photon limit, we define the following dimensionless kinematic variables:
\begingroup
\allowdisplaybreaks
\begin{align}
&\alpha_\gamma= (E_i^{N})_\mathrm{cm}/\sqrt{s}=\frac{s+M_N^2+Q^2}{2 s},   \\
&\alpha_\pi= (E_f^{N})_\mathrm{cm}/\sqrt{s} = \frac{s+M_N^2-m_\pi^2}{2 s},  \\              
 & \beta_\gamma = E^{\gamma}_\mathrm{cm}/\sqrt{s} = \frac{s-M_N^2-Q^2}{2 s} ,  \\
 & \beta_\pi= E^{\pi}_\mathrm{cm}/\sqrt{s} = \frac{s-M_N^2+m_\pi^2}{2 s} , \\
 &\lambda_\gamma  = |\vec{q}_i|_\mathrm{cm}/\sqrt{s}= \frac{\sqrt{(s-M_N^2 - Q^2)^2+4 s Q^2}}{2 s},  \\
&\lambda_\pi =  |\vec{q}_f|_\mathrm{cm}/\sqrt{s}  = \frac{\sqrt{(s-M_N^2 + m_\pi^2)^2-4 s m_\pi^2}}{2 s}.  \end{align} 
\endgroup
Here, $(E_i^{N})_\mathrm{cm}$  and $(E_f^{N})_\mathrm{cm}$ are the energies of the incoming and outgoing nucleon, $E^{\gamma}_\mathrm{cm}$ is the energy of the incoming photon, $E^{\pi}_\mathrm{cm}$ is the energy of the outgoing pion, all in the cm frame.

\small

\begin{flalign}
\sigma_{TT}^{(\pi^+ n)}=&\, -\frac{e^2 g_A^2 M_N^2}{64 \pi f_\pi^2 s^3 (s-M_N^2)^2 \lambda_\gamma^4}\Bigg\{ 4 s \lambda_\pi \lambda_\gamma \Big[ (M_N^2-s)(M_N^2 - Q^2 - s)(Q^2+ 2 s \beta_\gamma \beta_\pi) \nonumber &&\\
& + 2 s \Big(m_\pi^2 Q^2- (M_N^2 -s) \big(M_N^2 +s(-1+2 \beta_\gamma \beta_\pi)\big)\Big) \lambda_\gamma^2\Big]-2 (M_N^2 - s) (M_N^2 - Q^2 - s) \nonumber&& \\
& \times (Q^2 + 2 s \beta_\gamma \beta_\pi - 2 s \lambda_\pi \lambda_\gamma) \big(Q^2 + 2 s (\beta_\gamma \beta_\pi + \lambda_\pi \lambda_\gamma)\big) \arctanh\left( \frac{2 s \lambda_\gamma \lambda_\pi}{Q^2 + 2 s \beta_\gamma \beta_\pi }\right)\!\!\Bigg\}, &&
\end{flalign}

\begin{flalign}
\sigma_{TT}^{(\pi^0 p)}=&\,\frac{e^2 g_A^2 M_N^2 \lambda_\pi}{64 \pi f_\pi^2 s(s-M_N^2)^2 \lambda_\gamma}\Bigg\{ -4 m_\pi^2 Q^2 + 2(M_N^2 - s)\Bigg(2(s-M_N^2) + 4 s \beta_\gamma \beta_\pi \nonumber &&\\
&   + \frac{(s-M_N^2 + Q^2)\big(3(M_N^2 -  s) +2 s \beta_\gamma \beta_\pi \big)}{s \lambda_\gamma^2}    \nonumber &&\\
&+ \frac{2m_\pi^2 Q^2 (s-M_N^2)}{\big(M_N^2-s (1 - 2 \beta_\gamma \beta_\pi + 2 \lambda_\pi \lambda_\gamma)\big)\big(M_N^2-s (1 - 2 \beta_\gamma \beta_\pi - 2 \lambda_\pi \lambda_\gamma)\big)}\Bigg)+\frac{M_N^2-s}{s^2 \lambda_\pi \lambda_\gamma^3}\nonumber \\
&\times \Bigg(Q^2 s \bigg(4 \lambda_\gamma^2 m_\pi^2+s \left(-4 \beta_\gamma ^2 \beta_\pi ^2+8 \beta_\gamma  \beta_\pi +4 \lambda_\pi^2 \lambda_\gamma^2-3\right)\!\!\bigg)+3 M_N^6&&\nn\\
&+M_N^4 \bigg(s \left(8 \beta_\gamma  \beta_\pi +2 \lambda_\gamma^2-9\right)-3 Q^2\bigg)+s^3 \bigg(-4 \beta_\gamma ^2 \beta_\pi ^2+8 \beta_\gamma  \beta_\pi +2\left(1+2 \lambda_\pi^2\right) \lambda_\gamma^2-3\bigg)&&\nn\\
&+M_N^2 s \bigg(Q^2 (6-8 \beta_\gamma  \beta_\pi )+s \left(4 \beta_\gamma ^2 \beta_\pi ^2-16 \beta_\gamma  \beta_\pi -4 \left(1+\lambda_\pi^2\right) \lambda_\gamma^2+9\right)\!\bigg)\!\Bigg)&&\nn\\
&\times\arctanh\left(\frac{2s \lambda_\pi \lambda_\gamma}{M_N^2+s(2\beta_\ga \beta_\pi-1)}\right)\!\!\Bigg\},&&
\end{flalign}

\begin{flalign}
\sigma_{TT}^{(\pi^- p)}=&\,\frac{e^2 g_A^2 M_N^2}{16 \pi 
   f_{\pi }^2 s^2 \lambda _{\gamma }^3}\Bigg\{\frac{\lambda_\pi }{M_N^4-2M_N^2 s (1-2 \beta_\gamma  \beta_\pi )-s^2 \left(-4 \beta_\gamma ^2 \beta_\pi ^2+4 \beta_\gamma  \beta_\pi +4 \lambda_\pi^2 \lambda_\gamma^2-1\right)}&\nn\\
   &\times\Bigg[Q^2 s \left(-2 \lambda_\gamma^2 m_\pi^2+4 \beta_\gamma ^2 \beta_\pi ^2 s-4 \beta_\gamma  \beta_\pi  s-4 \lambda_\pi^2 \lambda_\gamma^2 s+s\right)-M_N^6&\nn\\
   &+M_N^4 \big(Q^2+s (3-4 \beta_\gamma  \beta_\pi )\big)+M_N^2 s \bigg(2Q^2 (2 \beta_\gamma  \beta_\pi -1)&&\nn\\
   &+s \left(-4 \beta_\gamma ^2 \beta_\pi ^2+8 \beta_\gamma  \beta_\pi +4 \lambda_\pi^2 \lambda_\gamma^2-3\right)\!\!\bigg)+s^3 \left(4 \beta_\gamma ^2 \beta_\pi ^2-4 \beta_\gamma  \beta_\pi -4 \lambda_\pi^2 \lambda_\gamma^2+1\right)\!\!\Bigg]&&\nn\\
   &+\frac{1}{2s\lambda_\gamma}\Bigg[\bigg(Q^4+4 \beta_\gamma  \beta_\pi  Q^2 s+4 s^2 \left(\beta_\gamma ^2 \beta_\pi ^2-\lambda_\pi^2 \lambda_\gamma^2\right)\!\!\bigg)\arctanh\left(\frac{2s\lambda_\pi \lambda_\gamma}{Q^2+2s\beta_\gamma\beta_\pi}\right)&&\nn\\
   &+\bigg(M_N^4-2 M_N^2 \left(Q^2-\lambda_\gamma^2 s+s\right)+s \Big(2Q^2 (1-2 \beta_\gamma  \beta_\pi )-4 \beta_\gamma ^2 \beta_\pi ^2 s+2\left(2 \lambda_\pi^2-1\right) \lambda_\gamma^2 s+s\Big)\!\bigg)&&\nn\\
   &\times\arctanh\left(\frac{2s\lambda_\pi \lambda_\gamma}{M_N^2+s(2\beta_\gamma\beta_\pi-1)}\right)\!\Bigg]\!\Bigg\}, && 
\end{flalign}

\begin{flalign}
\sigma_{LT}^{(\pi^+ n)}=&\,\frac{e^2 g_A^2 M_N^3 \lambda_\pi}{32 \pi f_\pi^2 Q\, s^3 (s-M_N^2)^2 \lambda_\gamma^4} \Bigg\{  2 s \lambda_\gamma \Bigg[    (M_N^2 - s) (Q^2 + 2 s \beta_\gamma^2) (Q^2 + 2 s \beta_\gamma \beta_\pi) &&\nonumber \\
&- 4 s \bigg(\!(M_N^2 - s)\big(Q^2 - 2 s ( \alpha_\pi-1) \beta_\gamma\big)  +Q^2 s \beta_\gamma \beta_\pi\bigg) \lambda_\gamma^2 + 8 s^3 ( \alpha_\pi-1) \lambda_\gamma^4\Bigg]&&&&\nn \\
&  + \frac{s-M_N^2 }{\lambda_\pi} \Bigg[ (Q^2+2 s \beta_\gamma^2)(Q^2 + 2 s \beta_\gamma \beta_\pi)^2 + 4 s^2 \bigg(2 (\alpha_\pi-1) \beta_\gamma (Q^2 + 2s \beta_\gamma \beta_\pi) &&\nonumber \\
&-Q^2 \lambda_\pi^2 \bigg)\lambda_\gamma^2+ 8 s^3 (\alpha_\pi -1 )^2\lambda_\gamma^4  \Bigg] \arctanh\left( \frac{2 s \lambda_\pi \lambda_\gamma}{Q^2 + 2 s \beta_\gamma \beta_\pi}\right)\!\!\Bigg\},&&
\end{flalign}

\begin{flalign}
\sigma_{LT}^{(\pi^0 p)}=&\,\frac{e^2 g_A^2 M_N^3 \lambda_\pi}{16 \pi f_\pi^2 Q\, s (s-M_N^2)^2 \lambda_\gamma} \Bigg\{ \frac{1}{-2 s(M_N^2+s(-1+2 \beta_\gamma \beta_\pi ))^2 \lambda_\gamma^2 + 8 s^3 \lambda_\pi^2 \lambda_\gamma^4}\Bigg[  -3M_N^8 (Q^2  \nonumber&&\\
&+2 s \beta_\gamma^2)+ 2 M_N^4 s^2 \bigg(-(Q^2+2 s \beta_\gamma^2)(2 \beta_\gamma \beta_\pi-3)(5 \beta_\gamma \beta_\pi-3)+\big(Q^2 ( 2 \beta_\gamma \beta_\pi +6 \lambda_\pi^2-3 )   \nonumber&&\\
&+ 2 s \beta_\gamma (12 \alpha_\pi + 2 \beta_\gamma \beta_\pi   -12 \alpha_\pi \beta_\gamma \beta_\pi + 4 \beta_\gamma \lambda_\pi^2-3)\big)\lambda_\gamma^2- 4 s (\alpha_\pi^2-1) \lambda_\pi^4\bigg)+ 2 M_N^2 s^3 \bigg[-(Q^2  \nonumber &&\\
&+ 2 s \beta_\gamma^2)(2 \beta_\gamma \beta_\pi -1) \big(6 + \beta_\gamma \beta_\pi (-9+ 2 \beta_\gamma \beta_\pi) \big) + \bigg(Q^2\big(3 -12 \lambda_\pi^2 +4 \beta_\gamma \beta_\pi( -1 + 2 \beta_\gamma \beta_\pi  \nonumber &&\\
&+ \lambda_\pi^2)\big)- 2 s \beta_\gamma \big(-3 +4 \alpha_\pi  (3 + 2 \beta_\gamma \beta_\pi(  \beta_\gamma \beta_\pi-3)) + 4 \beta_\gamma (\beta_\pi + (2 - \beta_\gamma \beta_\pi)  \nonumber &&\\
&\times \lambda_\pi^2 )\big)\!\bigg) \lambda_\gamma^2 + 8 s \big((\alpha_\pi-1)(1+\alpha_\pi-2 \beta_\gamma \beta_\pi)+ 2 \alpha_\pi \beta_\gamma \lambda_\pi^2\big) \lambda_\gamma^4 \bigg] + s^4 \bigg((Q^2 + 2 s \beta_\gamma^2)  \nonumber &&\\
&\times (1-2 \beta_\gamma \beta_\pi)^2 (2 \beta_\gamma \beta_\pi -3) + 2 \big(Q^2 (-1 + 6 \lambda_\pi^2 + 2 \beta_\gamma \beta_\pi ((1-2 \beta_\gamma \beta_\pi)^2 -2 \lambda_\pi^2  )) + 2 s \beta_\gamma  \nonumber&& \\
&\times (-1 + 2 \beta_\gamma \beta_\pi + 4 (-1 + \beta_\gamma \beta_\pi) (-\alpha_\pi + 2 \alpha_\pi \beta_\gamma \beta_\pi - \beta_\gamma \lambda_\pi^2)  )\big)  \lambda_\gamma^2 - 8 \big(s (\alpha_\pi-1) (\alpha_\pi &&\nn\\
&+ (1-2 \beta_\gamma \beta_\pi)^2) + 2 \beta_\gamma (2 s \alpha_\pi + Q^2 \beta_\pi ) \lambda_\pi^2 \big)\lambda_\gamma^4 +32 s (\alpha_\pi-1) \lambda_\pi^2 \lambda_\gamma^6 \bigg) + 2 M_N^6 s \bigg(Q^2 (6 \nonumber&& \\
&-7 \beta_\gamma \beta_\pi + \lambda_\gamma^2) + 2 s \beta_\gamma (\beta_\gamma(6 -7 \beta_\gamma \beta_\pi)+(1-4 \alpha_\pi) \lambda_\gamma^2) \bigg)    \Bigg]  + \frac{s-M_N^2}{4 s^2 \lambda_\pi \lambda_\gamma^3} \Bigg[  (Q^2 + 2 s \beta_\gamma^2)  \nonumber &&\\
&\times \big(3 M_N^2 + s (2 \beta_\gamma \beta_\pi -3)\big) \big(M_N^2 + s (2 \beta_\gamma \beta_\pi -1) \big)+2 s \big(-M_N^2 (Q^2 + 2 s \beta_\gamma - 8 s \alpha_\pi \beta_\gamma )\nonumber&& \\
&+ s (2 s \beta_\gamma (1-4 \alpha_\pi +4 \alpha_\pi \beta_\gamma \beta_\pi)  +Q^2 (1-2\lambda_\pi^2) ) \big)\lambda_\gamma^2 + 8 s^3 (\alpha_\pi^2-1) \lambda_\gamma^4    \Bigg] &&\nn\\
&\times\arctanh\left(   \frac{2 s \lambda_\pi \lambda_\gamma }{M_N^2 + s (2 \beta_\gamma \beta_\pi -1)}     \right)  \!\!  \Bigg\},&&
\end{flalign}

\begin{flalign}
\sigma_{LT}^{(\pi^- p)}=&\,\frac{e^2 g_A^2 M_N^3}{16 \pi 
   f_{\pi }^2 Q s^2 \lambda _{\gamma }^3}\Bigg\{\frac{\lambda_\pi }{\left(M_N^2+s (2 \beta_\gamma  \beta_\pi -2 \lambda_\pi \lambda_\gamma-1)\right) \left(M_N^2+s (2 \beta_\gamma  \beta_\pi +2 \lambda_\pi \lambda_\gamma-1)\right)}&\nn\\
   &\times\Bigg[2 \lambda_\gamma^2 s \bigg(s \Big(\!\!\left(1-2 \lambda_\pi^2\right) Q^2+2\beta_\gamma  s (2 \alpha_\pi-1)  (2 \beta_\gamma  \beta_\pi -1)\!\Big)-M_N^2 \Big(Q^2+2 \beta_\gamma  s(1-2 \alpha_\pi) \Big)\!\!\bigg)&&\nn\\
   &+\left(Q^2+2 \beta_\gamma ^2 s\right) \Big(M_N^2+s (2 \beta_\gamma  \beta_\pi -1)\Big)^2+8 (\alpha_\pi-1) \alpha_\pi \lambda_\gamma^4 s^3\!\Bigg]+\frac{1}{2s\lambda_\gamma(s-M_N^2+Q^2)}&&\nn\\
   &\times\Bigg[\bigg(4 \lambda_\gamma^2 s^2 \Big(2 (\alpha_\pi-1) \beta_\gamma  \left(Q^2+2 \beta_\gamma  \beta_\pi  s\right)-\lambda_\pi^2 Q^2\Big)+\left(Q^2+2 \beta_\gamma ^2 s\right) \left(Q^2+2 \beta_\gamma  \beta_\pi  s\right)^2&&\nn\\
   &+8 (\alpha_\pi-1)^2 \lambda_\gamma^4 s^3\!\bigg)\arctanh\left(\frac{2s\lambda_\pi \lambda_\gamma}{Q^2+2s\beta_\gamma\beta_\pi}\right)+\bigg(\!\!\left(Q^2+2 \beta_\gamma ^2 s\right) \Big(M_N^2+s (2 \beta_\gamma  \beta_\pi -1)\Big)&&\nn\\
   &\times\Big(\!-M_N^2+2 Q^2+s(2 \beta_\gamma  \beta_\pi  +1)\Big)+2 \lambda_\gamma^2 s \Big(\!-M_N^2 \left(Q^2+2 \beta_\gamma  s\right)+Q^4&&\nn\\
   &+Q^2 s \left(2(2 \alpha_\pi-1) \beta_\gamma -2 \lambda_\pi^2+1\right)+2 \beta_\gamma  s^2 (4 (\alpha_\pi-1) \beta_\gamma  \beta_\pi +1)\Big)+8 (\alpha_\pi-1)^2 \lambda_\gamma^4 s^3\!\bigg)&&\nn\\
   &\times\arctanh\left(\frac{2s\lambda_\pi \lambda_\gamma}{M_N^2+s(2\beta_\gamma\beta_\pi-1)}\right)\!\Bigg]\!\Bigg\}, && 
\end{flalign}

\subsection{$\Delta$-production channel}\label{DeltaXS}
\begin{subequations}
\bea
\ol S_1^{\Delta\text{-exch.}}(\nu,Q^2)&=& S_1^{\Delta\text{-pole}}(\nu,Q^2)+\widetilde S_1^{\Delta\text{-exch.}}(\nu,Q^2),\qquad\;\\
\nu \ol S_2^{\Delta\text{-exch.}}(\nu,Q^2)&=&\nu S_2^{\Delta\text{-pole}}(\nu,Q^2)+  \widetilde{\nu S_2}^{\Delta\text{-exch.}}(\nu,Q^2),
\eea
\end{subequations}
and similarly for the unpolarized VVCS amplitudes discussed in Ref.~\cite{Alarcon:2020wjg}. Here, we introduced the $\Delta$-pole contributions $S_i^{\Delta\text{-pole}}$ and the $\Delta$-non-pole contributions $\widetilde S_i^{\Delta\text{-exch.}}$. The former amplitudes feature a pole at the $\Delta(1232)$-production threshold, and thus, are proportional to:
\beq
\frac{1}{[s-M_\Delta^2][u-M_\Delta^2]}=\frac{1}{4M_N^2}\frac{1}{\nu_\Delta^2-\nu^2}. \eqlab{poleStruc}
\eeq
They can be reconstructed from the dispersion relations in \Eqref{genDRs} with the tree-level $\Delta$-production cross sections as input, cf.\ Fig.~10 in Ref.~\cite{Alarcon:2020wjg}:
\begin{subequations}
\eqlab{structurefunc}
\bea
\sigma_{TT}(\nu,Q^2)&=&\frac{\pi^2 \al}{M_N^2 M_+^2 \vert \vec{q}\,\vert}\Bigg\{-g_M^2M_N (M_+ +\nu)\vert \vec{q}\,\vert^2+\frac{g_E^2(\varDelta-\nu)(Q^2-M_N \nu)^2}{M_N}\\
&& +\frac{g_C^2 Q^4 s(\varDelta-\nu)}{M_N M_\Delta^2}-4g_M g_E (Q^2-M_N \nu)\vert \vec{q}\,\vert^2-4g_M g_C Q^2\vert \vec{q}\,\vert^2\nn\\
&&+\frac{2g_Eg_C Q^2\left[- M_N M_\Delta\ \vert \vec{q}\,\vert^2+s(Q^2+\varDelta \nu)\right]}{M_N M_\Delta} \Bigg\}\delta\!\left(\nu-\nu_\Delta\right)\nn\\
\sigma_{LT}(\nu,Q^2)&=&\frac{Q\pi^2 \al}{M_N^2 M_+^2 \vert \vec{q}\,\vert}\Bigg\{\frac{g_E^2(M_N \nu-Q^2)\left[M_\Delta(M_N+\nu)-s\right]}{M_N}\\
&&+\frac{g_C^2 Q^2 \left[M_N M_\Delta \vert \vec{q}\,\vert^2-s(Q^2+\varDelta \nu) \right]}{M_N M_\Delta^2}+g_M g_E M_\Delta \vert \vec{q}\,\vert^2- \frac{g_M g_C (Q^2-M_N \nu) \vert \vec{q}\,\vert^2}{M_\Delta}\nn\\
&&+\frac{g_E g_C(\nu-\varDelta)(M_N^2 \vert \vec{q}\,\vert^2-2 Q^2 s)}{M_N M_\Delta}\Bigg\}\delta\!\left(\nu-\nu_\Delta\right)\nn,
\eea
\end{subequations}
with  $\varDelta=M_\Delta - M_N$, $M_+=M_\Delta + M_N$ and the Mandelstam variable $s=M_N^2+2M_N \nu-Q^2$.
Analytical expressions for the spin structure functions $g_1(x,Q^2)$ and $g_2(x,Q^2)$ can be constructed from \Eqref{VVCSunitarity} with the flux factor $K(\nu,Q^2)=\vert \vec{q}\, \vert=\sqrt{\nu^2+Q^2}$.

In the $\Delta$-non-pole contributions to $S_1(\nu,Q^2)$ and $\nu S_2(\nu,Q^2)$, the pole in $\nu$ at the $\Delta(1232)$-production threshold has canceled out:
\begingroup
\allowdisplaybreaks
\begin{subequations}
\eqlab{noDpole}
\bea
\widetilde S_1^{\Delta\text{-exch.}}(\nu,Q^2)&=&\frac{\pi \al }{M_N M_+^2}\left[g_M^2 Q_+^2+g_E^2 \left(\varDelta ^2-3 Q^2\right)+\frac{4 g_C^2 Q^4}{M_\Delta^2}-8 g_M g_E M_\Delta \omega_-\right.\eqlab{S1nonpole}\\*
&&\qquad\qquad\left.-\frac{2 g_M g_C Q^2 (M_N-4 M_\Delta)}{M_\Delta}+\frac{2 g_E g_C Q^2 (3 M_N-2 M_\Delta)}{M_\Delta}\right],\qquad\nn\\
\widetilde{\nu S_2}^{\Delta\text{-exch.}}(\nu,Q^2)&=&\frac{2\pi \al }{M_N M_+^2}\bigg[g_E^2 \,M_\Delta \varDelta\, \omega_-+\frac{g_M^2\, M_N Q_+^2}{2}+\frac{g_C^2\,Q^2(Q^2-\varDelta^2)}{2M_\Delta}\eqlab{NUnonpoleS2}\\*
&&\qquad\qquad+g_E g_M \,M_\Delta (M_\Delta \omega_+-4 M_N \omega_-)-g_E g_C\, \varDelta (2Q^2+M_N \omega_+)\nn\qquad\\*
&&\qquad\qquad+g_M g_C \,Q^2(4M_N-\omega_+)\bigg]+\frac{\widetilde S_2^{\Delta\text{-exch.}}(\nu,Q^2)}{\nu}\left[\frac{M_\Delta^2 \,\omega_+^2}{M_N^2}+\nu^2\right],\nn
\eea
\end{subequations}
\endgroup
with $Q_+=\sqrt{(M_\Delta+M_N)^2+Q^2}$ and $\omega_\pm=(M_\Delta^2-M_N^2\pm Q^2)/2M_\Delta$, and the non-pole contribution to $S_2(\nu,Q^2)$:
\beq
\widetilde S_2^{\Delta\text{-exch.}}(\nu,Q^2)=-\frac{2\pi \al M_N \nu}{M_\Delta M_+^2}\big[g_M +g_E \big]g_C\,.\eqlab{nonpoleS2}
\eeq
These amplitudes, to the contrary, are not described by the tree-level $\Delta$-production cross sections in the standard dispersive approach \cite{Hagelstein:2018bdi}. 
This peculiarity has been previously missed, e. g., in the calculation of the $\Delta$-exchange contribution to the hydrogen hyperfine splitting in Ref.~\cite{Buchmann:2009uqa}. The importance of including the $\Delta$-non-pole contribution is also evident when considering the BC sum rule in Eq.~(\ref{BC}). The $\Delta$-pole terms by themselves violate the BC sum rule, but cancel exactly with the $\Delta$-non-pole terms:
\beq
\lim_{\nu \rightarrow 0}\nu S_2^{\Delta\text{-pole}}(\nu,Q^2)+\lim_{\nu \rightarrow 0}  \widetilde{\nu S_2}^{\Delta\text{-exch.}}(\nu,Q^2)=0.
\eeq

\section{Polarizabilities at $Q^2=0$}\label{App:PolarizabilitiesAll}

In this section, we give analytical expressions for the polarizabilities and their slopes at $Q^2=0$. In particular, we give the HB expansion of the $\pi N$-loop contributions and the $\Delta$-exchange contributions. The complete expressions, also for the $\pi \Delta$-loop contributions, can be found in the {\it Supplemented material}. Recall that $I_A(0)=I_1(0)=\bar d_2(0)=0$ and $\frac{\dd \bar d_2(Q
^2)}{\dd Q^2}\Big\vert_{Q^2=0}=0$.

\subsection{$\pi N$-loop  contribution}
\label{App:Polarizabilities}

Here, we give analytical expressions for the $\pi N$-loop contributions to the proton and neutron spin polarizabilities, expanded
in powers of $\mu = m_\pi/M_N$, viz., the HB expansion. Note that we
choose to expand here to a high order in $\mu$, the strict
HB expansion would only retain the leading term in an analogous
NLO calculation. 

 \small

\begin{itemize}
    \item Polarizabilities at $Q^2=0$:
\bea
\gamma_{0p} &=& \frac{e^2 g_A^2}{96\pi^3 f_\pi^2\,  m_\pi^2}\left\{ 1-\frac{21\pi \mu}{8} - \left(\frac{59}{2} + 26 \log\mu\right)\mu^2 + \frac{1875 \pi \mu^3}{64} \right.\nn\\
&&\left.+ 3\left(\frac{3}{2} + 26 \log\mu\right)\mu^4 + \dots \right\},\\
\gamma_{0n} &=& \frac{e^2 g_A^2}{48\pi^3 f_\pi^2\,  m_\pi^2}\left\{ \frac{1}{2}-\frac{9\pi \mu}{16} -  2\mu^2 \log\mu  + \frac{75 \pi \mu^3}{128} - \frac{3 \mu^4}{4}  + \dots \right\},\\
\delta_{LTp} &=&\frac{e^2 g_A^2}{192\pi^3 f_\pi^2\, m_\pi^2} \left\{ 1 - \frac{9 \pi \mu}{8} + \left(\frac{13}{2}-2\log\mu\right)\mu^2 - \frac{465 \pi \mu^3}{64}  \right.\nn\\
&&\left.- \left(\frac{47}{2}+ 42 \log\mu\right)\mu^4 +  \dots\right\} ,\\
\delta_{LTn} &=&\frac{e^2 g_A^2}{96\pi^3 f_\pi^2\, m_\pi^2} \left\{ \frac{1}{2} + \frac{3 \pi \mu}{16} + \left(1+2\log\mu\right)\mu^2 - \frac{105 \pi \mu^3}{128} + \frac{5\mu^4}{4}   + \dots\right\},\eea
\bea
\delta_{LTn} &=&\frac{e^2 g_A^2}{96\pi^3 f_\pi^2\, m_\pi^2} \left\{ \frac{1}{2} + \frac{3 \pi \mu}{16} + \left(1+2\log\mu\right)\mu^2 - \frac{105 \pi \mu^3}{128} + \frac{5\mu^4}{4}   + \dots\right\},\\
\bar \ga_{0p}&=&\frac{e^2g_A^2}{16 \pi^3 f_\pi^2 m_\pi^4}\left\{\frac{4}{45}-\frac{3\pi \mu}{16}+\frac{14\mu^2}{5}-\frac{1813 \pi \mu^3}{384}-\frac{192}{5}\left(1+\log \mu\right)\mu^4\right.\nn\\
&&\left.+\frac{80703 \pi \mu^5}{2048}+  \dots\right\},\\
\bar \ga_{0n}&=&\frac{e^2g_A^2}{16 \pi^3 f_\pi^2 m_\pi^4}\left\{\frac{4}{45}-\frac{5\pi \mu}{48}+\frac{4\mu^2}{5}-\frac{245 \pi \mu^3}{384}-\frac{32 \mu^4 \log \mu}{15}\right.\nn\\
&&\left.+\frac{1323 \pi \mu^5}{2048}+  \dots\right\}.
\eea
\item Slopes of polarizabilities at $Q^2=0$:
\begingroup
\allowdisplaybreaks
\bea
\left.\frac{\dd \gamma_{0p} (Q^2)}{\dd Q^2}\right|_{Q^2=0}&=& \frac{e^2 g_A^2}{1440 \pi^3 f_\pi^2\, m_\pi^4} \left\{ 2 - \frac{45 \pi \mu}{4} + 223 \mu^2 - \frac{28515 \pi \mu^3 }{64}\right.\nn\\* 
&&\left.- 9 \left( \frac{1953}{4} + 449 \log\mu\right)\mu^4 + \frac{570255 \pi \mu^5}{128} + \dots   \right\} ,\\
\left.\frac{\dd\gamma_{0n} (Q^2)}{\dd Q^2}\right|_{Q^2=0}&=& \frac{e^2 g_A^2}{1440 \pi^3 f_\pi^2\, m_\pi^4} \left\{ 2 - \frac{81 \pi \mu}{8} + 94 \mu^2 - \frac{2535 \pi \mu^3 }{32}   \right.\nn\\*
&&\left.- 3 \left( 1 + 90 \log\mu\right)\mu^4+ \frac{84315 \pi \mu^5}{1024} + \dots   \right\} ,\\
\left.\frac{\dd \delta_{LTp} (Q^2)}{\dd Q^2}\right|_{Q^2=0}&=& \frac{e^2 g_A^2}{2880 \pi^3 f_\pi^2 m_\pi^4} \left\{  -\frac{5}{2} - \frac{27 \pi \mu}{32} + 20 \mu^2 - \frac{5865 \pi \mu^3}{256} \right.\nn\\*
&&\left.+  3\left( \frac{617}{4} + 36 \log\mu \right)\mu^4 - \frac{2056845 \pi \mu^5}{4096} +\dots \right\},\\
\left.\frac{\dd\delta_{LTn} (Q^2)}{\dd Q^2}\right|_{Q^2=0}&=& \frac{e^2 g_A^2}{1440 \pi^3 f_\pi^2 m_\pi^4} \left\{  -\frac{5}{4} - \frac{81 \pi \mu}{64} -11  \mu^2 + \frac{10005 \pi \mu^3}{512} \right.\nn\\*
&&\left.+  \frac{15}{8}  \left( 11 + 48 \log\mu \right)\mu^4 - \frac{267015 \pi \mu^5}{8192} +\dots \right\},\\
\left.\frac{\dd I_{Ap} (Q^2)}{\dd Q^2}\right|_{Q^2=0}&=&\frac{g_A^2}{96 \pi^2 f_\pi^2 \mu^2 } \bigg\{ 1-\frac{15\pi\mu}{4}-\frac{1}{2}\left(115+88 \log \mu\right)\mu^2+\frac{1839\pi\mu^3}{32}\nn\\*
&&+5\left(5+34 \log \mu\right)\mu^4+\dots \bigg\},\\
\left.\frac{\dd I_{An} (Q^2)}{\dd Q^2}\right|_{Q^2=0}&=&  \frac{g_A^2}{48 \pi ^2 f_\pi^2 \mu ^2}\left\{\frac{1}{2}-\frac{11\pi\mu}{8}-\frac{1}{4}\left(1+20\log\mu\right)\mu^2+\frac{99\pi\mu^3}{64}\right.\nn\\*
&&\left.-\frac{25\mu^4}{12}+\dots\right\},\\
\left.\frac{\dd I_{1p} (Q^2)}{\dd Q^2}\right|_{Q^2=0}&=&\frac{g_A^2}{96 \pi^2 f_\pi^2 \mu } \left\{ \frac{3\pi}{8}+2\left(4+3\log \mu\right)\mu-\frac{537 \pi\mu^2}{64}\right.\nn\\*
&&\left.-\frac{1}{2}\left(15+56\log \mu\right)\mu^3+\dots\right\},\\
\left.\frac{\dd I_{1n} (Q^2)}{\dd Q^2}\right|_{Q^2=0}&=& \frac{g_A^2}{48 \pi ^2 f_\pi^2 \mu  }   \left\{-\frac{\pi}{16}+\frac{1}{4}\left(3+4\log \mu\right)\mu-\frac{57\pi \mu^2}{128}+\frac{2\mu^3}{3}+\dots \right\},\\
\frac{\dd\bar \ga_{0p}(Q^2)}{\dd Q^2}\Bigg\vert_{Q^2=0}&=&\frac{e^2g_A^2}{16 \pi^3 f_\pi^2 m_\pi^6}\left\{\frac{1}{105}-\frac{23\pi \mu}{256}+\frac{377\mu^2}{210}-\frac{15551 \pi \mu^3}{6144}+\frac{3371 \mu^4 }{105}\right.\nn\\*
&&\left.-\frac{1640457 \pi \mu^5}{32768}+  \dots\right\},\\
\frac{\dd\bar \ga_{0n}(Q^2)}{\dd Q^2}\Bigg\vert_{Q^2=0}&=&\frac{e^2g_A^2}{16 \pi^3 f_\pi^2 m_\pi^6}\left\{\frac{1}{105}-\frac{153\pi \mu}{1792}+\frac{69\mu^2}{70}-\frac{4615 \pi \mu^3}{6144}+\frac{172 \mu^4 }{35}\right.\nn\\*
&&\left.-\frac{120897 \pi \mu^5}{32768}+  \dots\right\}.
\eea
\endgroup
\end{itemize}

\subsection{ $\Delta$-exchange contribution}\seclab{DeltaTreePolarizabilities}

Here, we give analytical expressions for the  tree-level $\Delta$-exchange contributions to the nucleon spin polarizabilities and their slopes at $Q^2=0$. Note that the $\Delta$-exchange  contributes equally to proton and neutron polarizabilities. 
Recall that for the magnetic $\gamma^* N \Delta $ coupling we introduced a dipole form factor to mimic vector-meson dominance: $g_M \rightarrow g_M/(1+Q^2/\Lambda^2)^2$. 

\begin{itemize}
    \item Polarizabilities at $Q^2=0$:
\bea
\gamma_0&=&-\frac{e^2}{4\pi M_+^2}\left(\frac{g_M^2}{\varDelta^2}+\frac{g_E^2}{M_+^2}-\frac{4g_M g_E}{M_+\varDelta}\right),\\
\delta_{LT}&=&\frac{e^2 M_\Delta}{4\pi M_+^3}\left(\frac{g_E^2 }{M_N M_+}+\frac{g_M g_E }{\varDelta M_N}-\frac{g_E g_C}{M_\Delta^2}\right),\\
\bar \gamma_0&=&\frac{e^2M_N^2}{\pi \varDelta^2 M_+^4}\left(-\frac{g_M^2}{\varDelta^2}+\frac{g_E^2}{M_+^2}+\frac{4g_M g_E}{\varDelta M_+}\right).
\eea

\item Slopes of polarizabilities at $Q^2=0$:

\bea
\frac{\dd \gamma_0(Q^2)}{\dd Q^2}\Bigg\vert_{Q^2=0}&=&-\frac{e^2}{\pi M_+^2 \varDelta}\left(\frac{g_M^2}{\varDelta}\left[\frac{1}{4\varDelta^2}-\frac{1}{\varDelta M_+}+\frac{1}{2 M_+^2}\right]-\frac{1}{\Lambda^2}\frac{g_M^2}{ \varDelta}+\frac{g_E^2}{2 M_+^2}\left[\frac{1}{2\varDelta}-\frac{3}{M_+}\right]\right.\nn\\
&& \left.-\frac{g_M g_E}{M_+}\left[\frac{1}{\varDelta^2}-\frac{5}{\varDelta M_+}+\frac{1}{M_+^2}\right]+\frac{1}{\Lambda^2}\frac{2g_M g_E}{ M_+}+\frac{2 g_M g_C}{\varDelta M_+^2}-\frac{g_E g_C}{M_+^3}\right),
\eea
\bea
\frac{\dd\, \delta_{LT}(Q^2)}{\dd Q^2}\Bigg\vert_{Q^2=0}&=&\frac{e^2  M_\Delta \varDelta}{4
\pi M_NM_+^2}\left(\frac{g_E^2 }{\varDelta^2 M_+^2}\left[\frac{1}{\varDelta}-\frac{4}{M_+}\right]-\frac{g_C^2}{\varDelta M_\Delta^2 M_+^2}+\frac{g_M g_E }{\varDelta^2 M_+}\left[\frac{1}{\varDelta^2} \right.\right.\nn\\
&&\left.-\frac{3}{\varDelta M_+}+\frac{1}{M_+^2}\right]-\frac{2}{ \Lambda^2}\frac{g_M g_E }{\varDelta^2M_+}+\frac{g_M g_C}{\varDelta M_\Delta^2 }\left[\frac{1}{2\varDelta^2}-\frac{2}{\varDelta M_+}+\frac{1}{2 M_+^2}\right] \nonumber \\
&&\left.-\frac{g_E g_C}{2M_\Delta^2 M_+^2}\left[\frac{7}{\varDelta}+\frac{1}{M_+}\right]\right),\\
\frac{\dd I_A(Q^2)}{\dd Q^2}\Bigg\vert_{Q^2=0}&=&-\frac{M_N^2}{M_+^2}\left(\frac{g_M^2}{2 \varDelta ^2}+\frac{g_E^2}{M_NM_+}-\frac{2 g_M g_E}{\varDelta  M_+}-\frac{g_E g_C}{M_\Delta M_+}\right),\\
\frac{\dd I_1(Q^2)}{\dd Q^2}\Bigg\vert_{Q^2=0}&=&-\frac{M_\Delta M_N^2}{2M_+^3}\left(\frac{g_E^2}{M_N M_\Delta}-\frac{g_M g_E}{ \varDelta  M_N}
-\frac{g_E g_C}{M_\Delta^2 }\right),\\
\frac{\dd \bar \ga_0(Q^2)}{\dd Q^2}\Bigg\vert_{Q^2=0}&=&\frac{e^2 M_N^2}{\pi \varDelta^3 M_+^6}\left(\frac{g_M^2}{\varDelta}\left[\frac{3M_\Delta^2+2M_\Delta M_N-9M_N^2}{\varDelta^2}+\frac{4M_+^2}{\Lambda^2}\right]\right.\nn\\
&&-4g_M g_E \left[\frac{5M_\Delta^2-9M_N^2}{M_+ \varDelta^2}+\frac{2M_+}{\Lambda^2}\right]+\frac{g_E^2}{\varDelta M_+}\left[7M_\Delta-9M_N\right] \nn\\
&&\left.-\frac{8g_M g_C}{\varDelta}+\frac{4g_E g_C}{M_+}\right).
\eea
\end{itemize}

\small
\bibliography{lowQ}

\end{document}